\documentclass[aps,twocolumn,letterpaper,superscriptaddress]{revtex4-1}

\usepackage{amsmath}
\usepackage{amssymb}
\usepackage{amsfonts}
\usepackage{graphicx}
\usepackage{dcolumn}
\usepackage{float}
\usepackage{bm}
\usepackage{mathrsfs}
\usepackage{tabularx}
\usepackage{bigstrut}
\usepackage{epstopdf}
\usepackage{xfrac}
\usepackage{listings}

\usepackage{balance}

\usepackage[usenames,dvipsnames]{xcolor}
\usepackage[hyperindex,pdftex, breaklinks,colorlinks = true,linkcolor = blue,urlcolor=blue,citecolor=blue]{hyperref}

\usepackage{caption}
\usepackage{subcaption}
\usepackage{physics}
\usepackage[section]{placeins}

\begin{document}

\title{Quantum Interference Control of Localized Carrier Distributions in the Brillouin Zone}

\author{ Perry T. Mahon }
\email{pmahon@physics.utoronto.ca}
\affiliation{ Department of Physics, University of Toronto, Toronto, Ontario M5S 1A7, Canada }

\author{ Rodrigo A. Muniz }
\email{rodrigo.a.muniz@gmail.com}

\affiliation{ Department of Physics, University of Toronto, Toronto, Ontario M5S 1A7, Canada }
\affiliation{ Department of Electrical Engineering and Computer Science, University of Michigan, Ann Arbor, MI 48109, USA}

\author{ J. E. Sipe }
\email{sipe@physics.utoronto.ca}

\affiliation{ Department of Physics, University of Toronto, Toronto, Ontario M5S 1A7, Canada }

\date{\today }

\begin{abstract}
Using transition-metal dichalcogenides as an example, we show that
the quantum interference arising in two- and three-photon absorption processes can lead to controllable, highly localized carrier distributions in the Brillouin zone. We contrast this with the previously studied one- and two-photon absorption, and find qualitatively different features, including changes in the relevance of interband and intraband processes according to the excitation energy. Thus, the distribution of excitations arising under certain circumstances in two- and three-photon absorption can facilitate the study of far-from-equilibrium states that are initially well localized in crystal momentum space.
\end{abstract}

\maketitle

\section{Introduction}
Although nearly every technological device is based on systems in far-from-equilibrium states, our understanding of the properties of materials in such a regime is limited \cite{Bertini15, Kemper18}. 
This is the case even for intensively investigated materials such as semiconductors, which are the basis of digital technology. 
One of the main impediments in the study of materials far-from-equilibrium is the difficulty in creating quantum excitations in a controlled way. 
For example, we lack good methods for creating one of the simplest types of electronic excitation in a gapped material: the excitation of an electron from a valence to a conduction band at a given crystal momentum. A simple way to excite electrons from one band to another is via optical fields, but they usually excite carriers in almost every location of the Brillouin zone where the photon energy matches the energy difference of the electronic bands. 
However, because of quantum interference effects, optical fields of different frequencies can be used in combination to excite carriers with more local distributions in the Brillouin zone. 

Quantum interference between distinct processes that result in the same transition can lead to localized electronic excitations in crystal momentum space, as the different
processes interfere constructively in some parts of the Brillouin
zone and destructively in others. This interference can be 
manipulated by varying physically tunable parameters characterizing the excitation, such as a relative phase parameter of the optical fields and their intensities. Polar distributions of injected
carriers in the Brillouin zone lead to charge currents, and such ``injected
currents'' due to the interference of one- and two-photon absorption
processes (``1+2'' injection) have been observed in bulk \cite{Atanasov96,Hache97,Rioux12,Bas15,Bas16}
and two-dimensional materials \cite{Sun10,Rioux11,Rao12,Cui15}. The analogous injection of spin currents
has been detected in bulk semiconductors \cite{Bhat00,Stevens02,Stevens03,Hubner03,Zhao06}, and proposed in topological insulators \cite{Muniz14}; the injection of spin and valley
currents in transition-metal dichalcogenides (TMDs) has also been proposed \cite{Muniz15}.

This quantum interference control (QuIC) of carriers has been exploited
to determine the carrier-envelope phase of pulses short enough to contain
both the fundamental frequency and its second harmonic \cite{Roos04,Fortier04}; semiconductors 
with a relatively large band gap are of interest here to allow room
temperature operation. 
Recently the use of interference between two-
and three-photon absorption processes to inject currents in semiconductors
has also been studied both experimentally \cite{Cundiff18} and theoretically
\cite{Muniz18}. This ``2+3'' injection is of special interest for determining
the carrier-envelope phase of short pulses, since it can be used even
if the frequency spread of the short pulse does not span an octave. 
Coherent optical frequency combs can be used for studying even more general ``$N+M$'' QuIC of carriers in gapped materials.  

These interference processes have typically been studied by observing the net current they generate, either directly with the use of electrodes \cite{Hache97} or indirectly by the detection of the terahertz radiation resulting from the excitation and subsequent decay of that current \cite{Sun10}. These detection schemes are sensitive only to the first moment of the carrier distribution in the Brillouin
zone, and that is all that has been typically calculated. However,
recent advances in time-resolved angle-resolved photoemission spectroscopy (ARPES) 
\cite{Smallwood16} offer the promise
of detecting carrier distributions in the Brillouin zone as a function
of time. This would yield unparalleled insight into the relaxation
processes of such excitation distributions injected by QuIC, where carriers can be placed in
regions of the Brillouin zone far from those occupied by equilibrium
or near-equilibrium carrier distributions. It should also be convenient 
to use QuIC as part of pump--probe experiments, with QuIC used either to place carriers in a region of the Brillouin zone, or to detect them in a region of the Brillouin zone, or both. So detailed theoretical studies of the injected carrier distributions, and ultimately of their subsequent dynamics, are now in order.

In this first paper along these lines, we study the initial $\boldsymbol{k}$-space
carrier distributions in the TMD WSe$_2$ 
due to 1+2, 1+3, and 2+3 injection. An important result is that 2+3 injection can 
lead to more localized carrier distributions in the Brillouin zone than
1+2 injection, moving further towards the goal of coherent control
strategies that act as effective ``tweezers in the Brillouin zone''
for the placement of carriers where desired. The outline of the paper
is as follows. We begin in Sec. \ref{sec:2} by introducing a generic single-particle Hamiltonian, 
where the vector potential is included via minimal coupling, and derive
expressions for the first-, second-, and third-order perturbative coefficients. 
Using these coefficients we illustrate the origin of QuIC. In Sec. \ref{sec:3}  we 
introduce the quantities of interest, namely the 
electron excitation (carrier injection) rate, and the current injection rate. Following this,
in Sec. \ref{sec:4} we introduce a model Hamiltonian for TMDs and use this system
as a platform to compare features of different orders of photon absorption processes. We then analyze our findings: Sec. \ref{sec:5} contains the $\boldsymbol{k}$-space distributions of electronic excitations for various polarizations of incident light, and in Sec. \ref{sec:6} we plot the dependence of the carrier and current response tensors on the excitation energy. 

\section{Optical injection rates}
\label{sec:2}
We investigate the optical excitation of electrons by way of time-dependent 
perturbation theory (TDPT), using a second quantized Hamiltonian $\mathcal{H}(t)$ 
that follows from a single-particle Hamiltonian of the form 
\begin{align}
\mathscr{H}(\boldsymbol{x},\boldsymbol{p};t)&=\frac{1}{2m}\big(\boldsymbol{p}-e\boldsymbol{A}(t)\big)^{2}\label{eq:one-particle_H}\\
&\qquad+\mathscr{H}_{SO}\big(\boldsymbol{x},\boldsymbol{p}-e\boldsymbol{A}(t)\big)+\mathscr{V}_{lat}(\boldsymbol{x}),\nonumber
\end{align}
where $e=-\left|e\right|$ is the electronic charge, $\boldsymbol{x}$ and 
$\boldsymbol{p}$ are position and momentum operators, $\mathscr{H}_{SO}$
is the spin-orbit term, and $\mathscr{V}_{lat}(\boldsymbol{x})$ is
the periodic lattice potential energy. We have chosen a gauge in which the time-dependent 
scalar potential vanishes, and have assumed the external
electromagnetic field can be approximated as uniform, with an electric
field $\boldsymbol{E}(t)$ described solely by the vector potential
$\boldsymbol{A}(t)$. We only consider electrons and holes injected
at high enough energies to lead to currents, and so bound exciton
states are not relevant. The electron-hole interaction in the ionized
excitons that result can lead to phase shifts in the injected currents
at excitation energies close to the band gap \cite{Bhat05}; we ignore those
here, as well as other effects of electron-electron interactions.
The $O\big(A(t)^2\big)$ term arising in (\ref{eq:one-particle_H}) is solely a 
function of time and adds a global phase to the energy eigenstates, 
which has no consequence on the expectation values we compute. 
Then (\ref{eq:one-particle_H}) can be written as $\mathscr{H}_0+\mathscr{V}_{ext}(t)$, where $\mathscr{H}_0$ is the unperturbed Hamiltonian and the interaction term takes the form $\mathscr{V}_{ext}(t)=-e\boldsymbol{\mathfrak{v}}\cdot\boldsymbol{A}(t)$,
where $\boldsymbol{\mathfrak{v}}=i\hbar^{-1}\left[\mathscr{H}_0(\boldsymbol{x},\boldsymbol{p}),\boldsymbol{x}\right]$ is the velocity operator; this holds for any unperturbed single-particle Hamiltonian that is at most quadratic in the momentum. The only experimentally accessible parameters within $\mathcal{H}(t)$ enter through this external interaction; namely the intensity, polarization, and phase of the optical fields. It is therefore these parameters that can then be varied to tune the quantum interference between excitation processes.

The second quantized Hamiltonian can be written, in the Schrodinger picture, as $\mathcal{H}(t)=\mathcal{H}_{0}+\mathcal{V}_{ext}(t)$, where 
\begin{equation}
\begin{aligned}
\mathcal{H}_{0}&=\underset{n\boldsymbol{k}}{\sum} \hbar\omega_{n}(\boldsymbol{k})a_{n\boldsymbol{k}}^{\dagger}a_{n\boldsymbol{k}},\\
\mathcal{V}_{ext}(t)&=\underset{nm\boldsymbol{k}}{\sum}\mathscr{V}_{nm}(\boldsymbol{k},t)a_{n\boldsymbol{k}}^{\dagger}a_{m\boldsymbol{k}}.
\end{aligned}
\label{generalH}
\end{equation}
The crystal momentum wave vectors, $\boldsymbol{k}$, are summed over the first Brillouin zone, $n$ and $m$ label bands, and $\mathscr{V}_{nm}(\boldsymbol{k},t)=-e\boldsymbol{\mathfrak{v}}_{nm}(\boldsymbol{k})\cdot\boldsymbol{A}(t),$
with $\boldsymbol{\mathfrak{v}}_{nm}(\boldsymbol{k})\equiv\left\langle n\boldsymbol{k}|\boldsymbol{\mathfrak{v}}|m\boldsymbol{k}\right\rangle$.
We express (\ref{generalH}) in the basis of eigenstates of $\mathcal{H}_0$, where $\ket{n\boldsymbol{k}}\equiv a_{n\boldsymbol{k}}^\dagger\ket{vac}$ indicates a Bloch state
of band $n$ with crystal momentum $\hbar\boldsymbol{k}$ and energy
$\hbar\omega_{n}(\boldsymbol{k})$, and $\{a_{n\boldsymbol{k}}\}$ are the fermionic electron operators. To describe the applied optical fields, we take a vector potential of the form 
\begin{equation}
\begin{aligned}
\boldsymbol{A}(t)=\underset{\alpha}{\sum}\boldsymbol{A}_{\omega_{\alpha}}e^{-i(\omega_{\alpha}+i\epsilon)t}=-\underset{\alpha}{\sum}\frac{i}{\omega_{\alpha}}\boldsymbol{E}_{\omega_{\alpha}}e^{-i(\omega_{\alpha}+i\epsilon)t},
\end{aligned}
\end{equation}
where $\epsilon\rightarrow0^{+}$ describes the adiabatic turning on of the 
fields from $t=-\infty$, and for $N+M$ injection we sum over frequencies
$\omega_{\alpha}=\pm\Omega/N, \pm\Omega/M$,
where $\hbar\Omega$ identifies the total transition energy. To keep
track of the relative phases and the polarizations of the different
frequency components we write $\boldsymbol{E}_{\omega_{\alpha}}=E_{\omega_{\alpha}}e^{i\phi_{\omega_{\alpha}}}\boldsymbol{\hat{e}}_{\omega_{\alpha}}$,
where $E_{\omega_{\alpha}}$and $\phi_{\omega_{\alpha}}$are real valued
and $\boldsymbol{\hat{e}}_{\omega_{\alpha}}$ is a polarization vector satisfying
$\boldsymbol{\hat{e}}_{\omega_{\alpha}}^{*}\cdot\boldsymbol{\hat{e}}_{\omega_{\alpha}}=1$. 

The implementation of perturbation theory for problems of this type
has been previously discussed \cite{SipeBook,Rioux12}; here we simply summarize the
approach. We move to the interaction picture such that all operators evolve under $\mathcal{H}_0$, thus $a_{n\boldsymbol{k}}^{\dagger}(t)=e^{i\omega_n(\boldsymbol{k})t}a_{n\boldsymbol{k}}^{\dagger}$ and $a_{n\boldsymbol{k}}(t)=e^{-i\omega_n(\boldsymbol{k})t}a_{n\boldsymbol{k}}$, and all states evolve under the time-evolution operator
\begin{equation}
\begin{aligned}
\label{evolutionOperator}
\mathcal{U}(t)=1+\sum\limits_{N=1}^{\infty}\int\limits_{\text{   }-\infty}^{t}\frac{dt_N}{i\hbar}\mathcal{V}_{I}(t_N)\cdots\int\limits_{\text{   }-\infty}^{t_{2}}\frac{dt_1}{i\hbar}\mathcal{V}_{I}(t_1),
\end{aligned}
\end{equation}
where $\mathcal{V}_{I}(t)\equiv e^{i\mathcal{H}_0t/\hbar}\mathcal{V}_{ext}(t)e^{-i\mathcal{H}_0t/\hbar}$. Using the completeness relation for a multi-particle Hilbert space, a state $\left|\psi(t)\right\rangle$ can be written, as shown in Appendix \ref{PT}, as
\begin{equation}
\begin{aligned}
\left|\psi(t)\right\rangle=\mathcal{U}(t)\left|gs\right\rangle=\gamma_{0}(t)\left|gs\right\rangle +\underset{cv\boldsymbol{k}}{\sum}\gamma_{cv}(\boldsymbol{k},t)\left|cv\boldsymbol{k}\right\rangle +...
\end{aligned}
\end{equation}
where $\left|gs\right\rangle $ denotes the ground state of the unperturbed Hamiltonian, in which all
conduction bands (labeled $c$) are empty and all valence bands (labeled
$v$) are occupied, and $\left|cv\boldsymbol{k}\right\rangle \equiv a_{c\boldsymbol{k}}^{\dagger}a_{v\boldsymbol{k}}\left|gs\right\rangle$. Taking the applied optical field to be an ultrashort pulse, we consider only the contribution of single electron-hole pairs to the electronic charge and current density injection rates \cite{Rioux12}. These rates can can be found via the coefficients
\begin{equation}
\gamma_{cv}(\boldsymbol{k},t)=\left\langle cv\boldsymbol{k}|\mathcal{U}(t)|gs\right\rangle
\end{equation}
(see Appendix \ref{PT}). The perturbative calculation gives
\begin{equation}
 \gamma_{cv}(\boldsymbol{k},t)=\mathcal{R}_{cv}(\boldsymbol{k})\frac{e^{-i(\Omega-\omega_{cv}(\boldsymbol{k})+i\epsilon)t}}{\Omega-\omega_{cv}(\boldsymbol{k})+i\epsilon},
\end{equation}
where $\omega_{cv}(\boldsymbol{k})\equiv\omega_{c}(\boldsymbol{k})-\omega_{v}(\boldsymbol{k})$ such that $\hbar\omega_{cv}(\boldsymbol{k})$ is the energy difference between the bands $c$ and $v$ at $\boldsymbol{k}$, and 
\begin{equation}
\begin{aligned}
\mathcal{R}_{cv}(\boldsymbol{k})\equiv\sum_{N}\mathcal{R}_{cv}^{(N)}(\boldsymbol{k}).
\end{aligned}
\end{equation}
These $\mathcal{R}_{cv}^{(N)}(\boldsymbol{k})$ are transition amplitudes arising at different order, $N$, of TDPT and are of the form 
\begin{equation} 
 \mathcal{R}_{cv}^{(N)}(\boldsymbol{k})=\underset{\substack{a,...b;\alpha...\beta}}{\sum}R_{cv}^{(N)a\cdots b}(\boldsymbol{k};\omega_{\alpha},\ldots,\omega_{\beta})E_{\omega_{\alpha}}^{a}\cdots E_{\omega_{\beta}}^{b},\label{eq:Rscript_amplitudes}
\end{equation}
where the sum is over both frequencies and Cartesian components, the
latter indicated by superscript indices; at $N$th order of TDPT, 
there are $N$ frequencies in the list $(\omega_{\alpha},\ldots,\omega_{\beta})$ 
that add to $\Omega$, and $N$ Cartesian components in the list $a\cdots b$ that are accompanied by $N$ factors of the applied fields $E$. Here we are interested
in the lowest order amplitudes $\mathcal{R}_{cv}^{(1)}(\boldsymbol{k})$, $\mathcal{R}_{cv}^{(2)}(\boldsymbol{k})$,
and $\mathcal{R}_{cv}^{(3)}(\boldsymbol{k})$, which we identify with one-, two-, and three-photon absorption, respectively. We find
\begin{widetext}
\begin{align}  &R_{cv}^{\left(1\right)a}(\boldsymbol{k};\omega_{\alpha})=\dfrac{ie}{\hbar\omega_{\alpha}}\mathfrak{v}_{cv}^{a}(\boldsymbol{k}),\label{eq:R1coef-1} \\
&R_{cv}^{\left(2\right)ab}(\boldsymbol{k};\omega_{\alpha},\omega_{\beta})=-\dfrac{e^{2}}{\hbar^{2}\omega_{\alpha}\omega_{\beta}}\left(\underset{c^{\prime}}{\sum}\dfrac{\mathfrak{v}_{cc^{\prime}}^{a}(\boldsymbol{k})\mathfrak{v}_{c^{\prime}v}^{b}(\boldsymbol{k})}{\omega_{\beta}-\omega_{c^{\prime}v}(\boldsymbol{k})}-\underset{v^{\prime}}{\sum}\dfrac{\mathfrak{v}_{cv^{\prime}}^{b}(\boldsymbol{k})\mathfrak{v}_{v^{\prime}v}^{a}(\boldsymbol{k})}{\omega_{\beta}-\omega_{cv^{\prime}}(\boldsymbol{k})}\right),\label{eq:R2coef-1}\\
&R_{cv}^{\left(3\right)abd}(\boldsymbol{k};\omega_{\alpha},\omega_{\beta},\omega_{\delta})=
\dfrac{ie^{3}}{\hbar^{3}\omega_{\alpha}\omega_{\beta}\omega_{\delta}}\Bigg[\underset{c^{\prime}}{\sum}\dfrac{\mathfrak{v}_{cc^{\prime}}^{a}(\boldsymbol{k})}{\omega_{\alpha}-\omega_{cc^{\prime}}(\boldsymbol{k})}\left(\underset{c^{\prime\prime}}{\sum}\dfrac{\mathfrak{v}_{c^{\prime}c^{\prime\prime}}^{b}(\boldsymbol{k})\mathfrak{v}_{c^{\prime\prime}v}^{d}(\boldsymbol{k})}{\omega_{\delta}-\omega_{c^{\prime\prime}v}(\boldsymbol{k})}-\underset{v^{\prime}}{\sum}\dfrac{\mathfrak{v}_{c^{\prime}v^{\prime}}^{d}(\boldsymbol{k})\mathfrak{v}_{v^{\prime}v}^{b}(\boldsymbol{k})}{\omega_{\delta}-\omega_{c^{\prime}v^{\prime}}(\boldsymbol{k})}\right)\label{eq:R3coef-1}\\
&\qquad\qquad\qquad\qquad\qquad\qquad-\underset{v^{\prime}}{\sum}\left(\underset{c^{\prime}}{\sum}\dfrac{\mathfrak{v}_{cc^{\prime}}^{b}(\boldsymbol{k})\mathfrak{v}_{c^{\prime}v^{\prime}}^{d}(\boldsymbol{k})}{\omega_{\delta}-\omega_{c^{\prime}v^{\prime}}(\boldsymbol{k})}-\underset{v^{\prime\prime}}{\sum}\dfrac{\mathfrak{v}_{cv^{\prime\prime}}^{d}(\boldsymbol{k})\mathfrak{v}_{v^{\prime\prime}v^{\prime}}^{b}(\boldsymbol{k})}{\omega_{\delta}-\omega_{cv^{\prime\prime}}(\boldsymbol{k})}\right)\dfrac{\mathfrak{v}_{v^{\prime}v}^{a}(\boldsymbol{k})}{\omega_{\alpha}-\omega_{v^{\prime}v}(\boldsymbol{k})}\nonumber\\
&\qquad\qquad\qquad\qquad\qquad\qquad-\underset{c^{\prime}v^{\prime}}{\sum}\left(\dfrac{\mathfrak{v}_{cv^{\prime}}^{b}(\boldsymbol{k})\mathfrak{v}_{v^{\prime}c^{\prime}}^{a}(\boldsymbol{k})\mathfrak{v}_{c^{\prime}v}^{d}(\boldsymbol{k})}{\left(\omega_{\alpha}-\omega_{v^{\prime}c^{\prime}}(\boldsymbol{k})\right)\left(\omega_{\delta}-\omega_{c^{\prime}v}(\boldsymbol{k})\right)}+\dfrac{\mathfrak{v}_{cv^{\prime}}^{d}(\boldsymbol{k})\mathfrak{v}_{v^{\prime}c^{\prime}}^{a}(\boldsymbol{k})\mathfrak{v}_{c^{\prime}v}^{b}(\boldsymbol{k})}{\left(\omega_{\delta}-\omega_{cv^{\prime}}(\boldsymbol{k})\right)\left(\omega_{\alpha}-\omega_{v^{\prime}c^{\prime}}(\boldsymbol{k})\right)}\right)\Bigg].\nonumber
\end{align}
\end{widetext}

\begin{figure}[b!]
	\centering
	\begin{subfigure}{0.157\textwidth}
		\centering
		\includegraphics[width=1\textwidth]{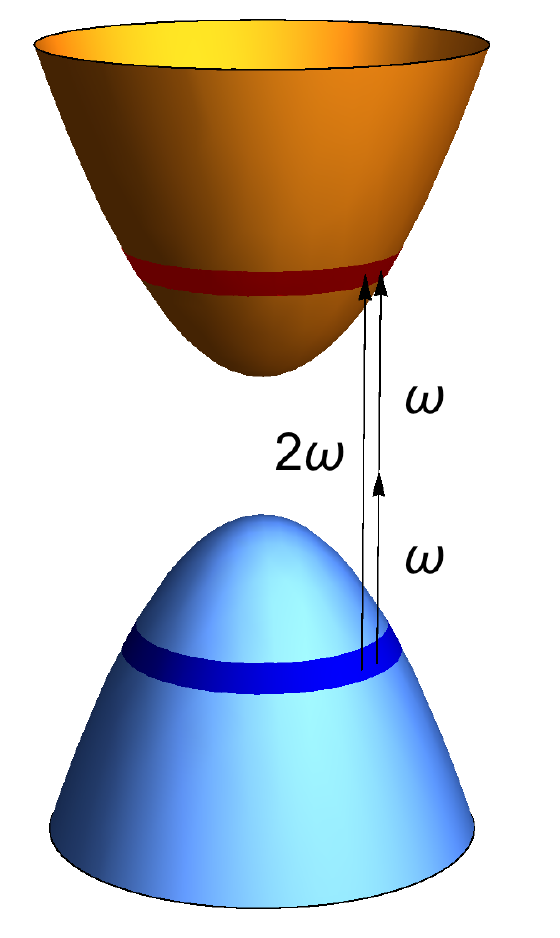}
		\caption{1+2 PA}
		\label{12Cartoon}
	\end{subfigure} %
	\begin{subfigure}{0.157\textwidth}
		\centering
		\includegraphics[width=1\textwidth]{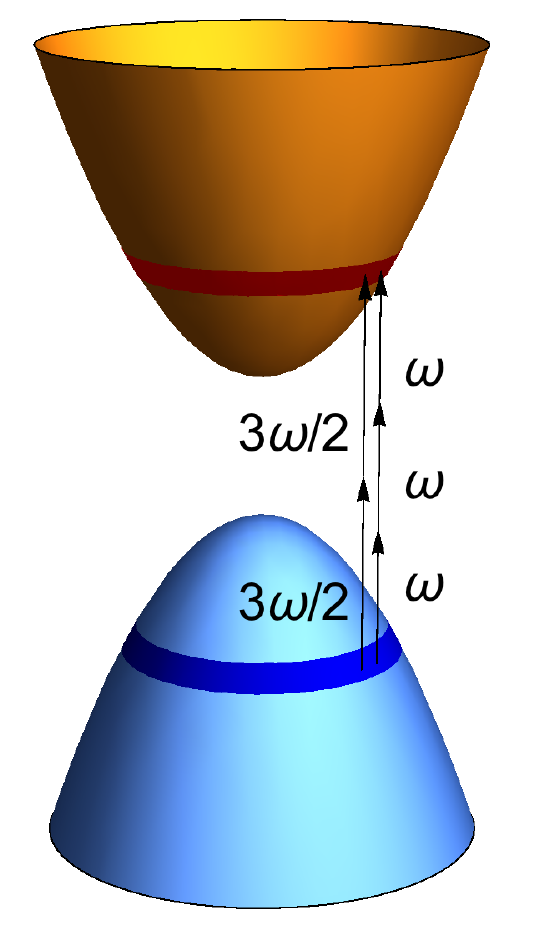}
		\caption{2+3 PA}
		\label{23Cartoon}
	\end{subfigure}
		\centering
	\begin{subfigure}{0.157\textwidth}
		\centering
		\includegraphics[width=1\textwidth]{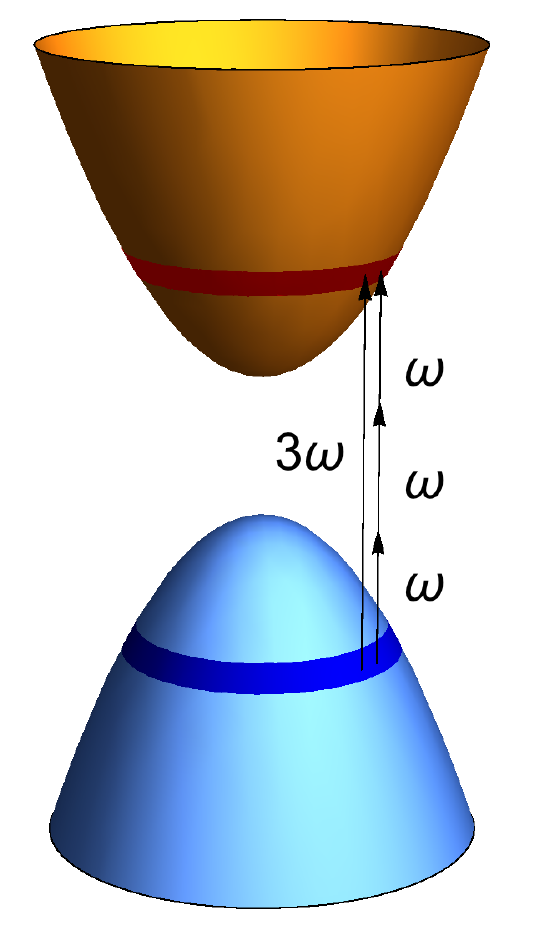}
		\caption{1+3 PA}
		\label{13Cartoon}
	\end{subfigure} %
	\caption{Schematic of the N+M photon absorption (PA) processes considered. We adopt the notation $\Omega=M\omega$.}
	\label{PACartoon}
\end{figure}
Illustrated in Fig.~\ref{PACartoon} are the pairs of simultaneous absorption processes we consider. The lowest-order terms comprising $\mathcal{R}_{cv}(\boldsymbol{k})$ that give nonvanishing contributions to the charge and current density injection rates, for applied fields containing particular frequencies, are those amplitudes $\mathcal{R}_{cv}^{(N)}(\boldsymbol{k})$ associated with each process in the pair. For instance, consider an applied field containing frequencies $\omega$ and $3\omega/2$ with corresponding energies less than the band gap energy. We say that such an applied field mediates 2+3 PA because, although $R_{cv}^{(1)a}(\boldsymbol{k};\omega)\neq0$ and $R_{cv}^{(1)a}(\boldsymbol{k};3\omega/2)\neq0$, these terms will not lead to injected carriers or currents; as it relates to the charge and current density injection rates, $\mathcal{R}_{cv}(\boldsymbol{k})$ is effectively $\mathcal{R}^{(2)}_{cv}(\boldsymbol{k})$+$\mathcal{R}^{(3)}_{cv}(\boldsymbol{k})$, as illustrated in Fig.~\ref{23Cartoon}. In our calculation below, we include only the dominant contributions to these amplitudes; for example, when investigating 1+2 PA, illustrated in Fig.~\ref{12Cartoon}, we neglect 2 PA involving a photon of energy $\hbar\omega$ and another of energy $2\hbar\omega$. This is justified because, in order for 2 PA mediated entirely by light of frequency $\omega$ to excite the same magnitude of carriers as 1 PA mediated by light of frequency $2\omega$, as required for QuIC, necessarily $E_{\omega}\gg E_{2\omega}$. From (\ref{eq:Rscript_amplitudes}) it is then clear that 2 PA by $\hbar\omega+2\hbar\omega$ is insignificant in comparison to 2 PA by $\hbar\omega+\hbar\omega$, and we include only the latter in the $\mathcal{R}_{cv}^{(2)}(\boldsymbol{k})$ implemented below.

\section{Carriers and Currents}
\label{sec:3}
We are interested in the injection rates of both conduction electrons
and current density arising from the distribution of electronic excitations in the Brillouin
zone just after the external fields have been removed. In the interaction picture, the operators
corresponding to the density of electrons in the conduction bands and the total electronic current density are given by
\begin{equation}
\begin{aligned}
n_{c}&=\frac{1}{L^{D}}\sum\limits _{c\boldsymbol{k}}a_{c\boldsymbol{k}}^{\dagger}a_{c\boldsymbol{k}},\\
\boldsymbol{\mathcal{J}}(t)&=\frac{1}{L^{D}}\sum\limits _{nm\boldsymbol{k}} e\boldsymbol{\mathfrak{v}}_{nm}(\boldsymbol{k})a_{n\boldsymbol{k}}^{\dagger}(t)a_{m\boldsymbol{k}}(t),
\end{aligned}
\label{conserved}
\end{equation}
respectively, where $L$ is a normalization length, $D$ is the spatial dimension of the
system, and in the latter term $n$ and $m$ range over all bands. 
To investigate the distribution of electronic excitations in crystal momentum space arising from various photon absorption processes, we resolve the injection rates through the Brillouin zone; for a $\boldsymbol{k}$-conserving, single-body operator $\mathcal{M}(t)$, which here will be the density $n_{c}$ or $\boldsymbol{\mathcal{J}}(t)$, we always have
\begin{equation} \begin{aligned}
\frac{d}{dt}\bra{\psi(t)}\mathcal{M}(t)\ket{\psi(t)}=\int_{BZ}\frac{d^Dk}{\left(2\pi\right)^{D}} \frac{d}{dt}\left\langle \mathcal{M}\left(\boldsymbol{k};t\right)\right\rangle,\label{eq:density_rewrite}
\end{aligned} \end{equation}
where $\frac{d}{dt}\left\langle\mathcal{M}\left(\boldsymbol{k};t\right)\right\rangle$ is the ``injected density rate'' in the Brillouin zone associated with the operator $\mathcal{M}$, in the limit of continuous $\boldsymbol{k}$ (Appendix \ref{PT}). It will be these
injected density rates that we later plot, primarily $\frac{d}{dt}\left\langle n_c(\boldsymbol{k})\right\rangle_{N+M}$, and are the main focus of this paper. 

Under our approximations $\frac{d}{dt}\left\langle\mathcal{M}\left(\boldsymbol{k};t\right)\right\rangle$ is independent of time and, in general, can be expressed as
\begin{align}
&\frac{d}{dt}\left\langle \mathcal{M}\left(\boldsymbol{k};t\right)\right\rangle=\label{eq:density_rewrite2}\\
&\underset{\substack{a,...b;N\\a',...b';N'}}{\sum}\mu_{NN'}^{a'\cdots b'a\cdots b}(\Omega;\boldsymbol{k})E_{-\Omega/N'}^{a'}\cdots E_{-\Omega/N'}^{b'}E_{\Omega/N}^{a}\cdots E_{\Omega/N}^{b},\nonumber
\end{align}
where we refer to $\mu_{NN'}^{a'\cdots b'a\cdots b}(\Omega;\boldsymbol{k})$ as the ``response coefficient density'' associated with $\mathcal{M}$ (for a particular absorption process). The general form of this object is given in Appendix \ref{PT}. As previously mentioned, we include only the dominant contributions to $\mathcal{R}^{(N)}_{cv}(\boldsymbol{k})$ as dictated by the conditions for QuIC; this restricts the optical frequencies included in (\ref{eq:density_rewrite2}). Furthermore, introducing response coefficients for the injection rate associated with $\mathcal{M}$ as
\begin{align}
&\frac{d}{dt}\expval{\mathcal{M}(t)}{\psi(t)}=\label{responseCoeff}\\
&\underset{\substack{a,...b;N\\a',...b';N'}}{\sum}\mu_{NN'}^{a'\cdots b'a\cdots b}(\Omega)E_{-\Omega/N'}^{a'}\cdots E_{-\Omega/N'}^{b'}E_{\Omega/N}^{a}\cdots E_{\Omega/N}^{b},\nonumber
\end{align} 
we find the response coefficient for a particular absorption process, $\mu_{NN'}^{a'\cdots b'a\cdots b}(\Omega)$, can be written as
\begin{align}
\mu_{NN'}^{a'\cdots b'a\cdots b}(\Omega)=\int_{BZ}\frac{d^Dk}{\left(2\pi\right)^{D}}\mu_{NN'}^{a'\cdots b'a\cdots b}(\Omega;\boldsymbol{k}).
\end{align}

\section{The two-band limit}
The model we later adopt is effectively a two-band model for electrons of a given spin about each valley; this simplifies (\ref{eq:R1coef-1}--\ref{eq:R3coef-1}) as well as the expressions for the electronic charge and current density injection rates (see Appendix \ref{PT}). Proceeding to take the limit $\epsilon\rightarrow0^+$ \cite{SipeBook}
\begin{equation}
	\begin{aligned}
		\frac{d}{dt}\big|\gamma_{cv}(\boldsymbol{k},t)\big|^{2}\rightarrow2\pi\big|\mathcal{R}_{cv}(\boldsymbol{k})\big|^{2}\delta\big(\Omega-\omega_{cv}(\boldsymbol{k})\big),
		\label{epsilonLimit}
	\end{aligned}
\end{equation}
which gives the rate of injection of electron-hole pairs at $\boldsymbol{k}$. It is the interference in this expression between the different $\mathcal{R}_{cv}^{(N)}(\boldsymbol{k})$
that contribute to $\mathcal{R}_{cv}(\boldsymbol{k})$ that allows the
possibility of quantum interference control; this is true even outside of this limit. We note that as the $\mathcal{R}_{cv}^{(N)}(\boldsymbol{k})$
coefficients are always accompanied by $\delta\big(\Omega-\omega_{cv}(\boldsymbol{k})\big)$
in the expression for the response, the substitution $\omega_{\alpha}+\omega_{\beta}+\omega_{\delta}=\omega_{cv}(\boldsymbol{k})$ can be made in $R_{cv}^{\left(3\right)abd}(\boldsymbol{k})$; in fact, this has been used to simplify (\ref{eq:R3coef-1}). 

The form of the response coefficient densities are simplified in this limit, and are defined to be (Appendix \ref{PT})
\begin{align}
\mu_{NN'}^{a'\cdots b'a\cdots b}(\Omega;\boldsymbol{k})&\equiv 2\pi\left(R^{(N')a'\cdots b'}_{cv}(\boldsymbol{k};\Omega/N',...,\Omega/N')\right)^*\nonumber\\
&\times R^{(N)a\cdots b}_{cv}(\boldsymbol{k};\Omega/N,...,\Omega/N)\label{responseCoeffDensity}\\
&\times L^D\bra{cv\boldsymbol{k}}\mathcal{M}\ket{cv\boldsymbol{k}}\delta\big(\Omega-\omega_{cv}(\boldsymbol{k})\big)\nonumber,
\end{align}
where $\mathcal{M}$ is an operator in the Schrodinger picture. An implication of the two-band limit is that only diagonal matrix elements of $\mathcal{M}$ appear in (\ref{responseCoeffDensity}); if more bands were to be included, off-diagonal elements would also appear.

\subsection{1+2 photon absorption}

Consider a system with a direct band gap $E_{g}$ as indicated
in Fig.~\ref{12Cartoon}, with an incident optical field composed of frequencies
$\omega$ and $2\omega$, where $\hbar\omega<E_{g}<2\hbar\omega$.
In such a system a total energy of at least $E_{g}$ is required to
excite electrons from valence to conduction bands, which can be satisfied
both by absorption of a single photon of energy $2\hbar\omega$ and
by the absorption of two photons each of energy $\hbar\omega$. We illustrate 
below how the interference of these two excitation pathways
gives rise to an injected current density in the system. Contributions
to the total injection rate corresponding to the conduction electron density, $\frac{d}{dt}\left\langle n_{c}\right\rangle _{1+2}$, are of the form
\begin{equation} \begin{aligned}
\frac{d}{dt}\left\langle n_{c}\right\rangle _{1}&=\xi_{1}^{ab}\left(2\omega\right)E_{-2\omega}^{a}E_{2\omega}^{b},\label{eq:1+2carriers_full}\\
\frac{d}{dt}\left\langle n_{c}\right\rangle _{1+2;i}&=\xi_{1+2}^{abd}\left(2\omega\right)E_{-\omega}^{a}E_{-\omega}^{b}E_{2\omega}^{d}+c.c., \\
\frac{d}{dt}\left\langle n_{c}\right\rangle _{2}&=\xi_{2}^{abde}\left(2\omega\right)E_{-\omega}^{a}E_{-\omega}^{b}E_{\omega}^{d}E_{\omega}^{e},
\end{aligned}\end{equation} 
as given in (\ref{responseCoeff}). In the above and below, the subscript $i$ is used in $\frac{d}{dt}\left\langle n_{c}\right\rangle _{N+M;i}$ to denote a contribution to the total rate that arises \textit{solely} from the interference of excitation processes. The contributions to the total injected density rate associated with each of the above contributions to the total injection rate then have the form
\begin{equation}
\begin{aligned}
\frac{d}{dt}\left\langle n_{c}(\boldsymbol{k})\right\rangle _{1}&=\xi_{1}^{ab}(2\omega;\boldsymbol{k})E_{-2\omega}^{a}E_{2\omega}^{b}, \\
\frac{d}{dt}\left\langle n_{c}(\boldsymbol{k})\right\rangle _{1+2;i}&=\xi_{1+2}^{abd}\left(2\omega;\boldsymbol{k}\right)E_{-\omega}^{a}E_{-\omega}^{b}E_{2\omega}^{d}+c.c., \\
\frac{d}{dt}\left\langle n_{c}(\boldsymbol{k})\right\rangle _{2}&=\xi_{2}^{abde}(2\omega;\boldsymbol{k})E_{-\omega}^{a}E_{-\omega}^{b}E_{\omega}^{c}E_{\omega}^{d}, 
\end{aligned}
\end{equation}
and the total injected density rate is given, from (\ref{eq:density_rewrite2}), by 
\begin{align}
&\frac{d}{dt}\left\langle n_{c}(\boldsymbol{k})\right\rangle _{1+2}=\\
&\qquad\frac{d}{dt}\left\langle n_{c}(\boldsymbol{k})\right\rangle _{1}+\frac{d}{dt}\left\langle n_{c}(\boldsymbol{k})\right\rangle _{2}+\frac{d}{dt}\left\langle n_{c}(\boldsymbol{k})\right\rangle _{1+2;i}.\nonumber
\end{align}
Implementing (\ref{responseCoeffDensity}), the response coefficient densities for the various absorption processes are found to be
\begin{align}
\xi_{1}^{ab}\left(2\omega;\boldsymbol{k}\right)&=2\pi R_{cv}^{(1)a}(\boldsymbol{k};2\omega)^*\label{eq:1+2carriers}\\
&\times R_{cv}^{(1)b}(\boldsymbol{k};2\omega)\delta\left(2\omega-\omega_{cv}(\boldsymbol{k})\right),\nonumber \\
\xi_{1+2}^{abd}\left(2\omega;\boldsymbol{k}\right)&=2\pi R_{cv}^{(2)ab}(\boldsymbol{k};\omega,\omega)^*\nonumber\\
&\times R_{cv}^{(1)d}(\boldsymbol{k};2\omega)\delta\left(2\omega-\omega_{cv}(\boldsymbol{k})\right), \nonumber\\
\xi_{2}^{abde}\left(2\omega;\boldsymbol{k}\right)&=2\pi R_{cv}^{(2)ab}(\boldsymbol{k};\omega,\omega)^*\nonumber\\
&\times R_{cv}^{(2)de}(\boldsymbol{k};\omega,\omega)\delta\left(2\omega-\omega_{cv}(\boldsymbol{k})\right)\nonumber.
\end{align}
Analogously, the total current injection
rate has the form 
\begin{equation}
\frac{d}{dt}\left\langle \mathcal{J}^{a}\right\rangle _{1+2}=\eta_{1+2}^{abde}\left(2\omega\right)E_{-\omega}^{b}E_{-\omega}^{d}E_{2\omega}^{e}+c.c.,\label{eq:1+2current_full}
\end{equation}
with the associated density
\begin{align}
\eta_{1+2}^{abde}(2\omega;\boldsymbol{k})&=2\pi e\Big[\mathfrak{v}_{cc}^{a}(\boldsymbol{k})-\mathfrak{v}_{vv}^{a}(\boldsymbol{k})\Big]R_{cv}^{(2)bd}(\boldsymbol{k};\omega,\omega)^*\nonumber\\
&\times R_{cv}^{(1)e}(\boldsymbol{k};2\omega)\delta\left(2\omega-\omega_{cv}(\boldsymbol{k})\right)
\label{eq:1+2currents}
\end{align}
for systems having current injected \textit{only} because of the interference of excitation processes. Generically this occurs in materials with center-of-inversion symmetry, where all odd-rank response tensors vanish; current is not injected by one- or two-photon absorption individually, but the interference of these excitation processes can lead to an injected current. We note that for crystals with high enough symmetry, such single color current injection can be forbidden even if the system lacks inversion symmetry.

\subsection{2+3 photon absorption}
Next suppose the optical field is composed of frequencies
$\omega$ and $3\omega/2$, with $2\hbar\omega<E_{g}<3\hbar\omega$.
Then two- and three-photon absorption processes can promote electrons from valence
to conduction bands, and also interfere; see Fig.~\ref{23Cartoon}. The contributions
to the total injection rate corresponding to the conduction electron density are
\begin{align}
 \frac{d}{dt}\left\langle n_{c}\right\rangle _{2}&=\xi_{2}^{abde}\left(3\omega\right)E_{-3\omega/2}^{a}E_{-3\omega/2}^{b}E_{3\omega/2}^{d}E_{3\omega/2}^{e},\nonumber\\
\frac{d}{dt}\left\langle n_{c}\right\rangle _{2+3;i}&=\xi_{2+3}^{abdef}\left(3\omega\right)E_{-\omega}^{a}E_{-\omega}^{b}E_{-\omega}^{d}E_{3\omega/2}^{e}E_{3\omega/2}^{f}+c.c.,\nonumber \\
\frac{d}{dt}\left\langle n_{c}\right\rangle _{3}&=\xi_{3}^{abdefg}\left(3\omega\right)E_{-\omega}^{a}E_{-\omega}^{b}E_{-\omega}^{d}E_{\omega}^{e}E_{\omega}^{f}E_{\omega}^{g}\label{eq:2+3carriers_full},  
\end{align}
where 
\begin{align}
\xi_{2}^{abde}\left(3\omega;\boldsymbol{k}\right)&=2\pi R_{cv}^{(2)ab}\Big(\boldsymbol{k};\frac{3\omega}{2},\frac{3\omega}{2}\Big)^*\label{eq:2+3carriers}\\
&\times R_{cv}^{(2)de}\Big(\boldsymbol{k};\frac{3\omega}{2},\frac{3\omega}{2}\Big)\delta\left(3\omega-\omega_{cv}(\boldsymbol{k})\right),\nonumber\\
\xi_{2+3}^{abdef}\left(3\omega;\boldsymbol{k}\right)&=2\pi R_{cv}^{(3)abd}(\boldsymbol{k};\omega,\omega,\omega)^*\nonumber\\
&\times R_{cv}^{(2)ef}\Big(\boldsymbol{k};\frac{3\omega}{2},\frac{3\omega}{2}\Big)\delta\left(3\omega-\omega_{cv}(\boldsymbol{k})\right), \nonumber\\
\xi_{3}^{abdefg}\left(3\omega;\boldsymbol{k}\right)&=2\pi R_{cv}^{(3)abd}(\boldsymbol{k};\omega,\omega,\omega)^*\nonumber\\
&\times R_{cv}^{(3)efg}(\boldsymbol{k};\omega,\omega,\omega)\delta\left(3\omega-\omega_{cv}(\boldsymbol{k})\right),\nonumber
\end{align}
while the current density injection rate is given by
\begin{equation}
\frac{d}{dt}\left\langle \mathcal{J}^{a}\right\rangle _{2+3}=\eta_{2+3}^{abdefg}\left(3\omega\right)E_{-\omega}^{b}E_{-\omega}^{d}E_{-\omega}^{e}E_{3\omega/2}^{f}E_{3\omega/2}^{g}+c.c.,\label{eq:2+3current_full}
\end{equation}
where 
\begin{align}
\eta_{2+3}^{abdefg}(3\omega;\boldsymbol{k})&=2\pi e\Big[\mathfrak{v}_{cc}^{a}(\boldsymbol{k})-\mathfrak{v}_{vv}^{a}(\boldsymbol{k})\Big]R_{cv}^{(3)bde}(\boldsymbol{k};\omega,\omega,\omega)^*\nonumber\\
&\times R_{cv}^{(2)fg}\Big(\boldsymbol{k};\frac{3\omega}{2},\frac{3\omega}{2}\Big)\delta\left(3\omega-\omega_{cv}(\boldsymbol{k})\right)\label{eq:2+3currents}
\end{align}
for systems in which current is injected only because of the interference of excitation processes.

\subsection{1+3 photon absorption}
Finally, consider an incident optical field consisting of frequencies
$\omega$ and $3\omega$, with $2\hbar\omega<E_{g}<3\hbar\omega$;
see Fig.~\ref{13Cartoon}. The contributions
to the total injection rate corresponding to the conduction electron density have the form 
\begin{equation} \begin{aligned}
\frac{d}{dt}\left\langle n_{c}\right\rangle _{1}&=\xi_{1}^{ab}\left(3\omega\right)E_{-3\omega}^{a}E_{3\omega}^{b},\label{eq:1+3carriers_full}\\
\frac{d}{dt}\left\langle n_{c}\right\rangle _{1+3;i}&=\xi_{1+3}^{abde}\left(3\omega\right)E_{-\omega}^{a}E_{-\omega}^{b}E_{-\omega}^{d}E_{3\omega}^{e}+c.c.,  \\
\frac{d}{dt}\left\langle n_{c}\right\rangle _{3}&=\xi_{3}^{abdefg}\left(3\omega\right)E_{-\omega}^{a}E_{-\omega}^{b}E_{-\omega}^{d}E_{\omega}^{e}E_{\omega}^{f}E_{\omega}^{g},  
\end{aligned} \end{equation}
where the densities of the response coefficients in the Brillouin zone are
given by 
\begin{align}
\xi_{1}^{ab}\left(3\omega;\boldsymbol{k}\right)&=2\pi R_{cv}^{(1)a}(\boldsymbol{k};3\omega)^*\label{eq:1+3carriers}\\
&\times R_{cv}^{(1)b}(\boldsymbol{k};3\omega)\delta\left(3\omega-\omega_{cv}(\boldsymbol{k})\right),\nonumber\\
\xi_{1+3}^{abde}\left(3\omega;\boldsymbol{k}\right)&=2\pi R_{cv}^{(3)abd}(\boldsymbol{k};\omega,\omega,\omega)^*\nonumber\\
&\times R_{cv}^{(1)e}(\boldsymbol{k};3\omega)\delta\left(3\omega-\omega_{cv}(\boldsymbol{k})\right), \nonumber\\
\xi_{3}^{abdefg}\left(3\omega;\boldsymbol{k}\right)&=2\pi R_{cv}^{(3)abd}(\boldsymbol{k};\omega,\omega,\omega)^*\nonumber\\
&\times R_{cv}^{(3)efg}(\boldsymbol{k};\omega,\omega,\omega)\delta\left(3\omega-\omega_{cv}(\boldsymbol{k})\right)\nonumber.  
 \end{align}
Similarly, the current density injection rate is given by 
\begin{equation}
\frac{d}{dt}\left\langle \mathcal{J}^{a}\right\rangle _{1+3}=\eta_{1+3}^{abdef}\left(3\omega\right)E_{-\omega}^{b}E_{-\omega}^{d}E_{-\omega}^{e}E_{3\omega}^{f}+c.c.,
\end{equation}
where 
\begin{align}
\eta_{1+3}^{abdef}(3\omega;\boldsymbol{k})&=2\pi e\Big[\mathfrak{v}_{cc}^{a}(\boldsymbol{k})-\mathfrak{v}_{vv}^{a}(\boldsymbol{k})\Big]R_{cv}^{(3)bde}(\boldsymbol{k};\omega,\omega,\omega)^*\nonumber\\
&\times R_{cv}^{(1)f}(\boldsymbol{k};3\omega)\delta\left(3\omega-\omega_{cv}(\boldsymbol{k})\right),
\end{align}
and again we are restricting ourselves to systems where the injected current arises entirely because of the interference of excitation processes.

\subsection{Qualitative functional behavior of the response coefficient densities}
\label{qual}

Before evaluating these expressions for a particular model,
we note some general features that can be expected from any system
with a direct band gap in which electron-electron and electron-phonon interactions are neglected, as well as possible small contributions from other bands.

In the one-photon absorption amplitude (\ref{eq:R1coef-1}) there is
a single interband velocity matrix element, $\boldsymbol{\mathfrak{v}}_{cv}(\boldsymbol{k})$,
which is in general nonzero for all $\boldsymbol{k}$ values. In
the two-photon absorption amplitude (\ref{eq:R2coef-1}) there are
products of inter- and intraband velocity matrix elements, while the
three-photon absorption amplitude contains terms involving one interband
and two intraband elements, as well as terms involving three interband
elements. 

The structure of these terms is vital for understanding the variation of the
absorption amplitudes through the Brillouin zone. Unlike the interband
velocity matrix element, the intraband matrix elements $\boldsymbol{\mathfrak{v}}_{cc}(\boldsymbol{k})$ $\big(\boldsymbol{\mathfrak{v}}_{vv}(\boldsymbol{k})\big)$ are zero at the conduction (valence) band minima (maxima), since they are directly related to
the slope of the bands at that $\boldsymbol{k}$ point; in particular, they both vanish
at the band gap, and have a much more significant crystal momentum
dependence than the interband matrix elements. Thus the two-photon
absorption amplitudes have more structure in the Brillouin zone than do the one-photon amplitudes.

This is even more dramatic for the component of the three-photon amplitude
involving two intraband terms, which also vanishes at a band extremum,
but varies more quickly in $\boldsymbol{k}$ than do the two-photon
amplitudes due to the appearance of two diagonal matrix elements
$\boldsymbol{\mathfrak{v}}_{nn}(\boldsymbol{k})$. In contrast to the 
two-photon amplitude, the three-photon amplitude has a component 
composed entirely of interband matrix
elements that is nonzero at the band gap, and generally has a slow variation
in the Brillouin zone, reminiscent of the one-photon absorption amplitude.
In this way, the three-photon absorption process contains qualitative
features of both the one- and two-photon processes. 

Since the two- and three-photon absorption amplitudes can generally
be expected to exhibit much more structure in the Brillouin zone than
the one-photon amplitude, it is not surprising that carriers injected
through a 2+3 absorption process, where there is interference between the two-
and three-photon absorption amplitudes, can be more localized in the
Brillouin zone than those injected from a 1+2 absorption process, involving interference
between one- and two-photon absorption amplitudes. 

Finally, for a material with center-of-inversion symmetry all full
response coefficients described by odd rank tensors vanish. For such
a material $\xi_{1+2}^{abc}(2\omega)$ and $\xi_{2+3}^{abdef}(3\omega)$
are identically zero. That is, in both 1+2 and 2+3 photon absorption
there is \textit{no} interference between the contributing processes that can
lead to QuIC of the total number of electron-hole pairs
created. Nonetheless, $\xi_{1+2}^{abc}(2\omega;\boldsymbol{k})$ and
$\xi_{1+2}^{abdef}(2\omega;\boldsymbol{k})$ are \textit{not} identically zero. That is, at
a particular point in the Brillouin zone there can be interference
between the contributing processes, with (say) more electron-hole
pairs created at a particular $\boldsymbol{k}_{o}$ than at $-\boldsymbol{k}_{o}$,
and indeed it is this interference that leads to the injected current
described by the even rank tensor $\eta_{1+2}^{abcd}(2\omega)$ (for
1+2 PA) or $\eta_{2+3}^{abcdef}(3\omega)$ (for 2+3 PA). 

The situation for $1+3$ photon absorption is
qualitatively different. For a material with center-of-inversion symmetry
there is no QuIC leading to an injected current, as the tensor
$\eta_{1+3}^{abcde}(3\omega)$ that describes it is of odd rank. However,
the tensor $\xi_{1+3}^{abcd}(3\omega)$ that describes QuIC related to the total number of electron-hole pairs injected is of even rank,
and so this QuIC survives.

We now illustrate these features with a calculation using a model
for TMDs.

\section{Low-Energy Model for transition-metal dichalcogenides}
\label{sec:4}
For the exciting optical fields we consider, only electron states near the band gap are relevant. We therefore consider an effective theory defined in the regions near the valleys, $\boldsymbol{K}$ and $\boldsymbol{K'}$, that describes the lowest-energy excitations of the system; 
we let $\boldsymbol{k}$ indicate the displacement
in the Brillouin zone from the nearby valley, writing 
\begin{equation} \begin{aligned}
  \boldsymbol{k}=k_{x}\boldsymbol{\hat{x}}+k_{y}\boldsymbol{\hat{y}}=k\Big(\boldsymbol{\hat{x}}\cos\theta+\boldsymbol{\hat{y}}\sin\theta\Big),
 \end{aligned}\end{equation}
where $k\equiv\left|\boldsymbol{k}\right|$ and $\theta$ is the angle
that $\boldsymbol{k}$ makes from the $\boldsymbol{\hat{x}}$ axis.
We adopt an exactly solvable model Hamiltonian introduced earlier \cite{Xiao12,Rostami13,Liu15,Silva16}, which includes terms allowed
by the symmetry of the lattice, and explicitly retains the intra-atomic 
spin-orbit coupling term: 
\begin{align}
{H}_0(\boldsymbol{k})&=\hbar \Xi\left(k_{x}\tau_{z}\otimes\sigma_{x}\otimes s_{0}+k_{y}\tau_{0}\otimes\sigma_{y}\otimes s_{0}\right)\\
&+\frac{\hbar\Delta}{2}\left(\tau_{0}\otimes\sigma_{z}\otimes s_{0}\right)+\frac{\hbar\lambda}{2}\left(\tau_{z}\otimes\left(\sigma_{0}-\sigma_{z}\right)\otimes s_{z}\right)\nonumber,
\end{align}
where the components of $\boldsymbol{\tau},$ $\boldsymbol{\sigma}$,
and $\boldsymbol{s}$ are the usual Pauli matrices, and with the index zero 
referring to the $2\times2$ identity matrix over the appropriate Hilbert space. 
For $\boldsymbol{\tau}$ that Hilbert space corresponds to the valley degree 
of freedom associated with the massive Dirac points,
for $\boldsymbol{\sigma}$ it corresponds to the pseudospin degree of freedom associated
with the inequivalent sublattice sites, and for $\boldsymbol{s}$ it corresponds to the spin degree of freedom. 

In general ${H}_0(\boldsymbol{k})$ is represented by an $8\times8$ 
matrix; but the valleys are not coupled, and about each valley the spin degrees
 of freedom also decouple. We
thus solve a general $2\times2$ eigenvalue equation for an input
$\left\{ \tau,s\right\} $, which we use to specify the valley and
spin for which we are solving: $\tau=1(-1)$ corresponds to the $\boldsymbol{K}$
($\boldsymbol{K'})$ valley, and $s_z=1(-1)$ corresponds to spin component in the
$z$ direction being up (down): 
\begin{equation}
{H}_{\tau s}(\boldsymbol{k})=\hbar\varpi_{\tau s}\sigma_{0}+\hbar\boldsymbol{d}_{\tau s}(\boldsymbol{k})\cdot\boldsymbol{\sigma},
\label{eq:H_work}
\end{equation}
where $\varpi_{\tau s}=\lambda\tau s_{z}$, $\boldsymbol{d}_{\tau s}(\boldsymbol{k})=\Xi\tau k_{x}\boldsymbol{\hat{x}}+\Xi k_{y}\boldsymbol{\hat{y}}+\Delta_{\tau s}\boldsymbol{\hat{z}}$,
and $\Delta_{\tau s}=(\Delta-\lambda\tau s_z)/2$. Here $\Xi$ is related to the hopping integral from site to site, $\hbar\Delta$ is the band gap 
energy if the spin-orbit interaction were to be neglected, and $\lambda$ characterizes the 
spin-orbit coupling; their values are given in Table \ref{table:parameters} for WSe$_2$ \cite{Xiao12}. Note that $\Delta_{\tau s}>0$. 

\begin{figure*}[hbt!]
	\centering
	\begin{subfigure}[b]{.33\textwidth}
		\centering
		\includegraphics[width=0.99\linewidth]{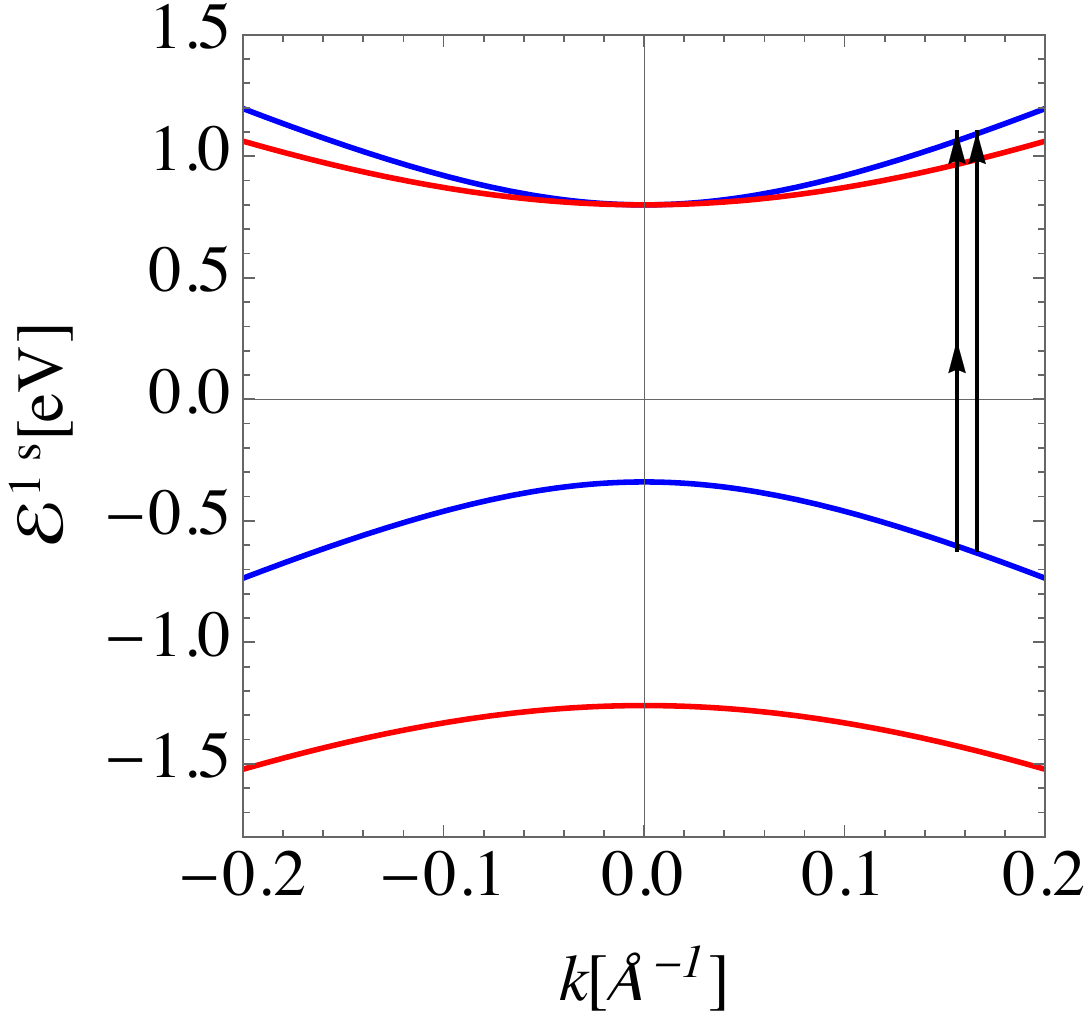}
		\caption{$\tau=+1$}
		\label{fig:energyK}
	\end{subfigure}%
	\begin{subfigure}[b]{.33\textwidth}
		\centering
		\includegraphics[width=0.99\linewidth]{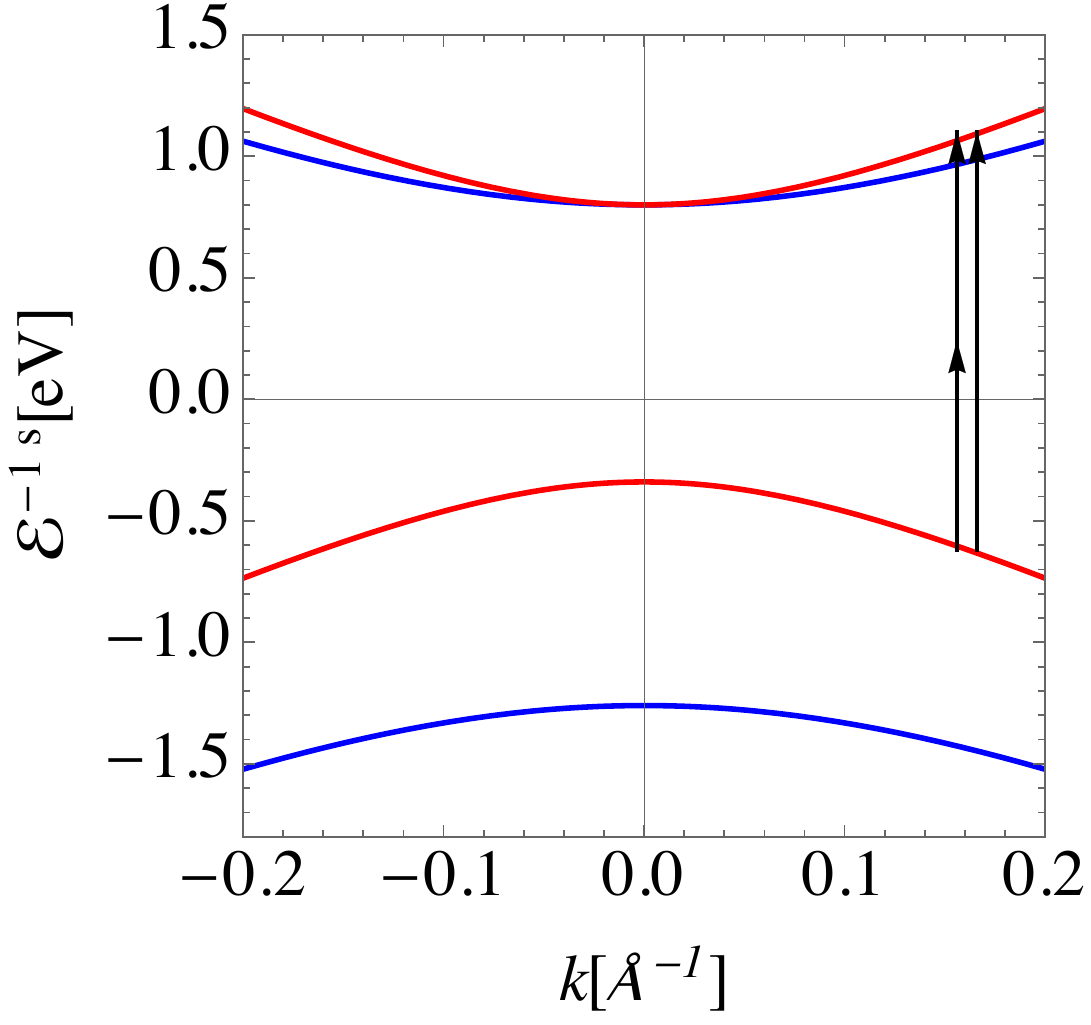}
		\caption{$\tau=-1$}
		\label{fig:energyK'}
	\end{subfigure}
	\begin{subfigure}[b]{.33\textwidth}
		\includegraphics[width=0.99\linewidth]{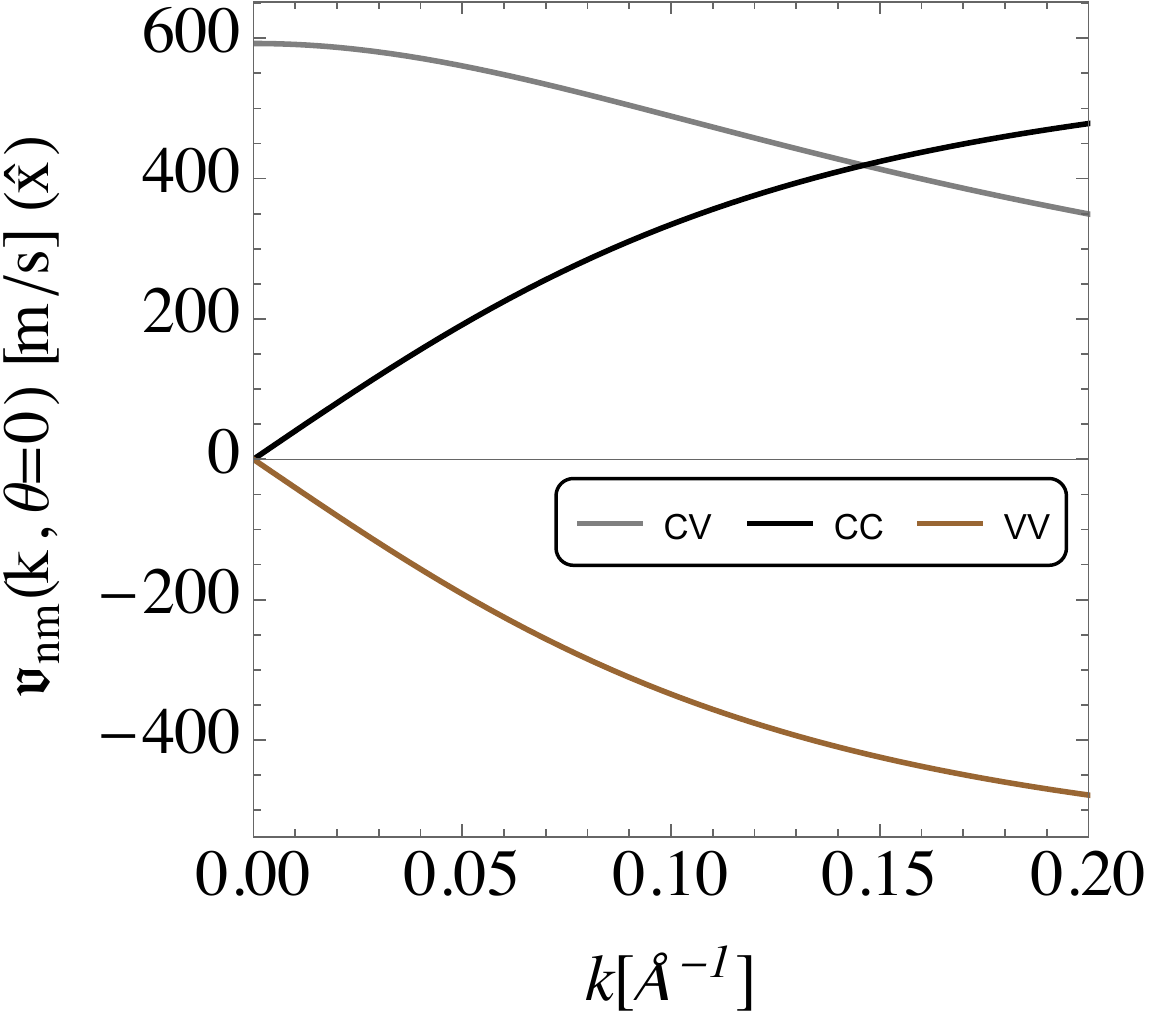}
		\caption{Velocity matrix elements for $\text{WSe}_2$.}
		\label{fig:velocityMatrixElements}
	\end{subfigure}
	\caption{(a)\&(b) Low-energy band structure of $\text{WSe}_2$ about the $\boldsymbol{K}$ ($\tau=+1$) and $\boldsymbol{K'}$ ($\tau=-1$) points, where only 1+2 photon absorption is indicated for clarity. Distinctly colored bands (red and blue) correspond to opposite spin projection ($s_z=-1$ and $s_z=+1$). (c) Velocity matrix elements for the upper valence band and lower conduction band about both valleys along the $\theta=0$ direction ($\boldsymbol{\hat{x}}$).}
\end{figure*}

One can diagonalize the Hamiltonian (\ref{eq:H_work}) to find energies
$\varepsilon^{\tau s}_\pm(\boldsymbol{k})=\hbar\varpi_{\tau s}\pm\hbar|\boldsymbol{d}_{\tau s}(\boldsymbol{k})|$
(see Fig.~\ref{fig:energyK} and~\ref{fig:energyK'}) for a given $\left\{ \tau,s\right\} $, leading
to a frequency difference 
\begin{equation}
\omega_{cv}^{\tau s}(\boldsymbol{k})=2\big|\boldsymbol{d}_{\tau s}(\boldsymbol{k})\big|
\end{equation}
between the bands $c$ and $v$ associated with $\{\tau,s\}$ at $\boldsymbol{k}$; note that
\begin{equation}
\big|\boldsymbol{d}_{\tau s}(\boldsymbol{k})\big|^{2}=\Xi^{2}k^{2}+\Delta_{\tau s}^{2}
\end{equation}
is independent of $\theta$, even though $\boldsymbol{d}_{\tau s}(\boldsymbol{k})$
is not, and the frequency difference function for a particular spin
in one valley is equal to the frequency difference function for the
opposite spin in the other valley. This is due to the combination of time-reversal symmetry
and because $\varepsilon^{\tau s}_{\pm}(\boldsymbol{k})=\varepsilon^{\tau s}_{\pm}(-\boldsymbol{k})$, which together imply $\varepsilon^{\tau s}_{\pm}(\boldsymbol{k})=\varepsilon^{-\tau -s}_{\pm}(\boldsymbol{k})$.

The only model dependent parameters that arise in the perturbative
expansion of the transition coefficient $\gamma_{cv}(\boldsymbol{k},t)$
are related to the velocity matrix elements \cite{Muniz15}. Writing $\boldsymbol{\hat{e}}_{\pm}=(\boldsymbol{\hat{x}}\pm i\boldsymbol{\hat{y}})/\sqrt{2}=\boldsymbol{\hat{e}}_{\mp}^{*}$,
the interband velocity matrix elements are given by 
\begin{align}
\boldsymbol{\mathfrak{v}}_{cv}(\boldsymbol{k})&=-i\Xi e^{-i\tau\theta}\left[\left(-\sin\theta+i\frac{\tau\Delta_{\tau s}}{\left|\boldsymbol{d}_{\tau s}(\boldsymbol{k})\right|}\cos\theta\right)\boldsymbol{\hat{x}}\right.\nonumber\\
&\qquad\qquad\qquad\quad+\left.\left(\cos\theta+i\frac{\tau\Delta_{\tau s}}{\left|\boldsymbol{d}_{\tau s}(\boldsymbol{k})\right|}\sin\theta\right)\boldsymbol{\hat{y}}\right]\nonumber\\
&=\frac{\Xi e^{-i\tau\theta}}{\sqrt{2}}\left[e^{-i\theta}\left(\frac{\tau\Delta_{\tau s}}{\left|\boldsymbol{d}_{\tau s}(\boldsymbol{k})\right|}-1\right)\boldsymbol{\hat{e}}_{-}^{*}\right.\nonumber\\
&\qquad\qquad\qquad\left.+e^{i\theta}\left(\frac{\tau\Delta_{\tau s}}{\left|\boldsymbol{d}_{\tau s}(\boldsymbol{k})\right|}+1\right)\boldsymbol{\hat{e}}_{+}^{*}\right],
\label{eq:interband}
\end{align}
and the intraband velocity matrix elements are given by
\begin{equation} \begin{aligned}
 \boldsymbol{\mathfrak{v}}_{cc}(\boldsymbol{k})=-\boldsymbol{\mathfrak{v}}_{vv}(\boldsymbol{k})=\frac{\Xi^{2}\boldsymbol{k}}{\left|\boldsymbol{d}_{\tau s}(\boldsymbol{k})\right|}.\label{eq:intraband}
\end{aligned} \end{equation}
The dominant contributions to the perturbative coefficients are found to be (see Sec. \ref{sec:2} for discussion)
\begin{align}
&R_{cv}^{(1)a}(\boldsymbol{k};\Omega)=\frac{ie}{\hbar}\frac{1}{\Omega}\mathfrak{v}_{cv}^{a}(\boldsymbol{k}),\label{TMDsCoeffs}\\
&R_{cv}^{(2)ab}\left(\boldsymbol{k};\frac{\Omega}{2},\frac{\Omega}{2}\right)= \frac{e^{2}}{\hbar^{2}}\frac{2^{3}}{\Omega^{3}}\Big(\mathfrak{v}_{cc}^{a}(\boldsymbol{k})-\mathfrak{v}_{vv}^{a}(\boldsymbol{k})\Big)\mathfrak{v}_{cv}^{b}(\boldsymbol{k}),\nonumber\\
&R_{cv}^{(3)abd}\left(\boldsymbol{k};\frac{\Omega}{3},\frac{\Omega}{3},\frac{\Omega}{3}\right)= -\frac{ie^{3}}{\hbar^{3}}\frac{3^{5}}{2\Omega^{5}}\Big[\Big(\mathfrak{v}_{cc}^{a}(\boldsymbol{k})-\mathfrak{v}_{vv}^{a}(\boldsymbol{k})\Big)\nonumber\\
&\qquad\qquad\quad\times \Big(\mathfrak{v}_{cc}^{b}(\boldsymbol{k})-\mathfrak{v}_{vv}^{b}(\boldsymbol{k})\Big)-\frac{1}{2}\mathfrak{v}_{cv}^{a}(\boldsymbol{k})\mathfrak{v}_{vc}^{b}(\boldsymbol{k})\Big]\mathfrak{v}_{cv}^{d}(\boldsymbol{k})\nonumber
\end{align}
and useful combinations of velocity matrix elements are thus
\begin{align}
 &\mathfrak{v}_{cc}^{i}(\boldsymbol{k})-\mathfrak{v}_{vv}^{i}(\boldsymbol{k})=\frac{2\Xi^{2}k^{i}}{\left|\boldsymbol{d}_{\tau s}(\boldsymbol{k})\right|},\\
&\mathfrak{v}_{cv}^{i}(\boldsymbol{k})\mathfrak{v}_{vc}^{j}(\boldsymbol{k})=\Xi^{2}\Bigg[\boldsymbol{\hat{i}}\cdot\boldsymbol{\hat{j}}+\frac{i\tau\Delta_{\tau s}}{\left|\boldsymbol{d}_{\tau s}(\boldsymbol{k})\right|}\boldsymbol{\hat{z}}\cdot\big(\boldsymbol{\hat{i}}\times\boldsymbol{\hat{j}}\big)-\frac{\Xi^{2}k^{i}k^{j}}{\left|\boldsymbol{d}_{\tau s}(\boldsymbol{k})\right|^{2}}\Bigg]\nonumber,  
\end{align}
which follow immediately from (\ref{eq:interband},\ref{eq:intraband}).
We also introduce $\boldsymbol{k}_{\tau s}(m\omega)$, the crystal momentum at which $\delta(m\omega-\omega_{cv}^{\tau s}\big(\boldsymbol{k})\big)$ is satisfied,
and the magnitude of which is given by 
\begin{equation}
k_{\tau s}(m\omega)=\Xi^{-1}\sqrt{\left(\frac{m}{2}\omega\right)^{2}-\Delta_{\tau s}^{2}}.
\end{equation}

The velocity matrix elements are plotted in Fig.~\ref{fig:velocityMatrixElements} along the direction
$\theta=0$ in the Brillouin zone. In the variation of these matrix
elements through the Brillouin zone, and in the way they combine in
the amplitudes (\ref{TMDsCoeffs}), one can easily identify the different
qualitative nature of the absorption amplitudes, as discussed in Sec. \ref{qual}. 
While the TMDs lack center-of-inversion
symmetry, for the configuration of optical fields at normal incidence to which we restrict
ourselves there are electric fields only in the $\boldsymbol{\hat{x}}$ and
$\boldsymbol{\hat{y}}$ directions, and we consider only current injections
in the plane defined by those two vectors. For this restricted set
of ``in-plane'' Cartesian components the TMD response coefficients
\textit{do} exhibit the selection rules that would follow from center-of-inversion
symmetry: All odd rank tensors vanish, QuIC of injected
current is possible only for $1+2$ and $2+3$ absorption, and QuIC of the number of injected electron-hole pairs is possible only for $1+3$ absorption. Expressions for the full set of nonvanishing
response coefficients (\ref{eq:1+2carriers_full},\ref{eq:1+2current_full},\ref{eq:2+3carriers_full},\ref{eq:2+3current_full},\ref{eq:1+3carriers_full})
for this model are given in Appendix \ref{Appendix}.

\begin{table}[h]
	\begin{ruledtabular}
		\renewcommand{\arraystretch}{1.7}
		\begin{tabular}{c c c c c c}
			$\hbar \Xi$ & $\hbar \lambda$ & $\hbar \Delta$ & $E_\omega$ \big($\frac{V}{m}$\big) & $E_{2\omega}$ \big($\frac{V}{m}$\big) & $E_{3\omega}$ \big($\frac{V}{m}$\big)\\
			\colrule
			3.9 \AA \space eV  &  0.46 eV  &  1.6 eV  &  $2\times10^8$  &  $1.3\times10^8$  &  $2\times10^7$\\
		\end{tabular}
	\end{ruledtabular}
	\caption{Model parameters for $\text{WSe}_2$ and upper bound of optical field amplitudes for single color absorption.}
	\label{table:parameters}
\end{table}

\section{Electronic Distribution in the Brillouin Zone}
\begin{figure*}[hbt!]
	\centering
	\begin{subfigure}[b]{.5\textwidth}
		\centering
		\includegraphics[width=0.99\linewidth]{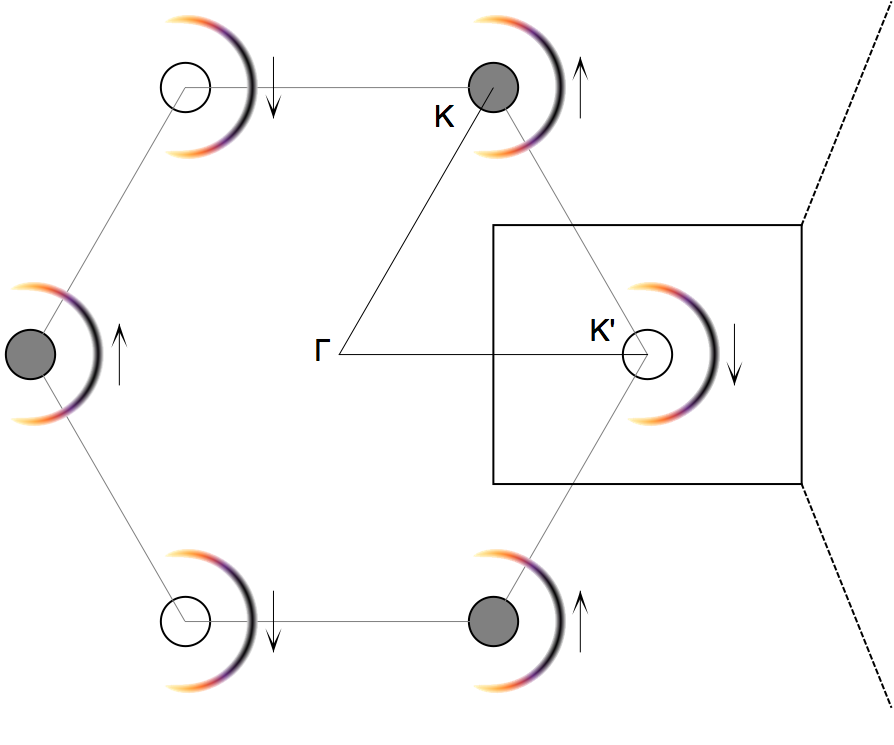}
	\end{subfigure}%
	\begin{subfigure}[b]{.5\textwidth}
		\centering
		\includegraphics[width=0.99\linewidth]{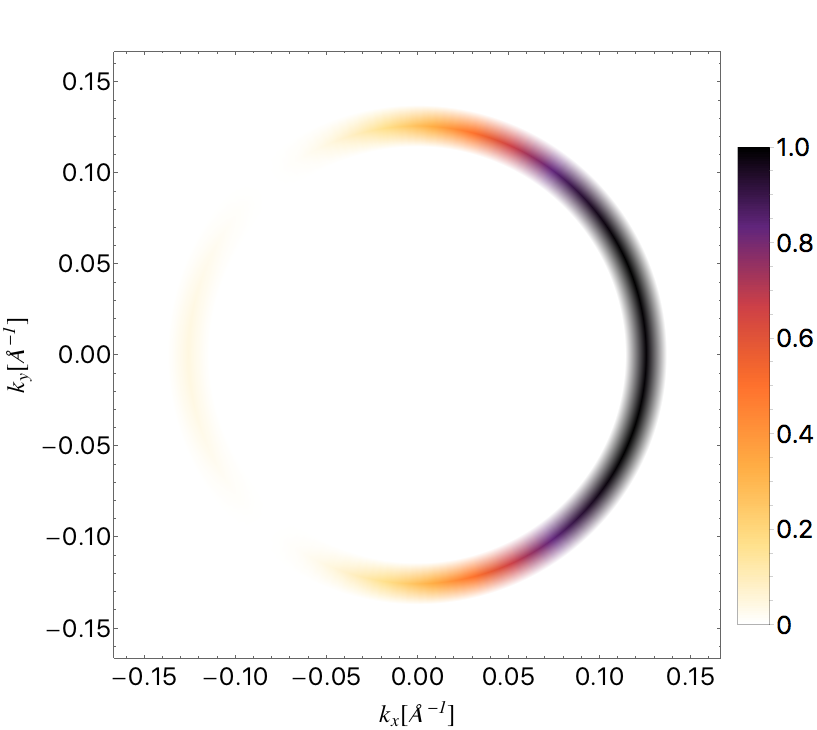}
	\end{subfigure}
	\caption{Distribution of injected carriers in crystal momentum space from an optical probe facilitating 1+2 photon absorption; the first Brillouin zone is composed of a single pair of $\boldsymbol{K}$ (filled circles) and $\boldsymbol{K'}$ (empty circles) points.}
	\label{fig:BZ}
\end{figure*}
\label{sec:5}
In what follows we focus primarily on an excitation energy
of $\hbar\Omega=1.5$ eV, which should be assumed unless otherwise
specified. To display the location of injected carriers in the Brillouin zone, we plot $\frac{d}{dt}\left\langle n_c(\boldsymbol{k})\right\rangle_{N+M}$ for $N+M$ being $1+2$, $2+3$, and $1+3$. These quantities are found by weighting the previously found densities of the carrier injection rate coefficients, $\xi(\Omega;\boldsymbol{k})$, by factors of the fields, and summing them appropriately; this was described in detail in Sec. \ref{sec:3}.

Since our calculations are done at the perturbative level, the field
amplitudes and pulse lengths must be such that only a small fraction of
carriers in any region of the valence band are excited into the conduction
band. Taking a nominal figure of 5\% for this limit, the rough upper
bounds for optical field amplitudes corresponding to one-, two- and
three-photon absorption at $3\omega$, $3\omega/2$,
and $\omega$, respectively, are given in Table \ref{table:parameters} for 50-fs full width at half maximum pulses normally incident
on WSe$_{2}$; these values were determined as described in detail
earlier \cite{Muniz15}. 

In $N+M$ injection the increased localization of excitations in the 
Brillouin zone results from interference of the $N$ and $M$ absorption 
amplitudes, and this increased localization is typically maximized for field strengths giving rise to 
total probabilities of $N$ and $M$ photon absorption that are nearly equal, 
$\frac{d}{dt}\langle n_{c}\rangle_{N}=\frac{d}{dt}\langle n_{c}\rangle_{M}$. For the 
model we consider, this is the explicit condition for
maximum interference between processes, which corresponds to 
the most localized excitations. In what follows the optical intensities
are \textit{always} set such that this holds. The field amplitudes listed in Table \ref{table:parameters} do \textit{not} correspond to the condition of maximum interference between any processes; these values specify the upper bound of the perturbative regime for one-, two-, or three-photon 
absorption processes individually. The values used for plotting are then approximately half of 
those listed, and scaled appropriately; the strength of the response, namely, the
number of excited electrons or magnitude of the injected current, depends intimately 
on the field strengths, however the qualitative features of Brillouin-zone densities 
are independent of these values so long as the interference is maximized. 

When considering excitation by incident optical fields, the quantum interference between pathways can be affected by adjusting the frequency, polarization, and phase shift of the fields. As previously mentioned, we consider initially an energy of $\hbar\Omega=1.5$ eV, and so it is the latter two parameters we initially vary. In what follows we then look at various relative polarizations of the fields, and for each combination of polarizations we vary the phase shift. The quantity we term the ``relative phase parameter'' arises as the natural parameter to vary; this contains information about phases of both fields. For $1+2$, $2+3$, and
$1+3$ absorption processes the relative phase parameters are given by 
\begin{equation} \begin{aligned}
  \Delta\phi_{12}&=\phi_{2\omega}-2\phi_{\omega},\\
  \Delta\phi_{23}&=2\phi_{3\omega/2}-3\phi_{\omega},\\
  \Delta\phi_{13}&=\phi_{3\omega}-3\phi_{\omega},
 \end{aligned}\end{equation}
respectively, or generally $\Delta\phi_{NM}=N\phi_{\Omega/N}-M\phi_{\Omega/M}$. 

\subsection{Co-linearly polarized incident fields}
Here the spin of the injected carriers is valley dependent, while the distribution of electronic excitations is valley independent. It is thus sufficient to show the injected carrier distribution about a single valley with the understanding that the same excited charge distribution is present at the other, composed of electrons of the opposite spin.

\subsubsection{1+2 absorption}
\begin{figure}[hbt!]
	\centering
	\begin{subfigure}{.155\textwidth}
		\centering
		\includegraphics[width=0.95\linewidth]{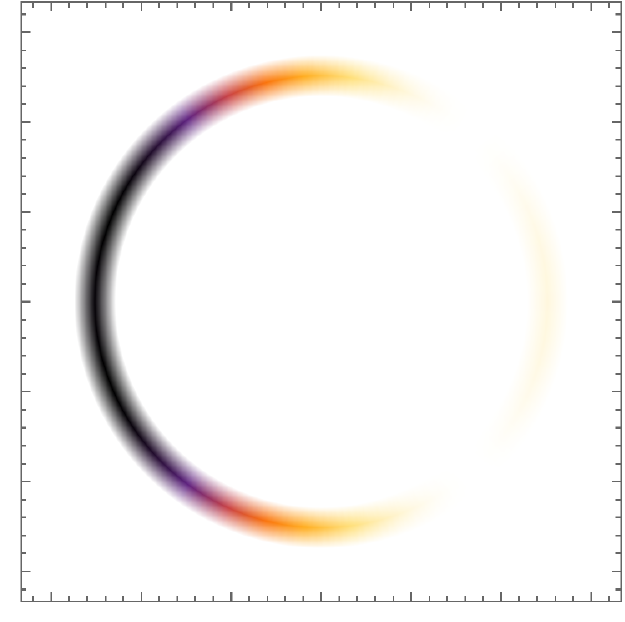}
		\caption{$\Delta\phi_{12}=\pi/2$}
		\label{fig:sub1}
	\end{subfigure}%
	\centering
	\begin{subfigure}{.155\textwidth}
		\centering
		\includegraphics[width=0.95\linewidth]{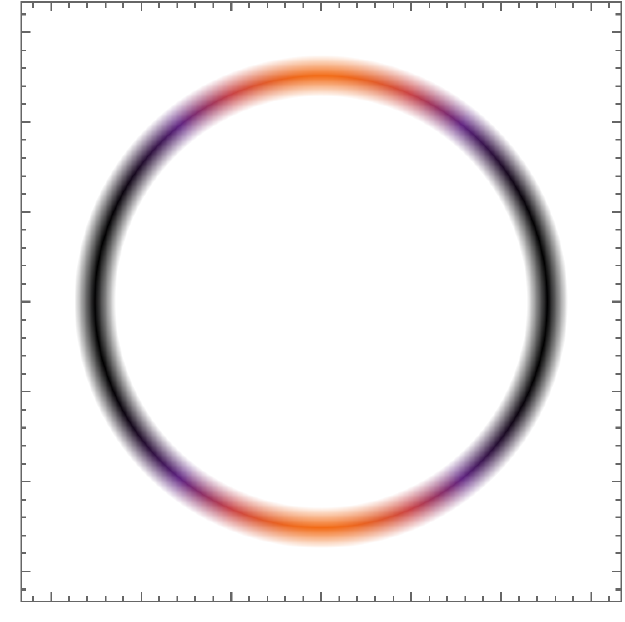}
		\caption{$\Delta\phi_{12}=0$}
		\label{fig:sub1}
	\end{subfigure}%
	\centering
	\begin{subfigure}{.155\textwidth}
		\centering
		\includegraphics[width=0.95\linewidth]{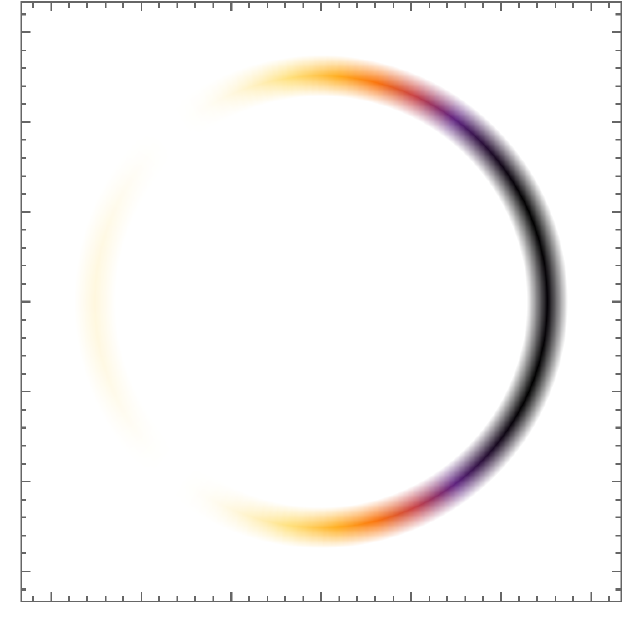}
		\caption{$\Delta\phi_{12}=3\pi/2$}
		\label{fig:sub2}
	\end{subfigure}
	\caption{Dependence of $\frac{d}{dt}\left\langle n_{c}(\boldsymbol{k})\right\rangle _{1+2}$ on $\Delta\phi_{12}$ for fields of frequency $\omega$ and $2\omega$ both polarized along $\boldsymbol{\hat{x}}$.}
	\label{fig:12CarrierInj}
\end{figure}

Yet to gain an overall perspective, in Fig.~\ref{fig:BZ} we plot the relative density of carriers injected in the
Brillouin zone for the fields at $\omega$ and $2\omega$ both linearly
polarized along the $\boldsymbol{\hat{x}}$ direction and $\Delta\phi_{12}=3\pi/2$,
for $2\hbar\omega=1.5$ eV. In the neighborhood of each
valley the carriers are injected with the same polar distribution,
indicating that a net current is injected, and that each valley contributes equally to the injected current.
In the inset of Fig.~\ref{fig:BZ} we show an enlarged
view of the carrier injection about one of the valleys; the width
of the arc is associated with the 50-fs pulse duration, and could be
made larger or smaller by considering shorter or longer pulses, respectively.
In the majority of the following figures we show a view that corresponds
to the inset of Fig.~\ref{fig:BZ}. Using such a view we show in Fig.~\ref{fig:12CarrierInj} how the injected carrier distribution changes with the relative phase parameter $\Delta\phi_{12}$, which controls the interference between the one- and two-photon absorption amplitudes.

\begin{figure}[hbt!]
	\centering
	\begin{subfigure}{.155\textwidth}
		\centering
		\includegraphics[width=0.95\linewidth]{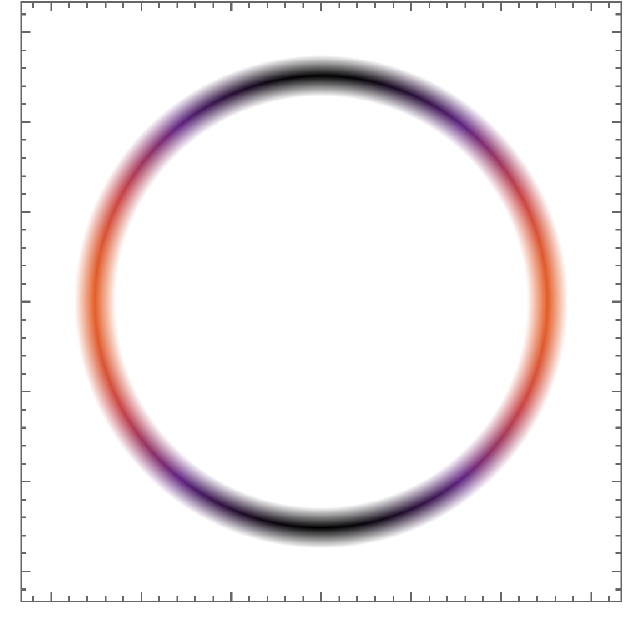}
		\caption{$\frac{d}{dt}\left\langle n_{c}(\boldsymbol{k})\right\rangle _{1}$}
		\label{fig:1PA}
	\end{subfigure}%
	\begin{subfigure}{.155\textwidth}
		\centering
		\includegraphics[width=0.95\linewidth]{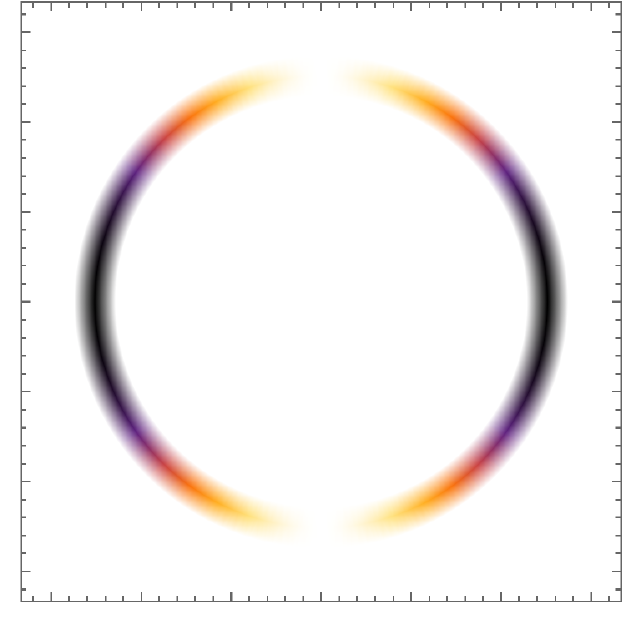}
		\caption{$\frac{d}{dt}\left\langle n_{c}(\boldsymbol{k})\right\rangle _{2}$}
		\label{fig:2PA}
	\end{subfigure}
	\begin{subfigure}{.155\textwidth}
		\centering
		\includegraphics[width=0.95\linewidth]{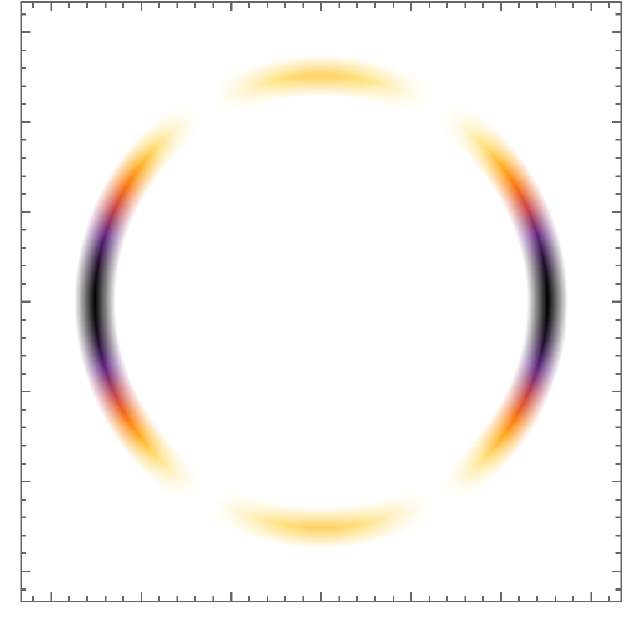}
		\caption{$\frac{d}{dt}\left\langle n_{c}(\boldsymbol{k})\right\rangle _{3}$}
		\label{fig:3PA}
	\end{subfigure}
	\caption{BZ resolved carrier injection rates arising from single color absorption, for fields polarized along $\boldsymbol{\hat{x}}$.}
	\label{fig:SinglePhoton}
\end{figure}

In Fig.~\ref{fig:SinglePhoton} we show the distribution of carriers injected from (solely)
one-, two-, or three-photon absorption, where the different field
amplitudes are again chosen so that there is equal total carrier injection
from each process. The results here are also valley independent,
with carriers of opposite spin being injected about the different
valleys. Although each of these carrier distributions is nonpolar, they
illustrate the general feature, noted above, that the higher-order
processes result in more localized regions of injected carriers in
the Brillouin zone. We now explain the qualitative differences.

The one-photon absorption, due solely to interband matrix elements,
is peaked in the $\pm\boldsymbol{\hat{y}}$ directions, perpendicular
to the direction of the electric field; this follows from (\ref{eq:interband}),
for from it we find 
\begin{equation} 
\begin{aligned}
\left|\boldsymbol{\mathfrak{v}}_{cv}(\boldsymbol{k})\cdot\boldsymbol{\hat{x}}\right|^{2}=\Xi^2\sin^{2}\theta+\frac{\Xi^2\Delta_{\tau s}^{2}}{\left|\boldsymbol{d}_{\tau s}(\boldsymbol{k})\right|^{2}}\cos^{2}\theta,\\
\left|\boldsymbol{\mathfrak{v}}_{cv}(\boldsymbol{k})\cdot\boldsymbol{\hat{y}}\right|^{2}=\Xi^2\cos^{2}\theta+\frac{\Xi^2\Delta_{\tau s}^{2}}{\left|\boldsymbol{d}_{\tau s}(\boldsymbol{k})\right|^{2}}\sin^{2}\theta,
\end{aligned}
\end{equation}
and thus at larger photon energy, or equivalently at larger $k$, as $\Delta_{\tau s}/\left|\boldsymbol{d}_{\tau s}(\boldsymbol{k})\right|$
becomes smaller there will be even less injection of carriers near
the $\pm\boldsymbol{\hat{x}}$ directions. The two-photon absorption
peaks in the directions $\pm\boldsymbol{\hat{x}}$ associated with
the oscillating electric field because of the presence of the intraband
matrix element, and exhibits more localization in the Brillouin zone
than that of the one-photon absorption. The strong maxima in the three-photon
absorption are also due to the presence of intraband matrix elements,
while the weaker maxima in the $\pm\boldsymbol{\hat{y}}$ directions
are due to the terms involving only interband matrix
elements. It is clear how the patterns displayed in Fig.~\ref{fig:12CarrierInj} result
from the interference of the one- and two-photon amplitudes responsible
for the plots shown in Fig.~\ref{fig:SinglePhoton}.

\subsubsection{2+3 absorption}
\begin{figure}[hbt!]
	\centering
	\begin{subfigure}{.155\textwidth}
		\centering
		\includegraphics[width=0.95\linewidth]{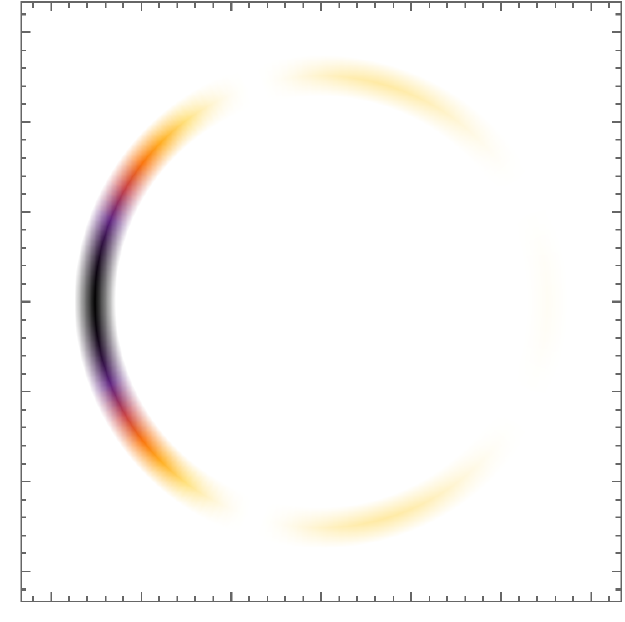}
		\caption{$\Delta\phi_{23}=\pi/2$}
		\label{fig:sub1}
	\end{subfigure}%
	\centering
	\begin{subfigure}{.155\textwidth}
		\centering
		\includegraphics[width=0.95\linewidth]{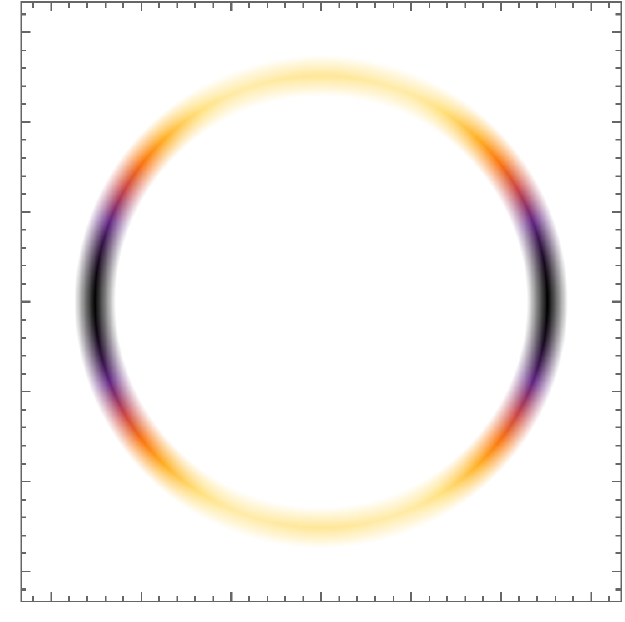}
		\caption{$\Delta\phi_{23}=\pi$}
		\label{fig:sub1}
	\end{subfigure}
	\begin{subfigure}{.155\textwidth}
		\centering
		\includegraphics[width=0.95\linewidth]{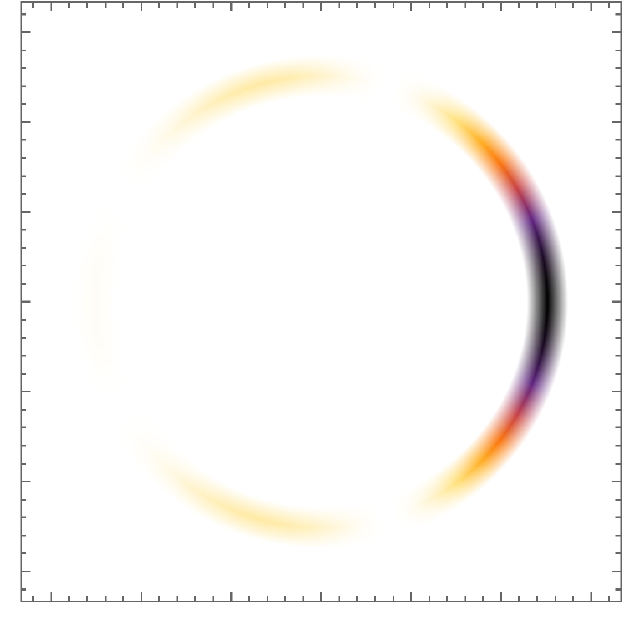}
		\caption{$\Delta\phi_{23}=3\pi/2$}
		\label{fig:sub2}
	\end{subfigure}
	\caption{Dependence of $\frac{d}{dt}\left\langle n_{c}(\boldsymbol{k})\right\rangle _{2+3}$ on $\Delta\phi_{23}$ for fields of frequency $\omega$ and $3\omega/2$ both polarized along $\boldsymbol{\hat{x}}$.}
	\label{fig:23CarrierInj}
\end{figure}
In Fig.~\ref{fig:23CarrierInj} we plot the carrier injection distributions from 2+3 absorption
corresponding to the carrier injection distributions from 1+2 absorption
shown in Fig.~\ref{fig:12CarrierInj}. As the relative phase parameters ($\Delta\phi_{23}$
in the former, $\Delta\phi_{12}$ in the latter) are varied, the two sets of
plots show the same qualitative behavior. However, as expected, we see stronger
localization of the injected carriers in the 2+3 process than in the 1+2 process. The location of these more localized excitations leads to a larger current injected in 2+3 than in 1+2 absorption at a transition energy of $1.5$ eV, as will be discussed in Sec. \ref{sec:6}.

\subsubsection{1+3 absorption}
\begin{figure}[hbt!]
	\centering
	\begin{subfigure}{.155\textwidth}
		\centering
		\includegraphics[width=0.95\linewidth]{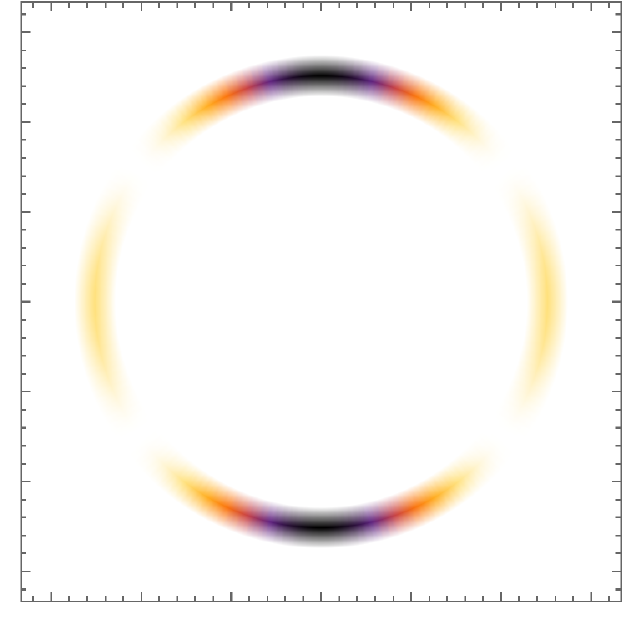}
		\caption{$\Delta\phi_{13}=0$}
		\label{}
	\end{subfigure}%
	\begin{subfigure}{.155\textwidth}
		\centering
		\includegraphics[width=0.95\linewidth]{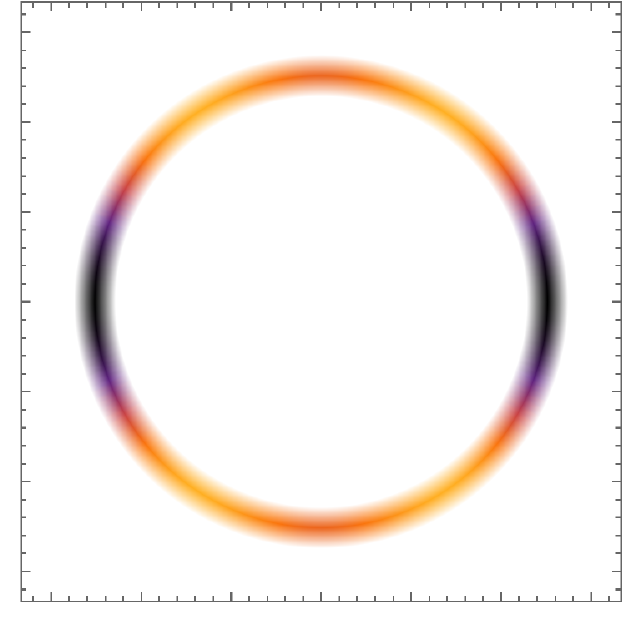}
		\caption{$\Delta\phi_{13}=\pi/2$}
		\label{}
	\end{subfigure}%
	\begin{subfigure}{.155\textwidth}
		\centering
		\includegraphics[width=0.95\linewidth]{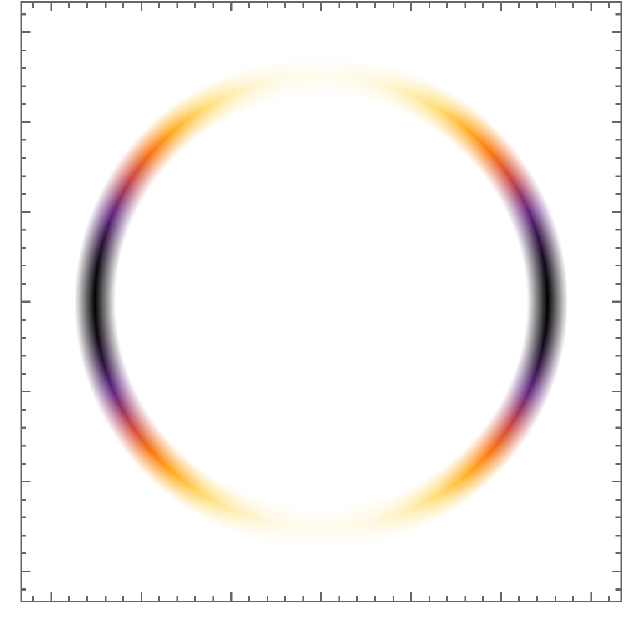}
		\caption{$\Delta\phi_{13}=\pi$}
		\label{}
	\end{subfigure}
	\caption{Dependence of $\frac{d}{dt}\left\langle n_{c}(\boldsymbol{k})\right\rangle _{1+3}$ on $\Delta\phi_{13}$ for fields of frequency $\omega$ and $3\omega$ both polarized along $\boldsymbol{\hat{x}}$.}
	\label{fig:13CarrierInj}
\end{figure}
The injected electronic distributions from 1+3 absorption are qualitatively
different than those from 1+2 and 2+3, and are shown in Fig.~\ref{fig:13CarrierInj}. Here
there is no net current injection, but rather a variation in the total
density of carriers injected as the relative phase parameter $\Delta\phi_{13}$
is varied. For $\Delta\phi_{13}=\pi/2$, the contributions to $\xi_{1+3}^{abde}(3\omega;\boldsymbol{k})$ from all $(\tau,s)$ vanish, and at any $\boldsymbol{k}$ point in the Brillouin zone the carriers injected are the sum
of those injected at that $\boldsymbol{k}$ point from one-photon absorption and at
three-photon absorption. There is constructive interference between
those two absorption processes at $\Delta\phi_{13}=\pi$, with about
$50\%$ more carriers injected than for the same intensities at $\Delta\phi_{13}=\pi/2$,
and there is destructive interference between the processes at $\Delta\phi_{13}=0$,
with about 50\% fewer carriers injected than for the same intensities
at $\Delta\phi_{13}=\pi/2$. 

\subsection{Cross-linearly polarized incident fields}

The spin of the injected carriers is again valley dependent. However, in contrast to the co-linear case, the injected electronic charge distributions also become valley dependent.

\subsubsection{1+2 absorption}
\begin{figure}[hbt!]
	\centering
	\begin{subfigure}{.155\textwidth}
		\centering
		\includegraphics[width=0.95\linewidth]{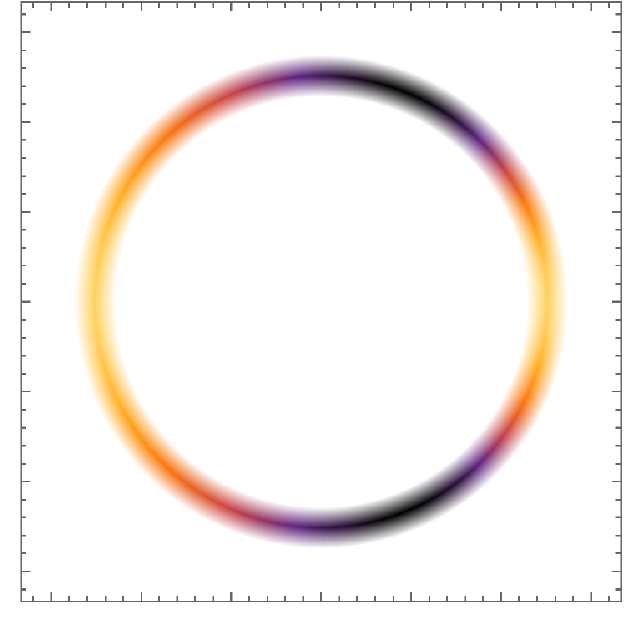}
		\caption{$\boldsymbol{K}$, $\Delta\phi_{12}=\frac{\pi}{2}$}
		\label{fig:12crossA}
	\end{subfigure}%
	\centering
	\begin{subfigure}{.155\textwidth}
		\centering
		\includegraphics[width=0.95\linewidth]{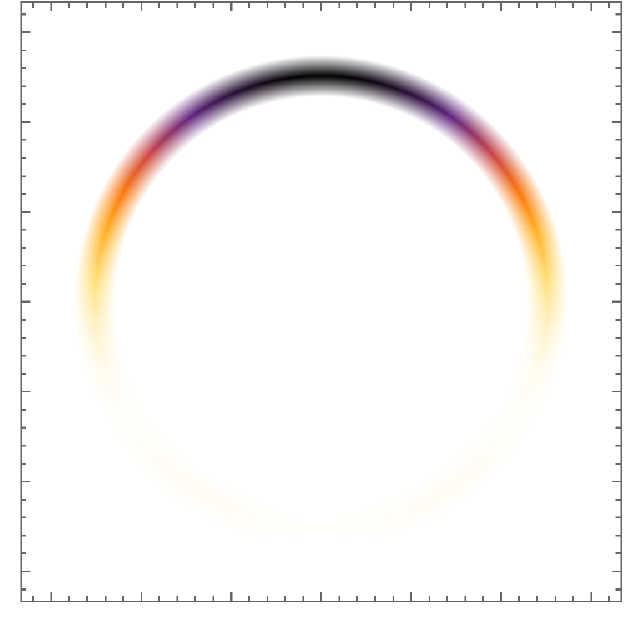}
		\caption{$\boldsymbol{K}$, $\Delta\phi_{12}=\pi$}
		\label{fig:12crossB}
	\end{subfigure}
	\begin{subfigure}{.155\textwidth}
		\centering
		\includegraphics[width=0.95\linewidth]{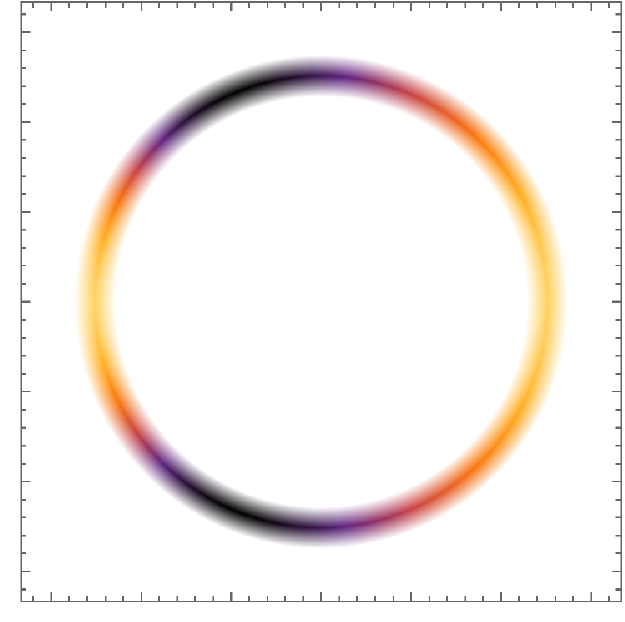}
		\caption{$\boldsymbol{K}$, $\Delta\phi_{12}=\frac{3\pi}{2}$}
		\label{fig:12crossC}
	\end{subfigure}
	\centering
	\begin{subfigure}{.155\textwidth}
		\centering
		\includegraphics[width=0.95\linewidth]{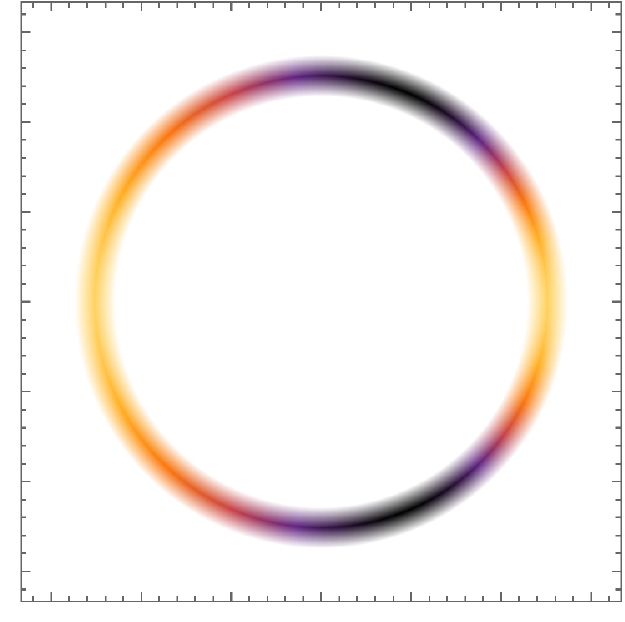}
		\caption{$\boldsymbol{K'}$, $\Delta\phi_{12}=\frac{\pi}{2}$}
		\label{fig:12crossD}
	\end{subfigure}%
	\centering
	\begin{subfigure}{.155\textwidth}
		\centering
		\includegraphics[width=0.95\linewidth]{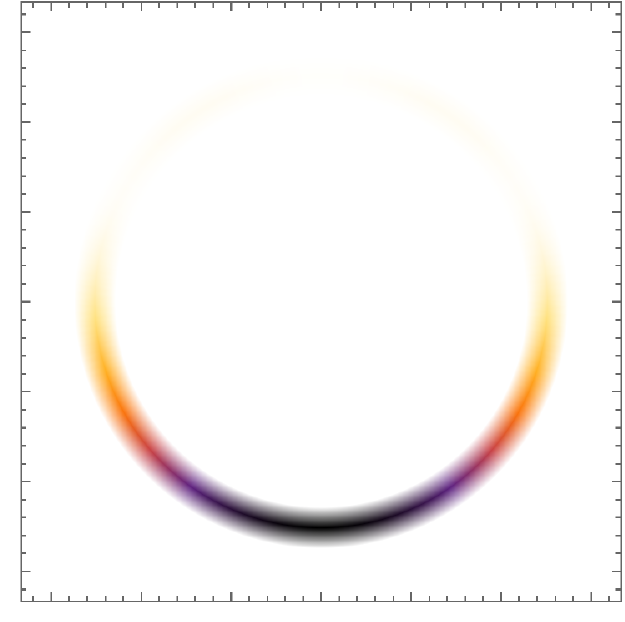}
		\caption{$\boldsymbol{K'}$, $\Delta\phi_{12}=\pi$}
		\label{fig:12crossE}
	\end{subfigure}
	\begin{subfigure}{.155\textwidth}
		\centering
		\includegraphics[width=0.95\linewidth]{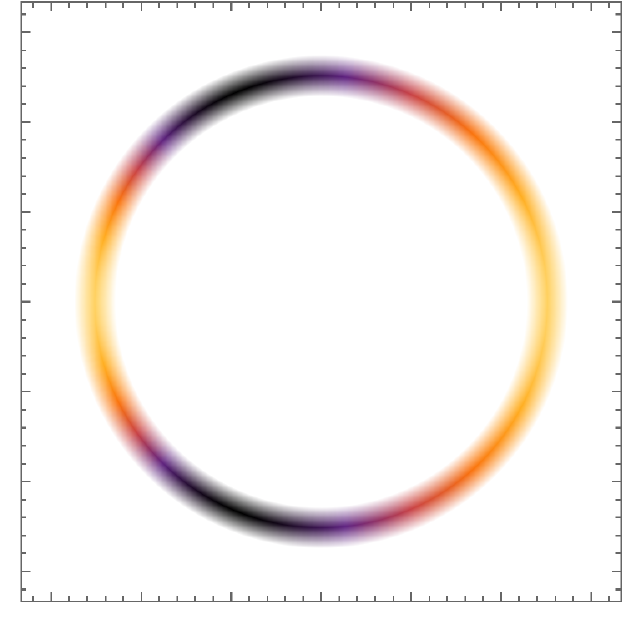}
		\caption{$\boldsymbol{K'}$, $\Delta\phi_{12}=\frac{3\pi}{2}$}
		\label{fig:12crossF}
	\end{subfigure}
	\caption{Dependence of $\frac{d}{dt}\left\langle n_{c}(\boldsymbol{k})\right\rangle _{1+2}$ on $\Delta\phi_{12}$ for fields orientated according to $\boldsymbol{\hat{e}}_{\omega}=\boldsymbol{\hat{y}}$ and $\boldsymbol{\hat{e}}_{2\omega}=\boldsymbol{\hat{x}}$.}
	\label{fig:12CarrierInjCross}
\end{figure}

In Fig.~\ref{fig:12CarrierInjCross} we plot the distribution of injected carriers
for $\boldsymbol{\hat{e}}_{3\omega}=\boldsymbol{\hat{x}}$ and $\boldsymbol{\hat{e}}_{3\omega/2}=\boldsymbol{\hat{y}}$.
For $\Delta\phi_{12}=\pi/2$ or $3\pi/2$ the distribution of injected carriers
about the different valleys is the same, and exhibits a current in the $\boldsymbol{-}\boldsymbol{\hat{x}}$
direction (for $\Delta\phi_{12}=\pi/2)$ or in the $\hat{\boldsymbol{x}}$
direction (for $\Delta\phi_{12}=3\pi/2$). However, more generally
the injected electronic distributions are valley dependent;
the most dramatic example is for $\Delta\phi_{12}=\pi$, where currents
are injected in the $-\boldsymbol{\hat{y}}$ and $\boldsymbol{\hat{y}}$
directions about $\boldsymbol{K}$ and $\boldsymbol{K'}$ respectively. Here there is no net
current injected in either direction. Due to the symmetries of the model, the only nonzero response tensor involved for the net current for the specified polarizations is $\eta_{1+2}^{xyyx}$, and so any nonvanishing
injected current is in the $\pm\boldsymbol{\hat{x}}$ direction. In
fact, regardless of the crystal axes associated with the cross-linear
polarizations, in this model it is always the direction associated with the field facilitating odd number photon absorption (i.e. 1 PA, 3 PA, etc.) that determines the direction of the net injected current, if there
is one. This can be shown analytically using the expressions provided, and is consistent with previously found results for the TMDs \cite{Muniz15}.

\subsubsection{2+3 absorption}
\begin{figure}[hbt!]
	\centering
	\begin{subfigure}{.155\textwidth}
		\centering
		\includegraphics[width=0.95\linewidth]{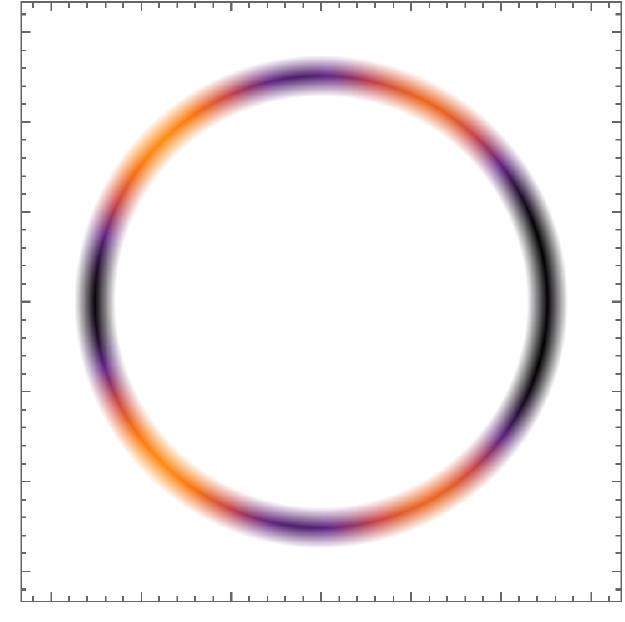}
		\caption{$\boldsymbol{K}$, $\Delta\phi_{23}=\pi/2$}
		\label{fig:23crossA}
	\end{subfigure}%
	\centering
	\begin{subfigure}{.155\textwidth}
		\centering
		\includegraphics[width=0.95\linewidth]{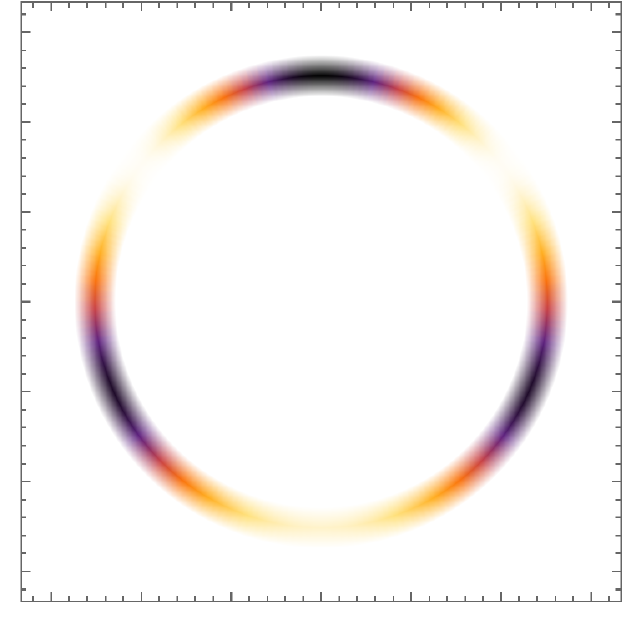}
		\caption{$\boldsymbol{K}$, $\Delta\phi_{23}=\pi$}
		\label{fig:23crossB}
	\end{subfigure}
	\begin{subfigure}{.155\textwidth}
		\centering
		\includegraphics[width=0.95\linewidth]{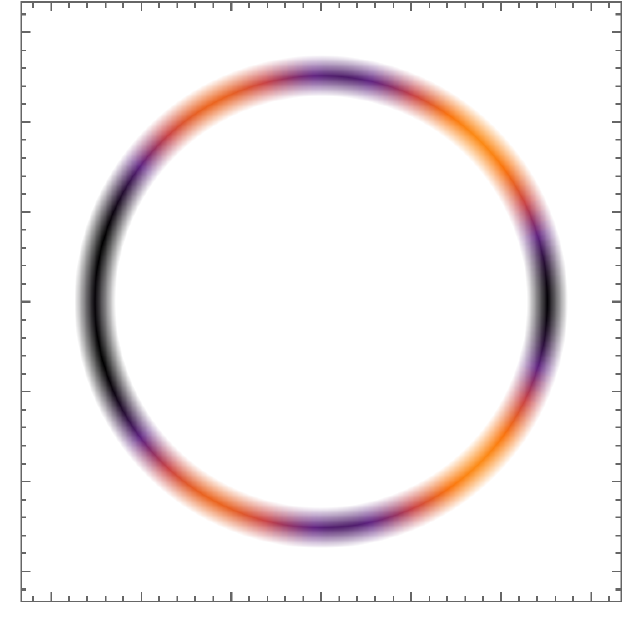}
		\caption{$\boldsymbol{K}$, $\Delta\phi_{23}=\frac{3\pi}{2}$}
		\label{fig:23crossC}
	\end{subfigure}
	\centering
	\begin{subfigure}{.155\textwidth}
		\centering
		\includegraphics[width=0.95\linewidth]{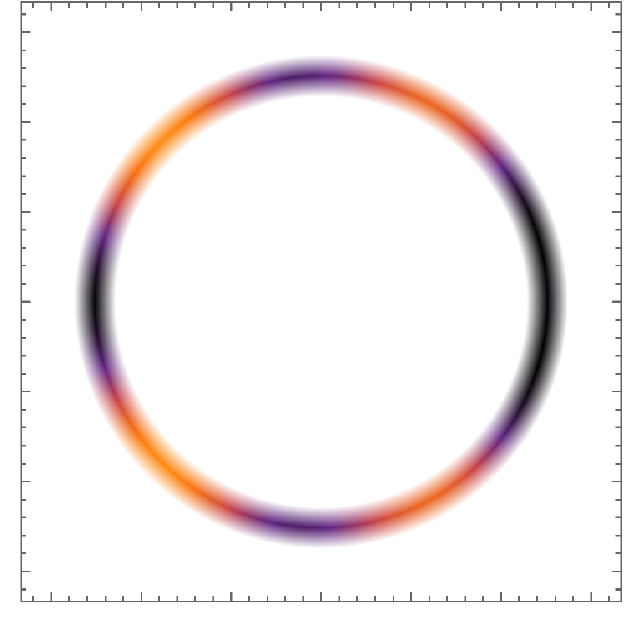}
		\caption{$\boldsymbol{K'}$, $\Delta\phi_{23}=\frac{\pi}{2}$}
		\label{fig:23crossD}
	\end{subfigure}%
	\centering
	\begin{subfigure}{.155\textwidth}
		\centering
		\includegraphics[width=0.95\linewidth]{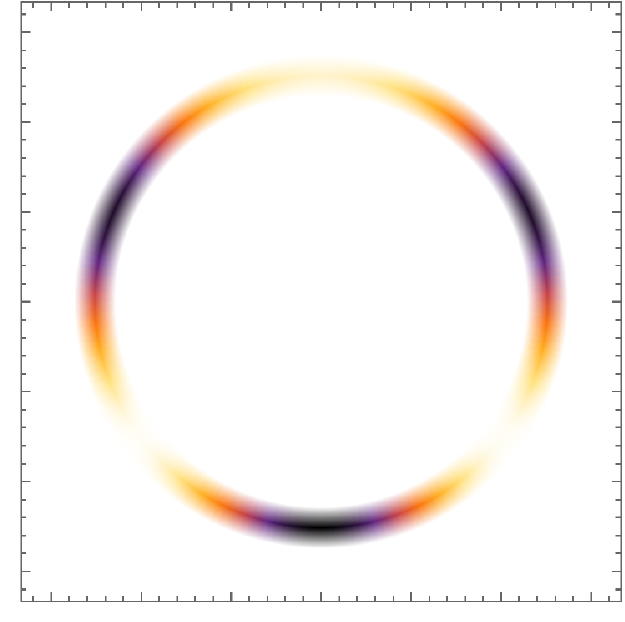}
		\caption{$\boldsymbol{K'}$, $\Delta\phi_{23}=\pi$}
		\label{fig:23crossE}
	\end{subfigure}
	\begin{subfigure}{.155\textwidth}
		\centering
		\includegraphics[width=0.95\linewidth]{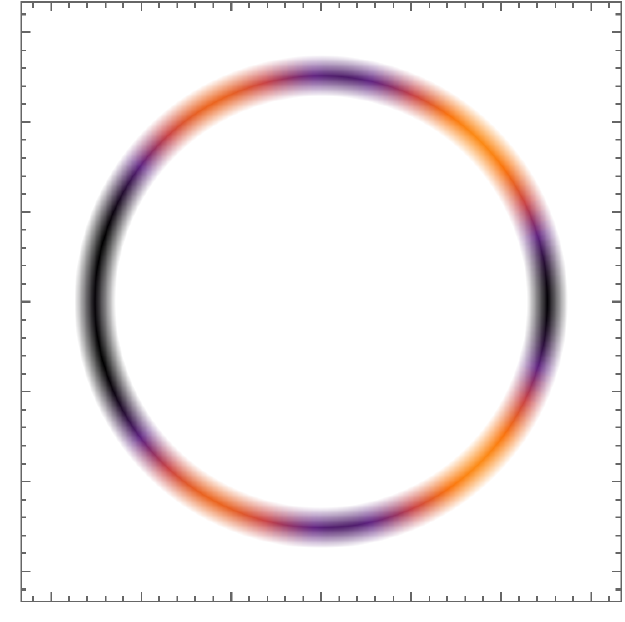}
		\caption{$\boldsymbol{K'}$, $\Delta\phi_{23}=\frac{3\pi}{2}$}
		\label{fig:23crossF}
	\end{subfigure}
	\caption{Dependence of $\frac{d}{dt}\left\langle n_{c}(\boldsymbol{k})\right\rangle _{2+3}$ on $\Delta\phi_{23}$ for fields orientated according to 
		$\boldsymbol{\hat{e}}_{\omega}=\boldsymbol{\hat{x}}$ and $\boldsymbol{\hat{e}}_{3\omega/2}=\boldsymbol{\hat{y}}$.}
	\label{fig:23CarrierInjCross}
\end{figure}

The situation is the same for $2+3$ absorption, where the direction
of the current injected by cross-linearly polarized light is always
determined by the direction of field facilitating three-photon absorption. In Fig.~\ref{fig:23CarrierInjCross} we
plot the injected carriers for $\boldsymbol{\hat{e}}_{\omega}=\boldsymbol{\hat{x}}$
and $\boldsymbol{\hat{e}}_{3\omega/2}=\hat{\boldsymbol{y}}$ ; the
corresponding response tensor component is $\eta_{2+3}^{xxxxyy}$. As in the $1+2$ case, at $\Delta\phi_{23}=\pi$ there is no net current injected. At $\Delta\phi_{23}=\pi/2$ and $\Delta\phi_{23}=3\pi/2$
the distributions of injected carriers about the $\boldsymbol{K}$ and $\boldsymbol{K'}$ points
are identical, mirroring the situation for $1+2$ absorption, and 
there is a weak current injected in the $-\boldsymbol{\hat{x}}$ direction
(for $\Delta\phi_{23}=\pi/2$) and in the $\boldsymbol{\hat{x}}$
direction (for $\Delta\phi_{23}=3\pi/2$). The carrier distributions
injected by cross-linearly polarized incident fields appear to be more strongly
localized in the Brillouin zone for 2+3 absorption than for 1+2 absorption,
as was seen for co-linearly polarized incident fields. Here, however, it is found that there is a larger current injected from 1+2 absorption than from 2+3 absorption at transition energy of $1.5$ eV (see Sec. \ref{sec:6}).

\subsubsection{1+3 absorption}
\begin{figure}[hbt!]
	\centering
	\begin{subfigure}{.155\textwidth}
		\centering
		\includegraphics[width=0.95\linewidth]{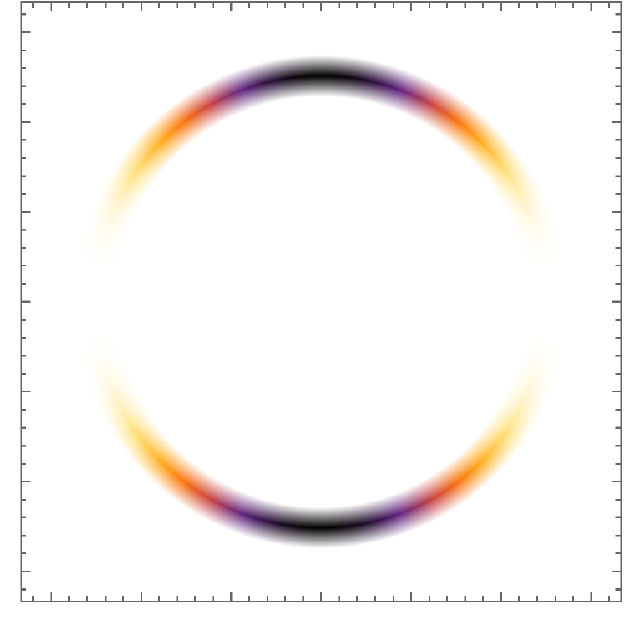}
		\caption{$\boldsymbol{K}$, $\Delta\phi_{13}=\frac{\pi}{2}$}
		\label{fig:13crossA}
	\end{subfigure}%
	\centering
	\begin{subfigure}{.155\textwidth}
		\centering
		\includegraphics[width=0.95\linewidth]{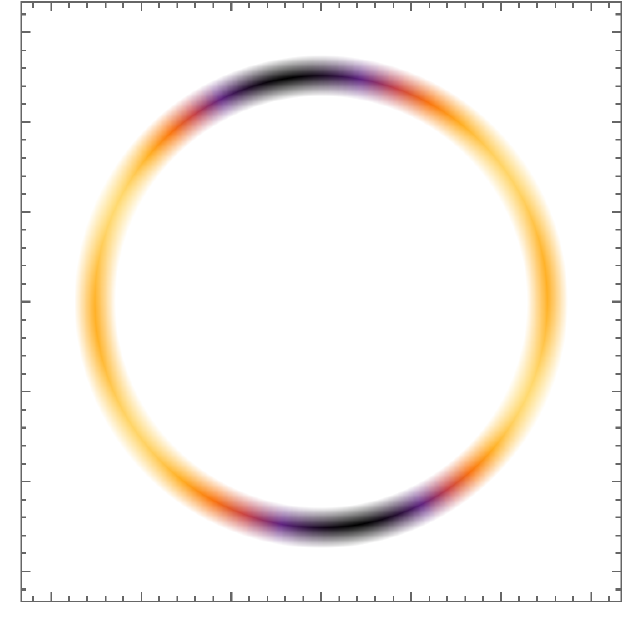}
		\caption{$\boldsymbol{K}$, $\Delta\phi_{13}=\pi$}
		\label{fig:13crossB}
	\end{subfigure}
	\begin{subfigure}{.155\textwidth}
		\centering
		\includegraphics[width=0.95\linewidth]{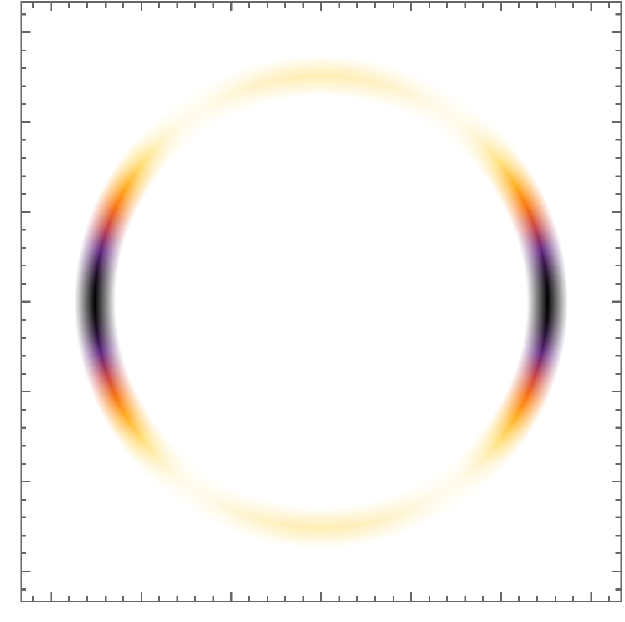}
		\caption{$\boldsymbol{K}$, $\Delta\phi_{13}=\frac{3\pi}{2}$}
		\label{fig:13crossC}
	\end{subfigure}
	\centering
	\begin{subfigure}{.155\textwidth}
		\centering
		\includegraphics[width=0.95\linewidth]{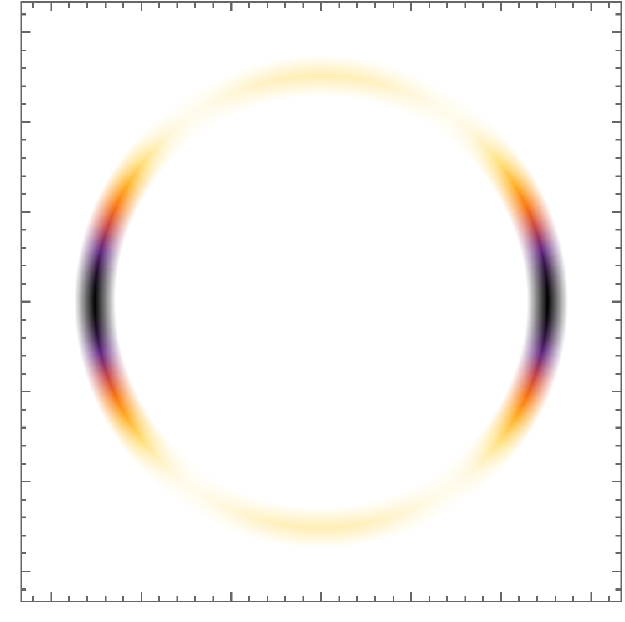}
		\caption{$\boldsymbol{K'}$, $\Delta\phi_{13}=\frac{\pi}{2}$}
		\label{fig:13crossD}
	\end{subfigure}%
	\centering
	\begin{subfigure}{.155\textwidth}
		\centering
		\includegraphics[width=0.95\linewidth]{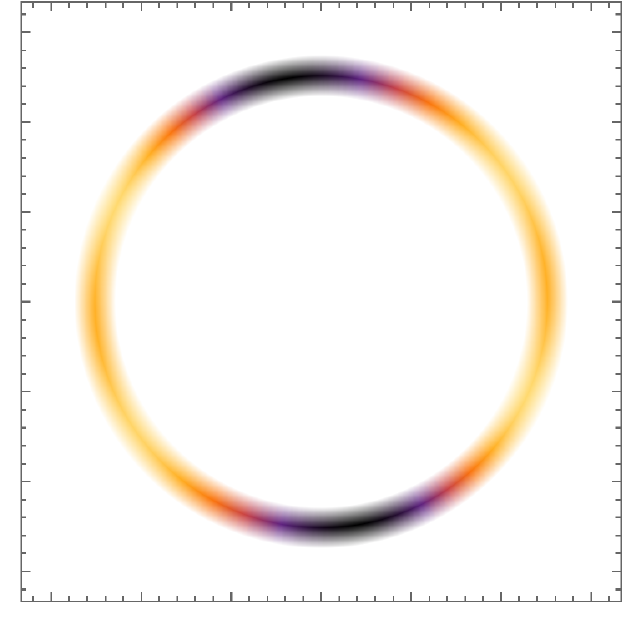}
		\caption{$\boldsymbol{K'}$, $\Delta\phi_{13}=\pi$}
		\label{fig:13crossE}
	\end{subfigure}
	\begin{subfigure}{.155\textwidth}
		\centering
		\includegraphics[width=0.95\linewidth]{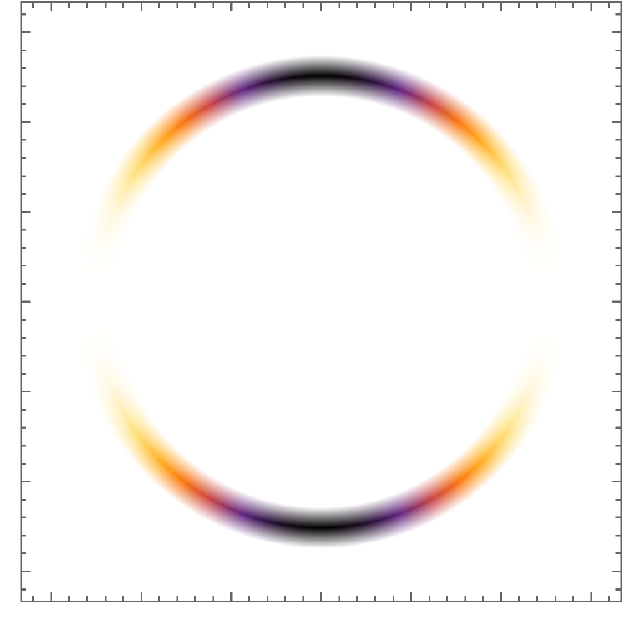}
		\caption{$\boldsymbol{K'}$, $\Delta\phi_{13}=\frac{3\pi}{2}$}
		\label{fig:13crossF}
	\end{subfigure}
	\caption{Dependence of $\frac{d}{dt}\left\langle n_{c}(\boldsymbol{k})\right\rangle _{1+3}$ on $\Delta\phi_{13}$ for fields orientated according to $\boldsymbol{\hat{e}}_{\omega}=\boldsymbol{\hat{y}}$ and $\boldsymbol{\hat{e}}_{3\omega}=\boldsymbol{\hat{x}}$.}
	\label{fig:13CarrierInjCross}
\end{figure}

Again, there is no current injection possible for 1+3 absorption.
For $\boldsymbol{\hat{e}}_{3\omega}=\boldsymbol{\hat{x}}$ and $\boldsymbol{\hat{e}}_{\omega}=\boldsymbol{\hat{y}}$
the injected carrier distributions are shown in Fig.~\ref{fig:13CarrierInjCross}. The distributions
about $\boldsymbol{K}$ and $\boldsymbol{K'}$ are the same for $\Delta\phi_{13}=\pi,$ but
different from each other generally. The result of the interference
is that one set of regions where injected carriers are localized (those in the $\pm\boldsymbol{\hat{y}}$
directions about $\boldsymbol{K}$) becomes less populated as the relative phase
parameter is increased from $\pi/2$ to $\pi$ to $3\pi/2$, and
another set of regions (those in the $\pm\boldsymbol{\hat{x}}$
directions about $\boldsymbol{K}$) becomes more populated. This parallels what
was seen for 1+3 absorption for co-linearly polarized incident fields.
As was found there, the number of carriers injected about each valley 
varies as the relative phase parameter $\Delta\phi_{13}$ is changed. Here, however,
the injected distributions are valley dependent and the result after summing
both valleys is that there is no net injection of carriers; this is consistent with what 
we display in Fig.~\ref{fig:13CarrierInjCross}.

\subsection{Circularly polarized incident optical fields}

Consider first the one-, two- and three-photon absorption processes
individually. For a given helicity, say $\boldsymbol{\hat{e}}_{+}$, the
distribution of carriers injected by one-photon absorption will show
no dependence on the angle $\theta$ about the nearby band gap. This
follows from (\ref{eq:interband}), which governs the one-photon absorption
rate, and from which we find 
\begin{equation} \begin{aligned}
 \boldsymbol{\mathfrak{v}}_{cv}(\boldsymbol{k})\cdot\boldsymbol{\hat{e}}_{+}=\frac{\Xi e^{i(1-\tau)\theta}}{\sqrt{2}}\left(\frac{\tau\Delta_{\tau s}}{\left|\boldsymbol{d}_{\tau s}(\boldsymbol{k})\right|}+1\right).\label{eq:interband_circular}
\end{aligned} \end{equation}
It is the absolute value squared of (\ref{eq:interband_circular}) that  enters in the one-photon absorption rate for $\boldsymbol{\hat{e}}_{+}$
polarized light, which is independent of $\theta$. Similar arguments hold
for the two- and three-photon absorption rates. Thus the localization of injected carriers resulting from a single color absorption processes using linearly polarized light (recall Fig.~\ref{fig:SinglePhoton})
is absent for excitation by circularly polarized light, and so the single color absorption 
processes do not help in establishing well-localized polar distributions
when interference effects are brought into play. Nonetheless, in another
sense the use of circular polarizations offers more control compared
to the use of linear polarizations, in that continuous 
variations in the relative phase parameter change the
direction of the injected current continuously; this is not the case when exciting electrons 
using linearly polarized fields.

Notice that near the $\boldsymbol{K}$ point ($\tau=1$), (\ref{eq:interband_circular}) will be larger than about the $\boldsymbol{K'}$ ($\tau=-1)$; indeed, as the excitation energy decreases to the band-gap energy and $\left|\boldsymbol{d}_{\tau s}(\boldsymbol{k})\right|\rightarrow\Delta_{\tau s}$ 
there will be no carriers injected about $\boldsymbol{K'}$ by light of this helicity. 

\subsubsection{Equal helicities}

\begin{figure}[hbt!]
	\centering
	\begin{subfigure}{.155\textwidth}
		\centering
		\includegraphics[width=0.95\linewidth]{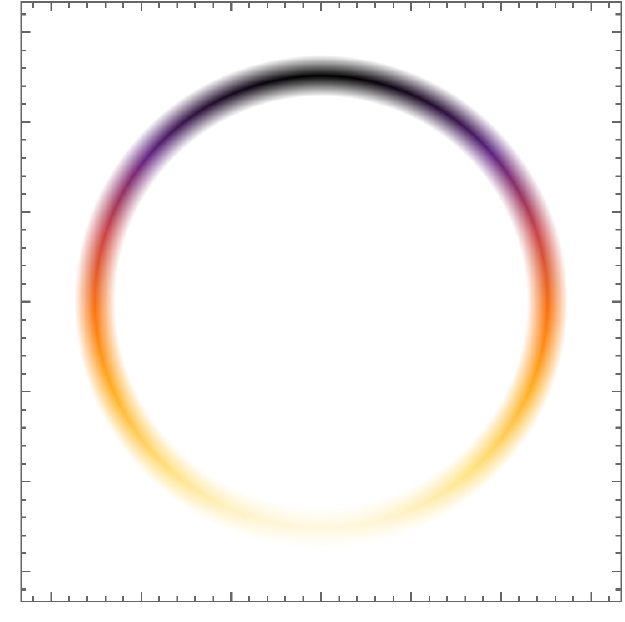}
		\caption{$\Delta\phi_{12}=0$}
		\label{fig:sub1}
	\end{subfigure}%
	\centering
	\begin{subfigure}{.155\textwidth}
		\centering
		\includegraphics[width=0.95\linewidth]{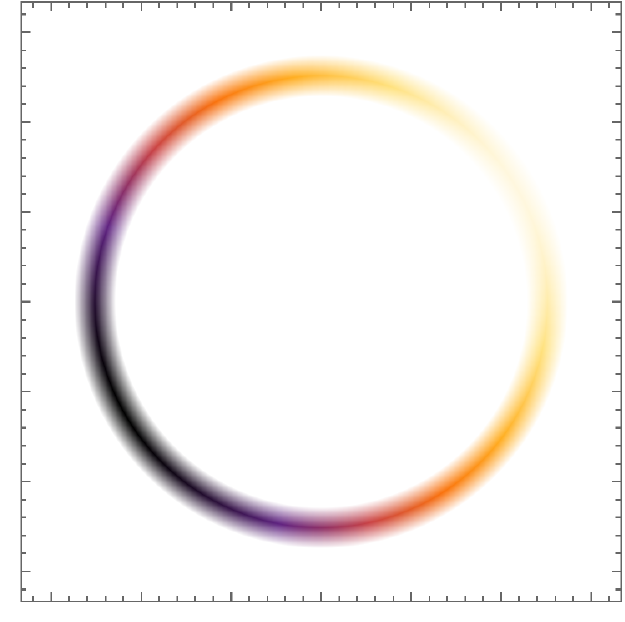}
		\caption{$\Delta\phi_{12}=2\pi/3$}
		\label{fig:sub1}
	\end{subfigure}
	\begin{subfigure}{.155\textwidth}
		\centering
		\includegraphics[width=0.95\linewidth]{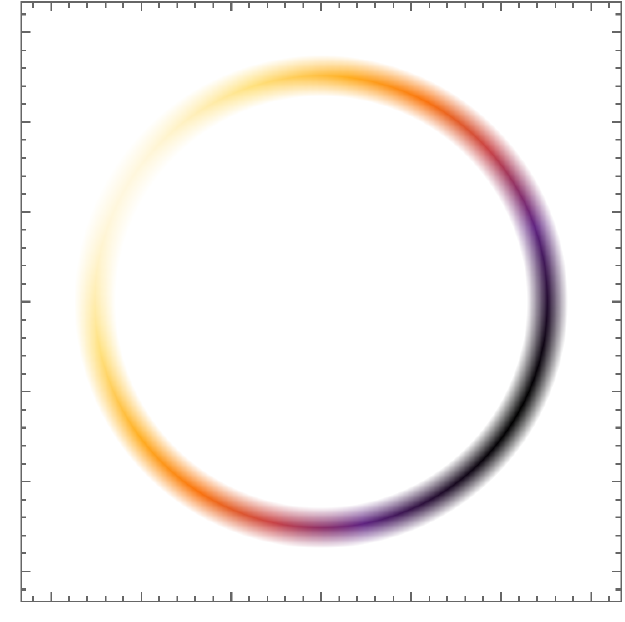}
		\caption{$\Delta\phi_{12}=4\pi/3$}
		\label{fig:sub2}
	\end{subfigure}
	\caption{Dependence of $\frac{d}{dt}\left\langle n_{c}(\boldsymbol{k})\right\rangle _{1+2}$ on $\Delta\phi_{12}$ for fields of frequency $\omega$ and $2\omega$ both circularly polarized in $\boldsymbol{\hat{e}}_+$.}
	\label{fig:12CarrierInjCirc}
\end{figure}

We first consider 1+2 absorption, with \textbf{$\boldsymbol{\hat{e}}_{+}$}
being the polarization for the fields at both frequencies. The injected
carrier distributions about the $\boldsymbol{K}$ point are shown in Fig.~\ref{fig:12CarrierInjCirc};
those about $\boldsymbol{K'}$ are qualitatively the same, but with far fewer carriers
injected. As the relative phase parameter $\Delta\phi_{12}$ is varied
from $0$ to $2\pi$ the direction of the injected current varies
continuously over the same angular range in real space. Since more
carriers are injected about the $\boldsymbol{K}$ valley than about the $\boldsymbol{K'}$ valley,
and as carriers injected in different valleys have different spins, the injected
current will be spin polarized. This was discussed previously \cite{Muniz15}. The scenario for 2+3 absorption is qualitatively the same and thus not included as a figure; as $\Delta\phi_{23}$ varies from $0$ to
$2\pi$ the direction of the injected spin current varies from $0$
to $2\pi$ in real space. 

\begin{figure}[hbt!]
	\centering
	\begin{subfigure}{.155\textwidth}
		\centering
		\includegraphics[width=0.95\linewidth]{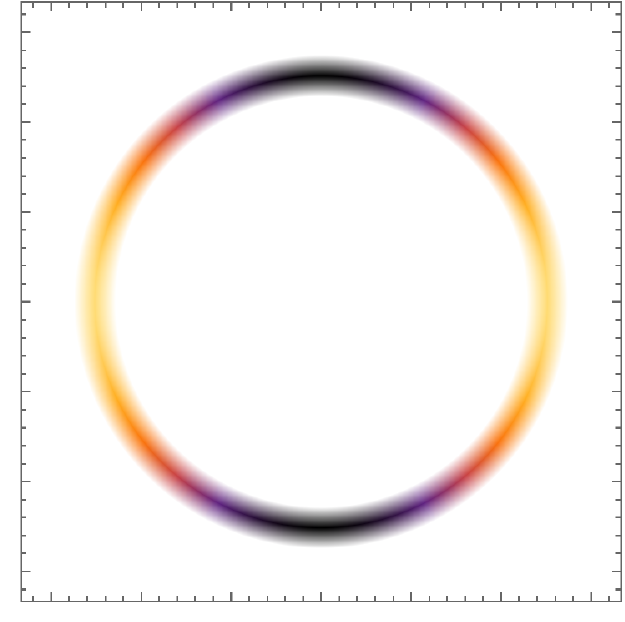}
		\caption{$\Delta\phi_{13}=0$}
		\label{fig:sub1}
	\end{subfigure}%
	\centering
	\begin{subfigure}{.155\textwidth}
		\centering
		\includegraphics[width=0.95\linewidth]{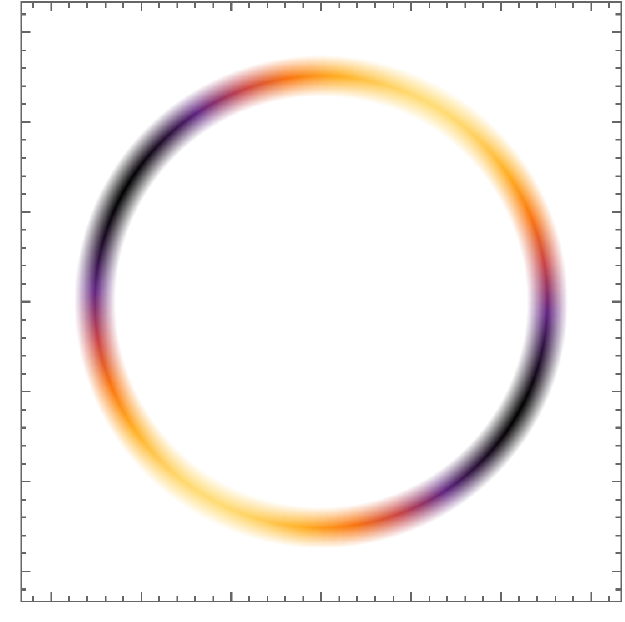}
		\caption{$\Delta\phi_{13}=2\pi/3$}
		\label{fig:sub1}
	\end{subfigure}
	\begin{subfigure}{.155\textwidth}
		\centering
		\includegraphics[width=0.95\linewidth]{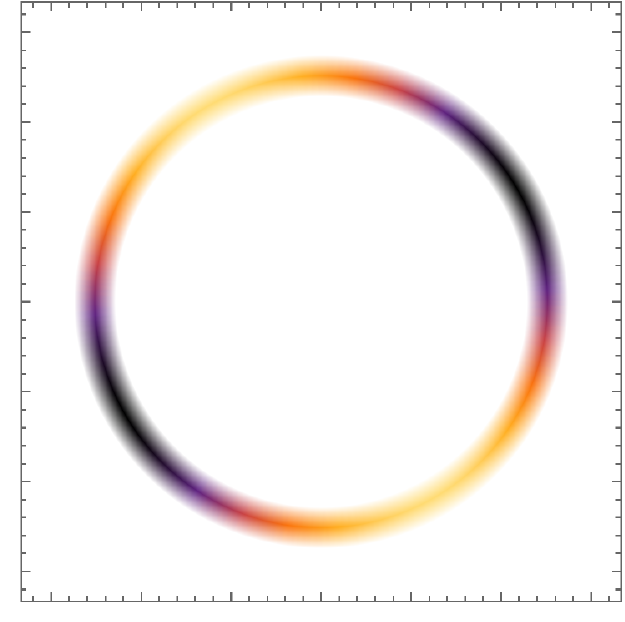}
		\caption{$\Delta\phi_{13}=4\pi/3$}
		\label{fig:sub1}
	\end{subfigure}
	\caption{Dependence of $\frac{d}{dt}\left\langle n_{c}(\boldsymbol{k})\right\rangle _{1+3}$ on $\Delta\phi_{13}$ for fields of frequency $\omega$ and $3\omega$ both circularly polarized in $\boldsymbol{\hat{e}}_+$.}
	\label{fig:13CarrierInjCirc}
\end{figure}

For 1+3 absorption there is no injected current,
as expected, but there is a nonuniform distribution of injected carriers
in the Brillouin zone that rotates as the relative phase parameter $\Delta\phi_{13}$
varies; this is shown in Fig.~\ref{fig:13CarrierInjCirc}. Here the total number of carriers
injected does not vary with $\Delta\phi_{13}$, as was possible for linearly
polarized excitation, but since there are more carriers
injected around $\boldsymbol{K}$ than $\boldsymbol{K'}$, and since the spin of the injected
carriers is correlated with the valley, the injected carriers are
spin polarized.

The scenario for $\boldsymbol{\hat{e}}_{-}$ polarized light is essentially
the same for all interference processes, but with the predominant
valley and spin of the injected carriers reversed. 

\subsubsection{Opposite helicities}
\begin{figure}[hbt!]
	\centering
	\begin{subfigure}{.155\textwidth}
		\centering
		\includegraphics[width=0.95\linewidth]{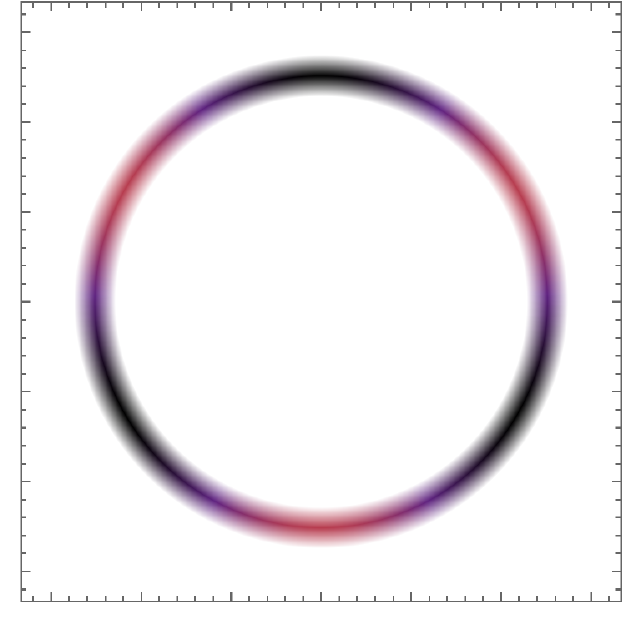}
		\caption{$\Delta\phi_{12}=0$}
		\label{fig:sub1}
	\end{subfigure}%
	\centering
	\begin{subfigure}{.155\textwidth}
		\centering
		\includegraphics[width=0.95\linewidth]{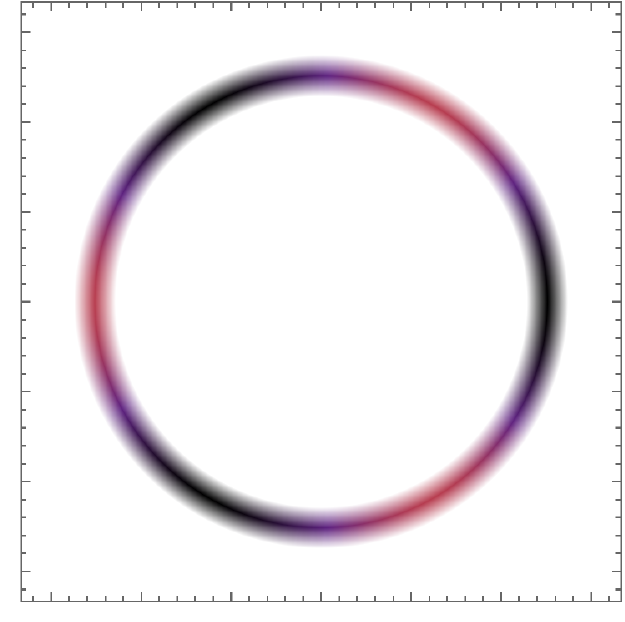}
		\caption{$\Delta\phi_{12}=\pi/2$}
		\label{fig:sub1}
	\end{subfigure}
	\begin{subfigure}{.155\textwidth}
		\centering
		\includegraphics[width=0.95\linewidth]{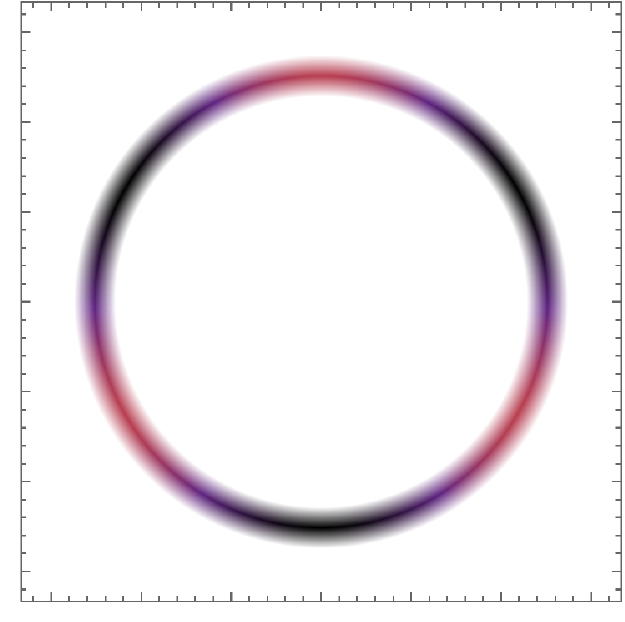}
		\caption{$\Delta\phi_{12}=\pi$}
		\label{fig:sub2}
	\end{subfigure}
	\caption{Dependence of $\frac{d}{dt}\left\langle n_{c}(\boldsymbol{k})\right\rangle _{1+2}$ on $\Delta\phi_{12}$ for fields orientated according to  $\boldsymbol{\hat{e}}_{\omega}=\boldsymbol{\hat{e}}_+$ and $\boldsymbol{\hat{e}}_{2\omega}=\boldsymbol{\hat{e}}_-$.}
	\label{fig:23CarrierInjCircOpp}
\end{figure}
For a transition precisely at the band gap facilitated by fields having opposite helicity,
carriers are injected into each valley by only one of the two frequencies, namely, the frequency 
associated with the helicity of light that couples to that particular valley, and of course
there is no interference. For higher excitation energies interference
does arise because each polarization injects carriers into both
valleys, although more into one than into the other. Yet in our model
for the TMDs there is no variation in the number of carriers injected
as the appropriate phase parameter is varied, nor is there net current
injected for $any$ of the interference processes. However, anisotropic carrier distributions
are injected that rotate as with variation in the relative phase parameter.
For 1+2 absorption the anisotropic distribution has three-fold rotational
symmetry, for 1+3 absorption it has four-fold rotational symmetry,
and for 2+3 interference it has five-fold symmetry. We show the first
of these for our excitation energy of $\hbar\Omega=1.5$ eV in Fig.~\ref{fig:23CarrierInjCircOpp}. 

\section{Frequency Dependence of Injection Coefficients}
\label{sec:6}
\begin{figure*}[t]
	\centering
	\begin{subfigure}{.45\textwidth}
		\centering
		\includegraphics[width=0.95\linewidth]{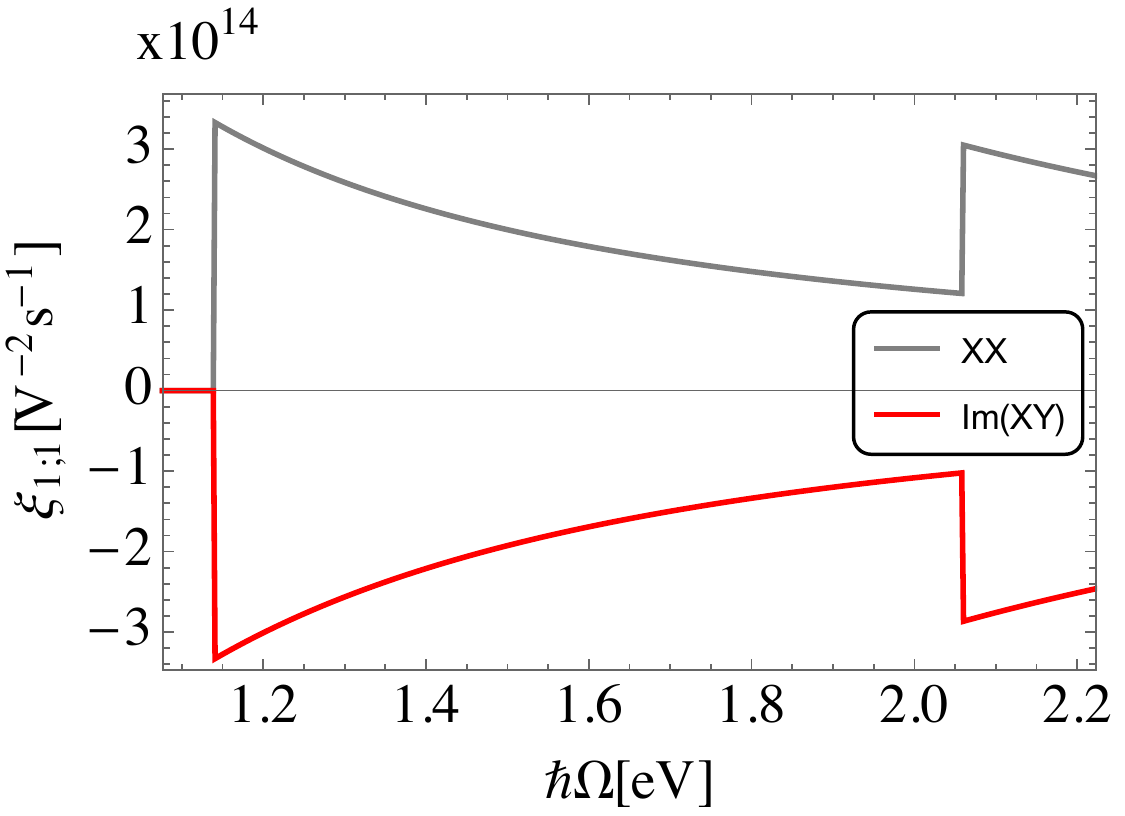}
		\label{fig:12crossA}
	\end{subfigure}%
	\centering
	\begin{subfigure}{.45\textwidth}
		\centering
		\includegraphics[width=0.95\linewidth]{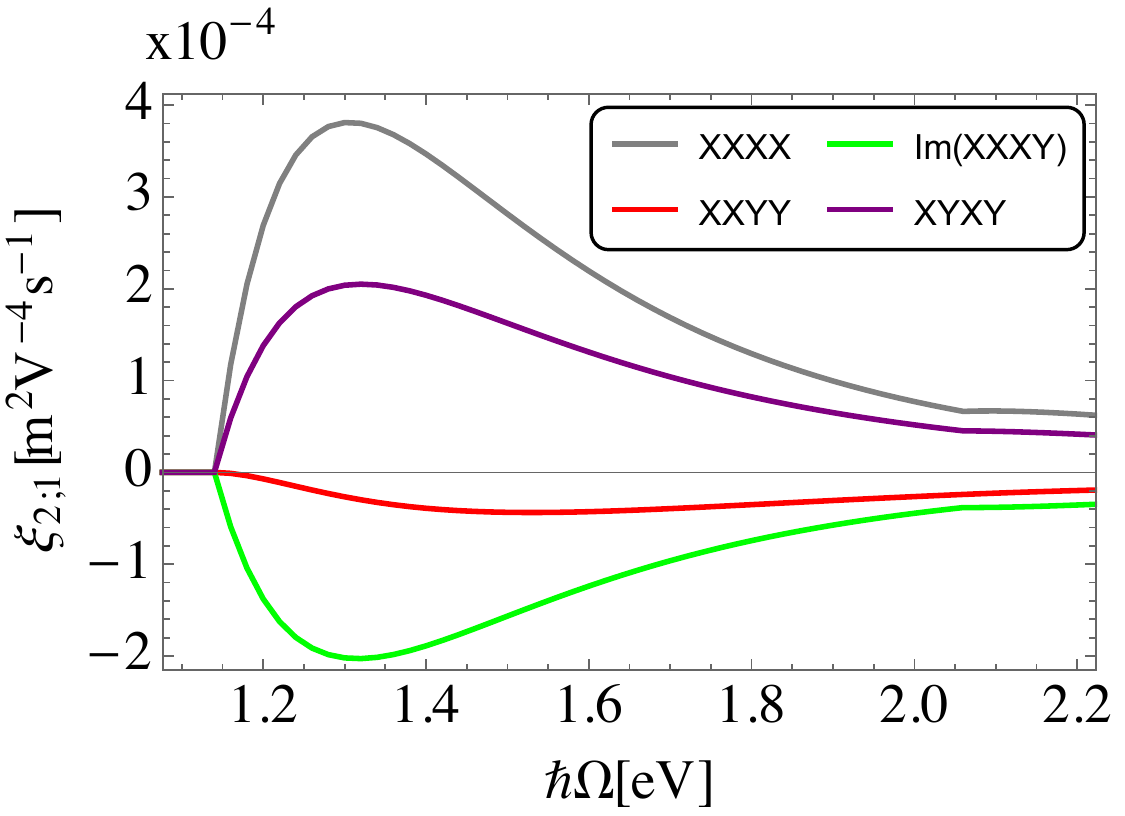}
		\label{fig:12crossC}
	\end{subfigure}
	\centering
	\begin{subfigure}{.45\textwidth}
		\centering
		\includegraphics[width=0.95\linewidth]{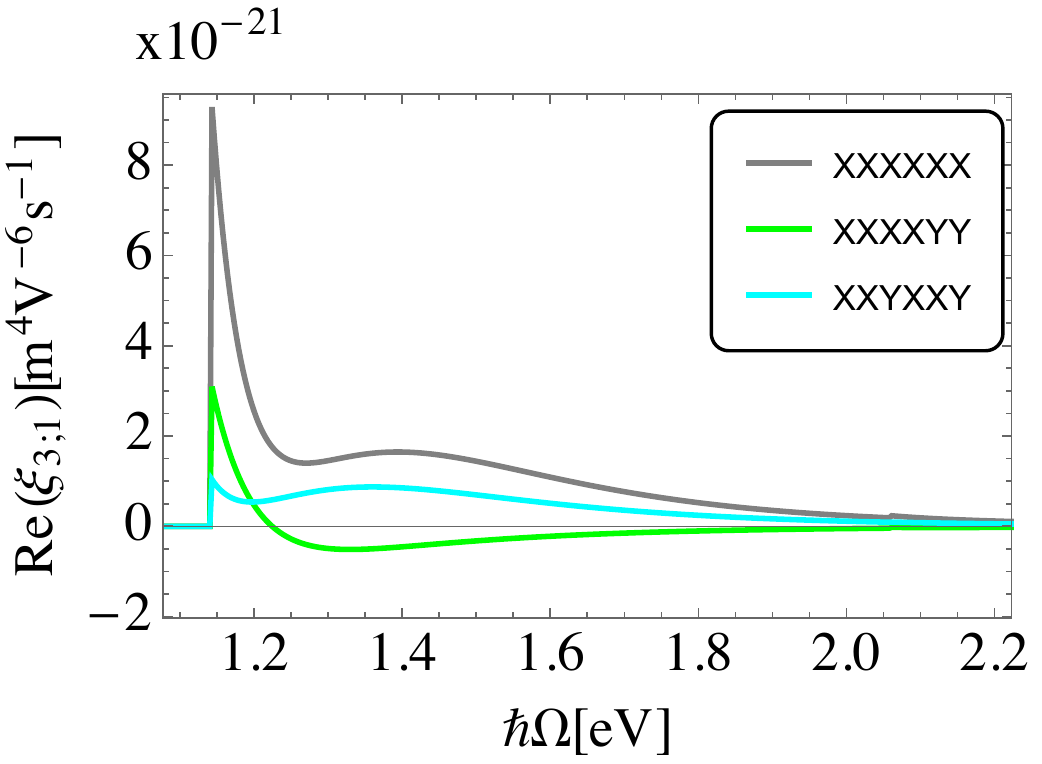}
		\label{fig:12crossD}
	\end{subfigure}%
	\centering
	\begin{subfigure}{.45\textwidth}
		\centering
		\includegraphics[width=0.95\linewidth]{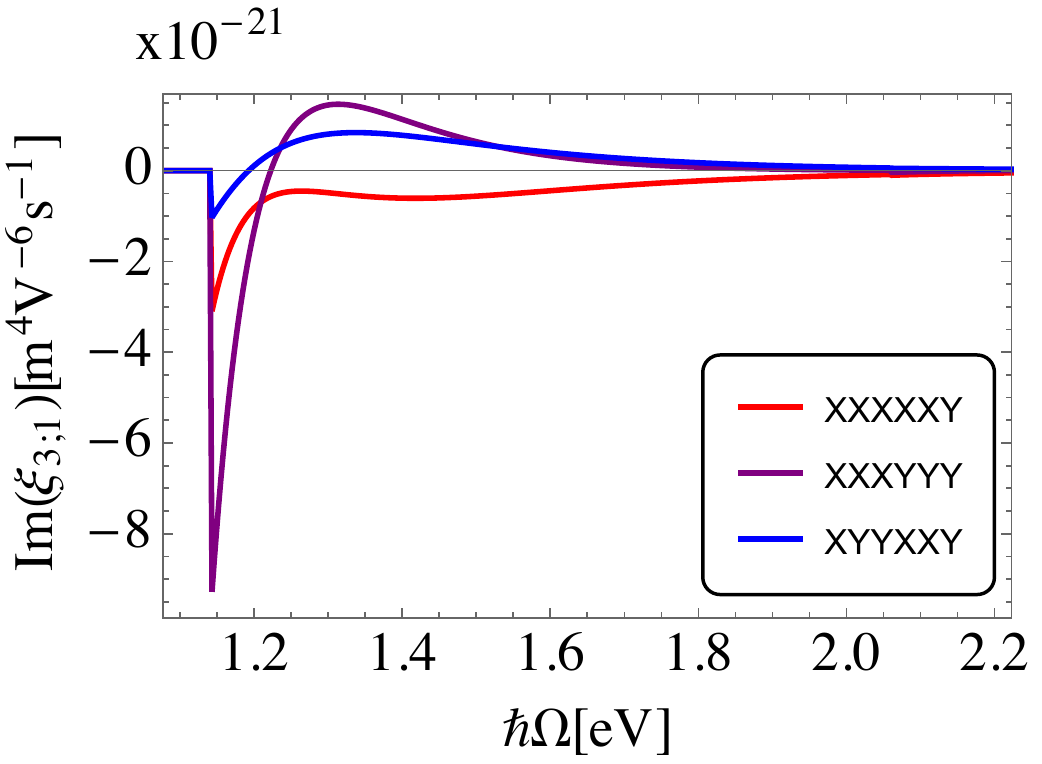}
		\label{fig:InjectionCoeffs}
	\end{subfigure}
	\caption{Excitation energy dependence of the independent components of the carrier injection response tensor for single color photon absorption processes. We plot nonzero components about the $\boldsymbol{K}$ valley.}
	\label{fig:123CarrierInjCoeffs}
\end{figure*}
As the excitation energy $\hbar\Omega$ is varied, the carrier and current injection distributions
shown above will change. For an overview of this we look at plots of the injection
coefficients themselves. 

\begin{figure*}
	\centering
	\begin{subfigure}[b]{.45\textwidth}
		\centering
		\includegraphics[width=0.95\linewidth]{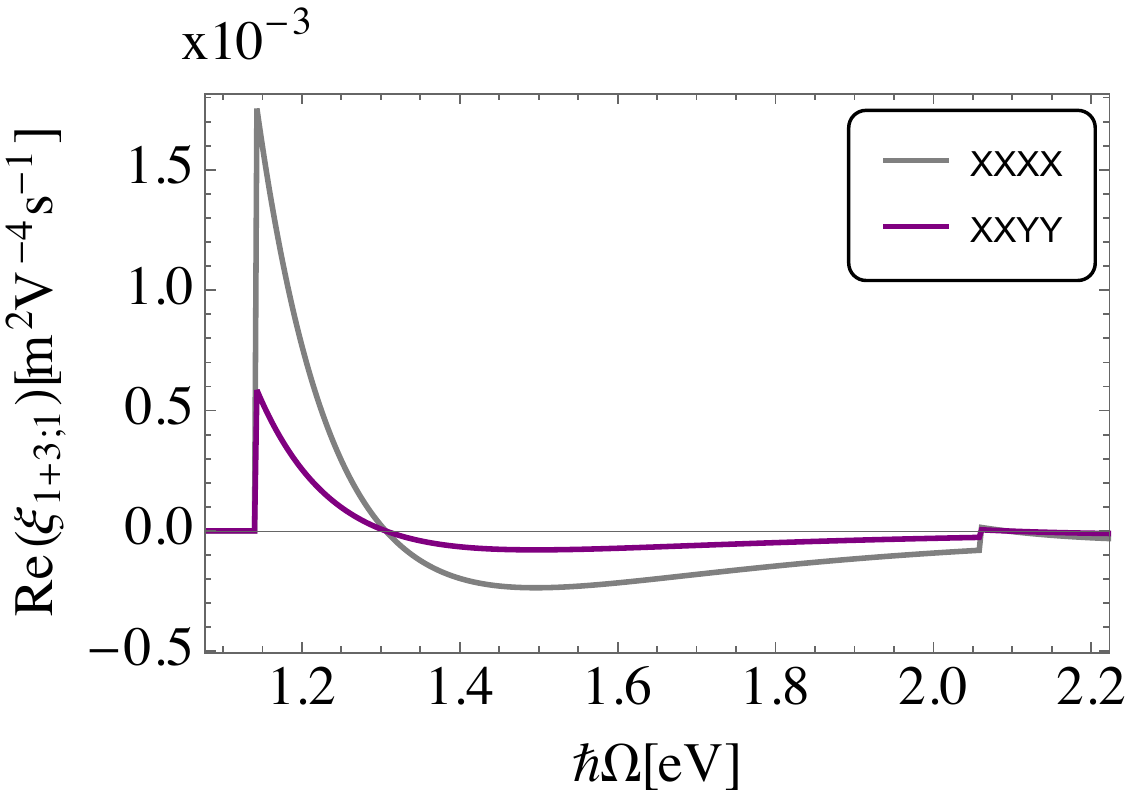}
		\label{fig:12crossA}
	\end{subfigure}%
	\centering
	\begin{subfigure}[b]{.45\textwidth}
		\centering
		\includegraphics[width=0.95\linewidth]{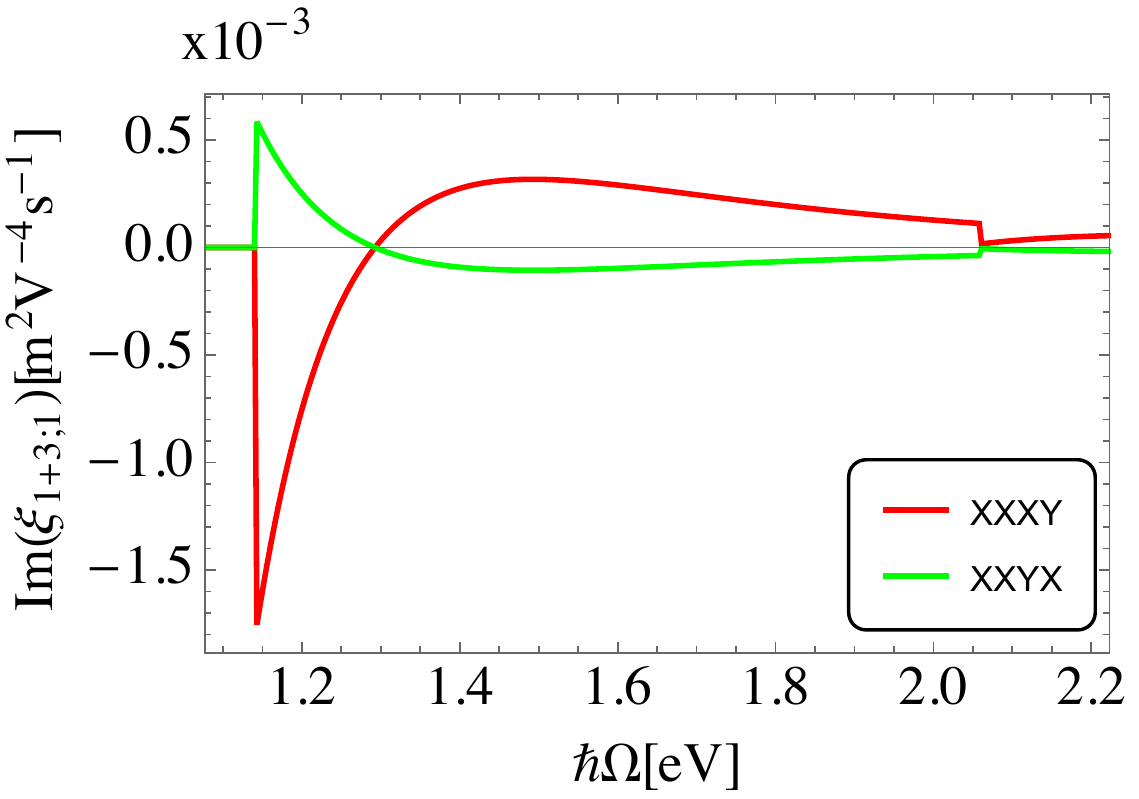}
		\label{fig:12crossC}
	\end{subfigure}
	\caption{Excitation energy dependence of the independent components of the carrier injection response tensor arising from the interference of photon absorption processes. We plot the nonzero components about the $\boldsymbol{K}$ valley.}
	\label{fig:CarrierInjCoeffs}
\end{figure*}

\subsection{Carrier injection coefficients}
In Fig. \ref{fig:123CarrierInjCoeffs} we show contributions to the carrier injection rate about the $\boldsymbol{K}$ point arising solely from single color absorption:  $\xi_{1}$, $\xi_{2}$, and $\xi_{3}$. The total coefficient, $\xi_{n}$,
is given as a sum over contributions $\xi_{n;\tau s}$ associated with valley $\tau$ and spin $s$, $\xi_{n}=\sum_{\tau s}\xi_{n;\tau s}$; however, to label the plots we use the notation $\xi_{n;\tau}=\sum_{s}\xi_{n;\tau s}$. In Fig.~\ref{fig:CarrierInjCoeffs} we display the interference terms that give rise to a nonvanishing carrier injection rate. As it happens, this term is only nonvanishing in the case of $1+3$ absorption.

We find the real valued components of the response tensor to be valley independent, 
while the imaginary parts differ by a sign between $\boldsymbol{K}$ and $\boldsymbol{K'}$. So the real part characterizes the total carrier injection, while the imaginary part characterizes the imbalance of injected carriers between the valleys. For one-photon absorption we have $\xi^{xx}=\xi^{yy}$ and they are both real; here and below we only show independent components. The cross
term $\xi^{xy}$ is imaginary, reflecting the structure of the interband
matrix element (\ref{eq:interband}), and close but not identically
equal in magnitude to $\xi^{xx}$. There is the usual step-like increase
in the one-photon absorption coefficient at the band gap because the
matrix element (\ref{eq:interband}) is finite there, and a second
step-like increase in magnitude at the onset of absorption from the
lower valence band. 

Unlike the one-photon carrier injection coefficients, the two-photon
injection coefficients have a smooth initial onset because of the
presence of intraband matrix elements (\ref{eq:intraband}) appearing in the
expression (\ref{TMDsCoeffs}) for the $R_{cv}^{(2)ab}$, which arises
in the expression (\ref{eq:2+3carriers}) for $\xi_{2}$; these intraband
elements vanish at the band gap and change continuously as one moves away from it. 
The onset of absorption from the
lower valence band is also smooth for the same reason. The overall
magnitudes of the $\xi_{2}$ coefficients drop off faster with increasing
excitation energy than do those of the $\xi_{1}$ coefficients because
a larger number of frequencies appear in the denominator of the coefficients
(\ref{eq:Rscript_amplitudes}) $\mathcal{R}_{cv}^{(N)}(\boldsymbol{k})$
as $N$ is increased. 
\begin{figure*}[hbt!]
	\centering
	\begin{subfigure}[b]{.45\textwidth}
		\centering
		\includegraphics[width=0.95\linewidth]{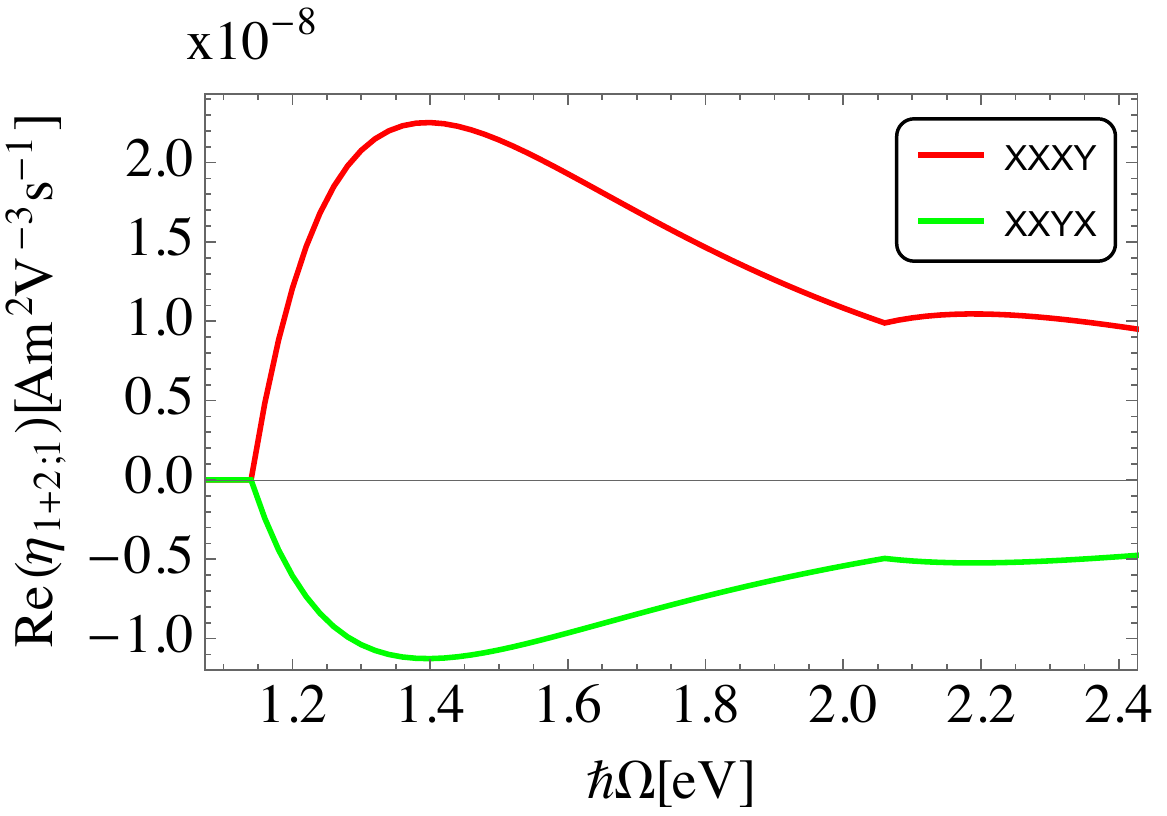}
	\end{subfigure}%
	\begin{subfigure}[b]{.45\textwidth}
		\centering
		\includegraphics[width=0.95\linewidth]{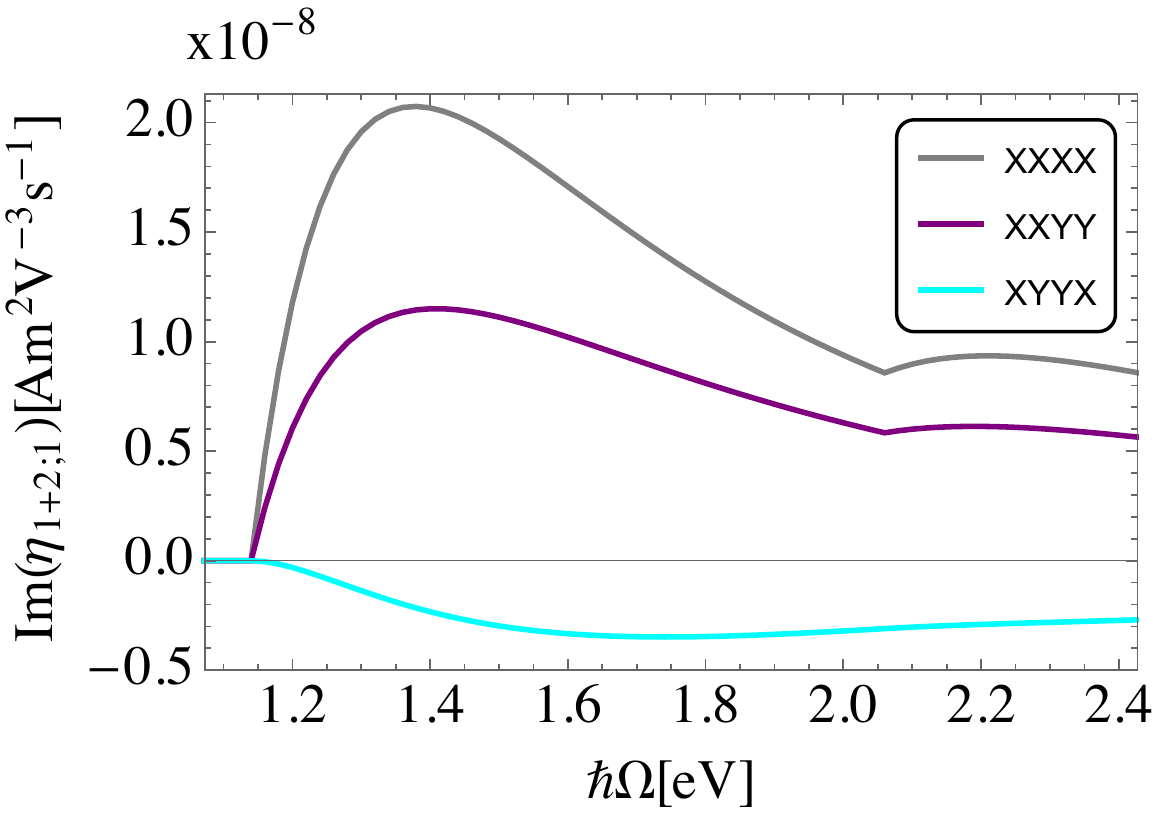}
	\end{subfigure}
	\caption{Excitation energy dependence of the independent components of the current injection response tensor for 1+2 photon absorption processes. We plot components about a single valley, $\boldsymbol{K}$, and omit vanishing components.}
	\label{fig:12CurrentInjCoeffs}
\end{figure*}
\begin{figure*}
	\centering
	\begin{subfigure}[b]{.45\textwidth}
		\centering
		\includegraphics[width=0.95\linewidth]{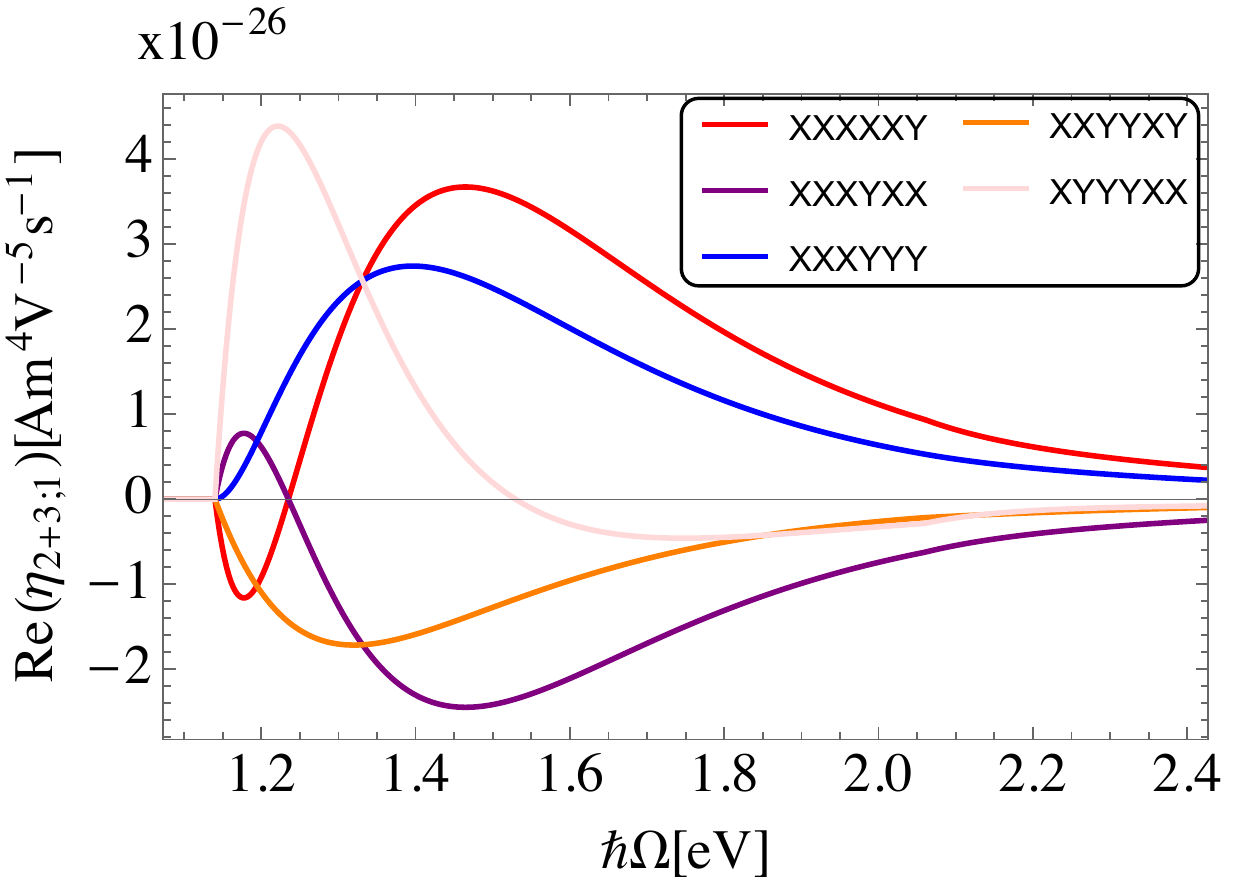}
	\end{subfigure}%
	\begin{subfigure}[b]{.45\textwidth}
		\centering
		\includegraphics[width=0.95\linewidth]{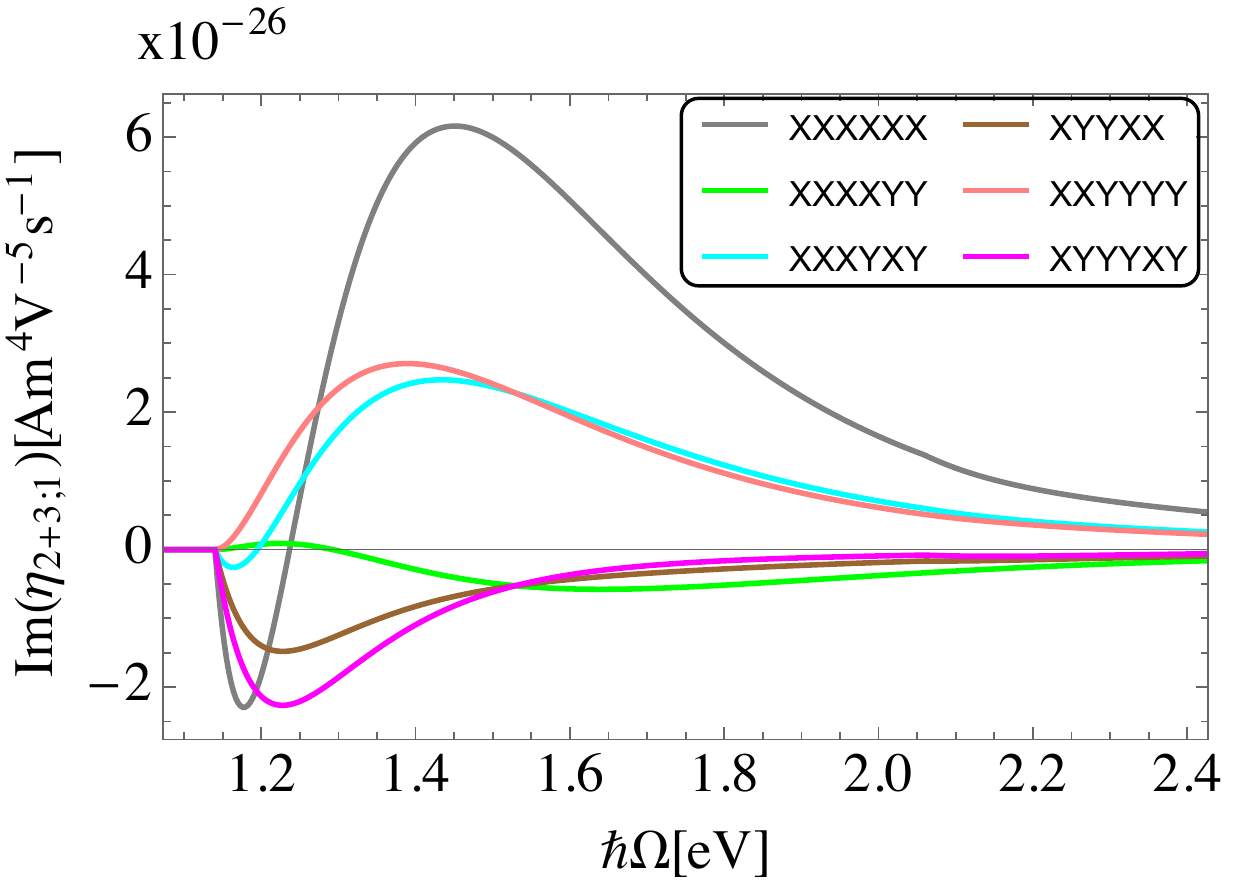}
	\end{subfigure}
	\caption{Excitation energy dependence of the independent components of the current injection response tensor for 2+3 photon absorption processes. We plot components about a single valley, $\boldsymbol{K}$, and omit vanishing components.}
	\label{fig:23CurrentInjCoeffs}
\end{figure*}

The three-photon carrier injection coefficients $\xi_{3}$ share some
of the features of the $\xi_{1}$ and some of the $\xi_{2}$, since
$\mathcal{R}_{cv}^{(3)}(\boldsymbol{k})$ contains pure interband (PR)
contributions (as does $\mathcal{R}_{cv}^{(1)}(\boldsymbol{k})$) and
interband-intraband contributions (RA) that involve both types of
matrix elements (as does $\mathcal{R}_{cv}^{(2)}(\boldsymbol{k})$);
this also leads to the coefficients $\xi_{3}$ exhibiting a more complicated
energy dependence than the elements of either $\xi_{1}$ or $\xi_{2}$.
For example, the sharp onset at the band edge of the different components
of $\xi_{3}$ is due to the PR contribution, while the later, smoother
rise (or fall) in the different components is due to an increasing
contribution from the RA contributions as the magnitude of $\boldsymbol{\mathfrak{v}}_{cc}(\boldsymbol{k})-\boldsymbol{\mathfrak{v}}_{vv}(\boldsymbol{k})$
increases at larger $k$; certain components
of $\xi_{3}$ can actually vanish as the PR and RA contributions cancel.

The interference coefficient $\xi_{1+3}$ again shows the presence
of PR and RA contributions. The first leads to the step-like increase
at the band edge, and another step-like change at the onset of absorption
from the lower valence band. The second leads to the smoother change
with increasing excitation energy that actually produces a sign reversal 
of the nonvanishing coefficients as the PR and RA contributions
cancel one another. Note that all components of Re$\left(\xi_{1+3}\right)$,
which govern the net carrier injection, vanish at a particular excitation
energy $\hbar\Omega=1.31$ eV.
Thus the destructive interference between the PR and RA contributions
(see (\ref{TMDsCoeffs}) for $R_{cv}^{(3)abd}$) leads
to a frequency region where there is very small coherent control of
the carrier injection rate. At the energy that Re$\left(\xi_{1+3}\right)$
vanishes Im$\left(\xi_{1+3}\right)$ is not strictly vanishing, and
therefore there will be carrier injection interference in both the
$\boldsymbol{K}$ and $\boldsymbol{K'}$ valleys, but the interference effects will cancel when
producing the total carrier injection rate. Nonetheless, at this energy
Im$\left(\xi_{1+3}\right)$ is very small, and so even the ``valley-by-valley''
interference will be very small.
\begin{figure*}
	\centering
	\begin{subfigure}[b]{.45\textwidth}
		\centering
		\includegraphics[width=0.95\linewidth]{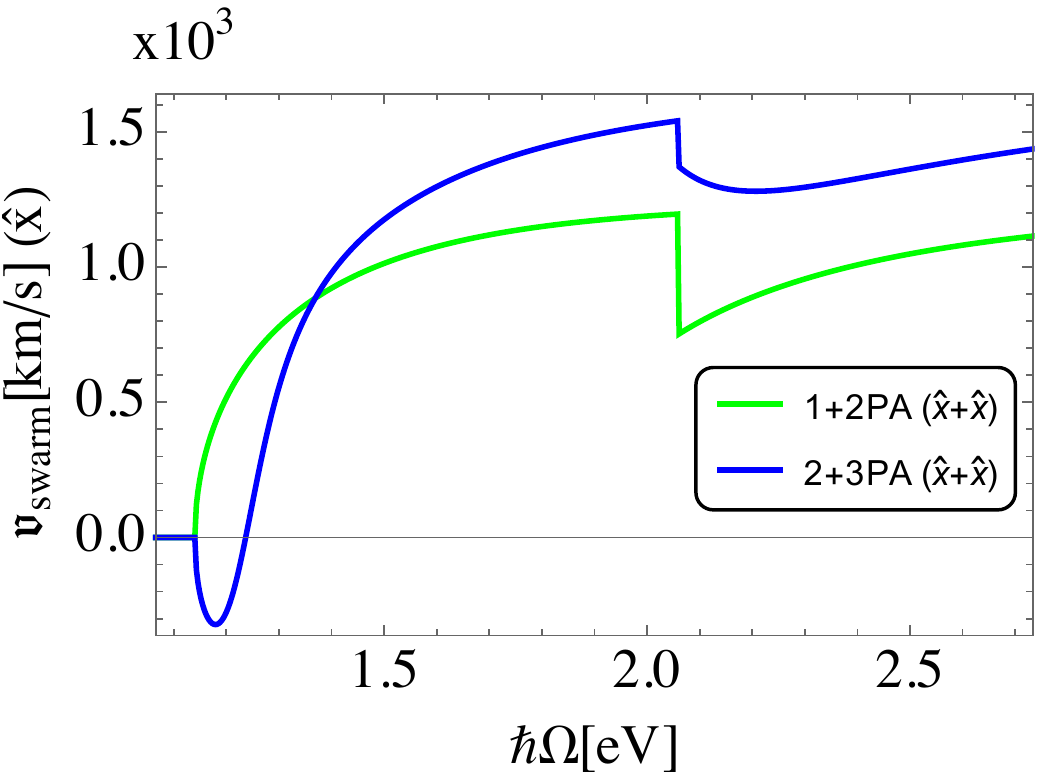}
		\caption{Co-linearly polarized incident fields.}
		\label{fig:vSwarmCo}
	\end{subfigure}%
	\begin{subfigure}[b]{.45\textwidth}
		\centering
		\includegraphics[width=0.95\linewidth]{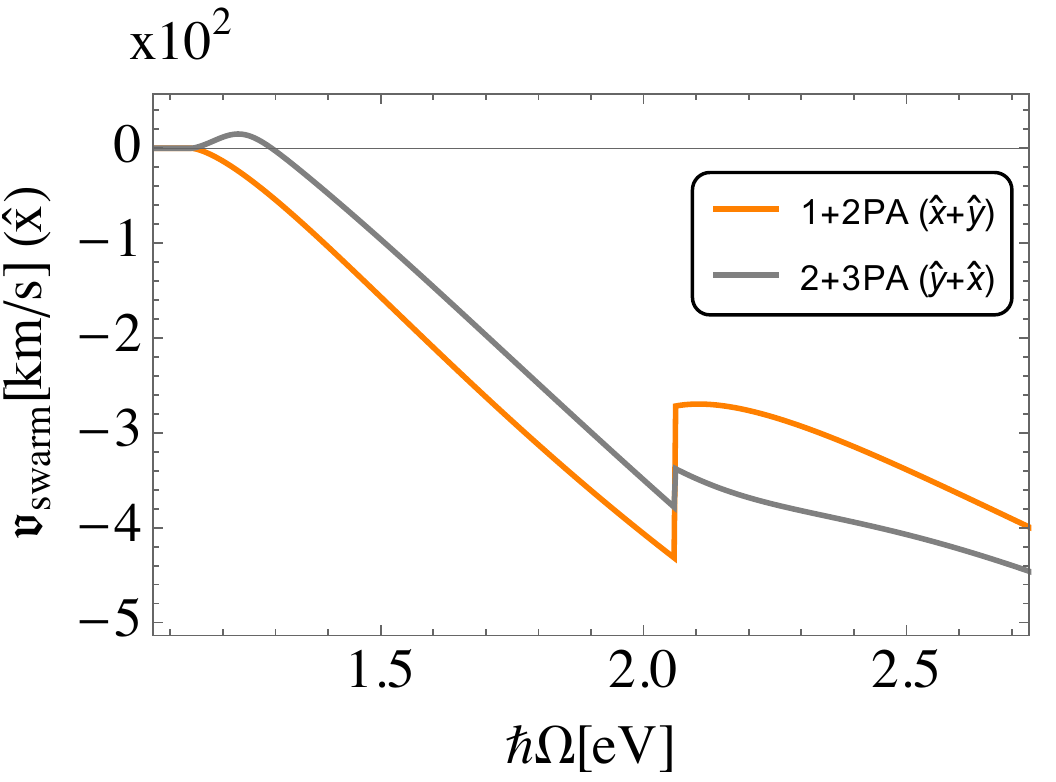}
		\caption{Cross-linearly polarized incident fields.}
		\label{fig:vSwarmCross}
	\end{subfigure}
	\caption{Swarm velocities for co- and cross-linearly polarized optical fields. The relative polarizations of the fields are indicated in the legend, and combinations of polarizations that lead to a vanishing swarm velocity in the $\pm\boldsymbol{\hat{x}}$ direction are omitted.}
	\label{fig:vSwarm}
\end{figure*}

\subsection{Current injection coefficients}
The contributions to the current injection coefficients $\eta_{1+2}$
and $\eta_{2+3}$ from the $\boldsymbol{K}$ valley are shown in Fig.~\ref{fig:12CurrentInjCoeffs} and \ref{fig:23CurrentInjCoeffs}, respectively, as a
function of excitation energy; an analogous notation to that used for plotting 
carrier injection rates is adopted here. In contrast to the carrier injection, here we find the imaginary valued components of the response tensor to be valley independent, while the real parts differ by a sign between $\boldsymbol{K}$ and $\boldsymbol{K'}$, so it is the imaginary parts that completely characterize the charge current injected into the system. Earlier work \cite{Bhat05} on simpler systems showed that including the
Coulomb interaction between injected electrons and holes led to the
prediction of a phase shift introduced in the response, and thus to
an expected maximum injection current occurring at a relative phase
parameter different than $\Delta\phi_{12}$ (or $\Delta\phi_{23})$
= $\pi/2$ and $3\pi/2$. That should be expected here as well; we
plan to investigate the inclusion of this effect in a later publication.

Neither $\eta_{1+2}$ nor $\eta_{2+3}$ shows step-like behavior as
the energy crosses band gaps because of the involvement of intraband
matrix elements in each term of the expression, again arising through the  $R_{cv}^{(2)ab}$ term. Likewise there is
no step-like behavior at the onset of the absorption from the lower
valence band. Again $\eta_{2+3}$ is suppressed more at excitation
energies greater than $\eta_{1+2}$ because of frequency factors in the denominators.
The nature of the excitation energy dependence of $\eta_{2+3}$ is
much more complicated than that of $\eta_{1+2}$, and arises again
primarily because of the combination of PR and RA terms in the third
order response. Note that the sign of many of the imaginary components
of $\eta_{2+3}$ changes as a function of excitation energy, exhibiting
the interplay between those two terms in the third order response.
Since it is these components that characterize the total current injection
in 2+3 absorption, we can expect interesting consequences in the excitation
energy dependence of the injected current, which we consider next.

\subsection{Swarm velocities}
We now characterize the average velocity of the injected carriers
by considering the ``swarm velocity'', which is the current injection
rate divided by the total charge injection rate 
\footnote{ Notice that for constant injection rates, the expression for the swarm velocity reduces to $\bm{\mathfrak{v}}_{swarm}(\hbar\Omega) = < \bm{J} > / e <n_{c}>$}. 
For 1+2 absorption it is given by 
\begin{equation} \begin{aligned}
 \boldsymbol{\mathfrak{v}}_{swarm}(\hbar\Omega)=\left[\frac{\frac{d}{dt}\left\langle \boldsymbol{\mathcal{J}}\right\rangle _{1+2}}{e\left(\frac{d}{dt}\left\langle n_{c}\right\rangle _{1}+\frac{d}{dt}\left\langle n_{c}\right\rangle _{1+2;i}+\frac{d}{dt}\left\langle n_{c}\right\rangle _{2}\right)}\right]_{max},\label{eq:swarm1+2}
\end{aligned} \end{equation}
and for 2+3 absorption it is given by 
\begin{equation} \begin{aligned}
\boldsymbol{\mathfrak{v}}_{swarm}(\hbar\Omega)=\left[\frac{\frac{d}{dt}\left\langle \boldsymbol{\mathcal{J}}\right\rangle _{2+3}}{e\left(\frac{d}{dt}\left\langle n_{c}\right\rangle _{2}+\frac{d}{dt}\left\langle n_{c}\right\rangle _{2+3;i}+\frac{d}{dt}\left\langle n_{c}\right\rangle _{3}\right)}\right]_{max},
\label{eq:swarm2+3}
\end{aligned} \end{equation} 
where the subscript
$max$ indicates that the relative amplitudes of the fields appearing,
and the relative phase parameter ($\Delta\phi_{12}$ and $\Delta\phi_{23}$
respectively), are set to guarantee that the magnitude of the swarm
velocity is a maximum. In both 1+2 and 2+3 excitation the phase parameter
that does this can be $\Delta\phi=\pi/2$ or $\Delta\phi=3\pi/2$,
and we choose the latter.

We look at the examples of co- and cross-linearly polarized light
for the excitation scenarios, where the current injected is in the
$\pm\boldsymbol{\hat{x}}$ direction. Recall that in the co-linear case
this arises with all fields polarized in the $\boldsymbol{\hat{x}}$
direction, while in the cross-linear case it arises for $\boldsymbol{\hat{e}}_{3\omega}=\boldsymbol{\hat{x}}$
and $\boldsymbol{\hat{e}}_{3\omega/2}=\boldsymbol{\hat{y}}$ in 1+2
absorption, and for $\boldsymbol{\hat{e}}_{\omega}=\boldsymbol{\hat{x}}$
and $\boldsymbol{\hat{e}}_{3\omega/2}=\hat{\boldsymbol{y}}$ in 2+3
absorption. As the excitation energy
is increased we expect the magnitude of the swarm velocity to increase,
simply because the carriers injected will have larger velocities.

Comparing two swarm velocities at the same excitation energy gives
a measure of how well localized the carriers are in the Brillouin
zone. Under the specified excitation conditions we plot the swarm velocities
for excitations facilitated by co-linearly polarized light in Fig.~\ref{fig:vSwarmCo}, and by cross-linearly
polarized light in Fig.~\ref{fig:vSwarmCross}. The former are generally larger
than the latter, and in particular the latter are very small near an excitation energy of $1.5$ eV,
especially for 2+3 absorption as noted in our discussion above;
these velocities do increase at larger excitation energy. In general
the swarm velocities for 2+3 absorption become larger than those of 1+2
absorption for high enough excitation energy. At low excitation
energies we see a reversal of the swarm velocity for 2+3 absorption
with increasing excitation energy, again due to interplay between
the PR and RA contributions to the three-photon absorption amplitude, and as a result its magnitude is
smaller than the swarm velocity for 1+2 absorption. The PR contribution to the amplitude is finite at the band gap, but the RA contribution vanishes. As the excitation energy moves away
from the gap, the RA contribution becomes finite with net opposite sign of the PR term; at an excitation energy of $\hbar\Omega=1.24$ eV 
for co-linearly polarized excitation, and $\hbar\Omega=1.29$ eV for cross-linearly
polarized light, the contributions cancel, leading to no net current
injection. Within the model adopted here, this adds another level of control over the injected current,
above and beyond what can be done by adjusting field intensities and
the relative phases parameters, and only arises for 2+3 absorption. Also, all the swarm velocity plots show a discontinuous jump as the energy
crosses the second band gap, as would be expected from the behavior
of the injection coefficients. 

\section{Results and Discussion}
We have shown that the quantum interference arising from 2+3 photon absorption can give rise to significantly more localized distributions of electronic excitations in the Brillouin zone than
the 1+2 counterpart (see Fig.~\ref{fig:vSwarm}). The primary reason for this is the increased number of intraband velocity matrix elements in the transition coefficients at third order perturbation
theory. The increased localization of these distributions is most apparent for co-linearly polarized incident optical fields, and it is also this orientation of fields that lends itself most to the idea of using QuIC as ``tweezers in the Brillouin zone'' (Fig.~\ref{fig:23CarrierInj}), using quantum interference of excitation processes as a mechanism to place carriers where one desires in $\boldsymbol{k}$ space. Studying the subsequent dynamics of such injected distributions is of interest from both theoretical and experimental perspectives, driven by the recent advances in time-resolved ARPES. In principle, one could directly implement the QuIC mechanism into pump--probe strategies to study nonequilibrium dynamics; 2+3 photon absorption could be used to set the system in a far-from equilibrium state with carriers well localized in $\boldsymbol{k}$ space, or to probe carriers in a well-localized region of $\boldsymbol{k}$ space, or both.

We have also shown that, akin to 1+2 absorption, the quantum interference arising in 2+3 absorption can be manipulated by varying a relative phase parameter; by doing so we can, for both 1+2 and 2+3 processes, change the direction of the injected current. Additionally, for transitions at sufficiently large energies, there will be a larger current injected from 2+3 absorption than from 1+2 absorption.

\section{Acknowledgements}
We thank Steve Cundiff, Kai Wang, David Jones, Andrea Damascelli, and Sergey Zhdanovich for useful discussions. This work was supported by the Natural Sciences and Engineering Research Council of Canada, including a scholarship awarded to P. T. M.

\appendix

\section{Injection Rates}
\label{PT}
In the interaction picture, a state $\left|\psi(t)\right\rangle$ evolves under $\left|\psi(t)\right\rangle=\mathcal{U}(t)\left|gs\right\rangle$, where $\mathcal{U}(t)$ is given by (\ref{evolutionOperator}). Using this, in combination with the resolution of identity in a multi-particle Hilbert space
\begin{equation*}
\begin{aligned}
1=\left|gs\right\rangle\left<gs\right|+\sum_{cv\boldsymbol{k}}\left|cv\boldsymbol{k}\right\rangle\left<cv\boldsymbol{k}\right|+...,
\end{aligned}
\end{equation*}
where $\left|gs\right\rangle$ is the filled Fermi sea and $\left|cv\boldsymbol{k}\right\rangle\equiv a_{c\boldsymbol{k}}^{\dagger}a_{v\boldsymbol{k}}\left|gs\right\rangle$ are eigenstates of $\mathcal{H}_0$, we re-express $\left|\psi(t)\right\rangle$ as 
\begin{align}
\label{tEvo}
\left|\psi(t)\right\rangle&=\left(\left|gs\right\rangle\left<gs\right|+\sum_{cv\boldsymbol{k}}\left|cv\boldsymbol{k}\right\rangle\left<cv\boldsymbol{k}\right|+...\right)\mathcal{U}(t)\left|gs\right\rangle\nonumber\\
&=\left|gs\right\rangle\left<gs\right|\mathcal{U}(t)\left|gs\right\rangle+\sum_{cv\boldsymbol{k}}\left|cv\boldsymbol{k}\right\rangle\left<cv\boldsymbol{k}\right|\mathcal{U}(t)\left|gs\right\rangle+... \nonumber\\
&=\gamma_{0}(t)\left|gs\right\rangle +\underset{cv\boldsymbol{k}}{\sum}\gamma_{cv}(\boldsymbol{k},t)\left|cv\boldsymbol{k}\right\rangle+...,
\end{align}
where $\gamma_{0}(t)\equiv\left<gs\right|\mathcal{U}(t)\left|gs\right\rangle$, $\gamma_{cv}(\boldsymbol{k},t)\equiv\left<cv\boldsymbol{k}\right|\mathcal{U}(t)\left|gs\right\rangle$. Then, the expectation value of a general single-body operator $\mathcal{M}(t)$ can be written, neglecting higher-order electron-hole excitations, as
\begin{align*}
&\expval{\mathcal{M}(t)}{\psi(t)}=\\
&\qquad|\gamma_{0}(t)|^2\expval{\mathcal{M}(t)}{gs}\\
&\qquad+\sum_{cv\boldsymbol{k}}\gamma_{0}(t)^*\gamma_{cv}(\boldsymbol{k},t)\bra{gs}\mathcal{M}(t)\ket{cv\boldsymbol{k}}\nonumber\\
&\qquad+\sum_{cv\boldsymbol{k}}\gamma_{cv}(\boldsymbol{k},t)^*\gamma_{0}(t)\bra{cv\boldsymbol{k}}\mathcal{M}(t)\ket{gs}\nonumber\\
&\qquad+\sum_{\substack{cv\boldsymbol{k}\\c'v'\boldsymbol{k}'}}\gamma_{c'v'}(\boldsymbol{k}',t)^*\gamma_{cv}(\boldsymbol{k},t)\bra{c'v'\boldsymbol{k}'}\mathcal{M}(t)\ket{cv\boldsymbol{k}}.\nonumber
\end{align*}
The operators we are interested in are the densities given in (\ref{conserved}), and since $\ket{gs}$ is taken to be an equilibrium state, it does not contribute to these quantities. Also, since (\ref{conserved}) are $\boldsymbol{k}$-conserving and $\mathcal{M}(t)=e^{i\mathcal{H}_0t/\hbar}\mathcal{M}e^{-i\mathcal{H}_0t/\hbar}$, we find
\begin{equation*}
\begin{aligned}
&\bra{c'v'\boldsymbol{k}'}\mathcal{M}(t)\ket{cv\boldsymbol{k}}=\\
&\qquad e^{i\omega_{c'v'}(\boldsymbol{k})t}e^{-i\omega_{cv}(\boldsymbol{k})t}\bra{c'v'\boldsymbol{k}}\mathcal{M}\ket{cv\boldsymbol{k}}\delta_{\boldsymbol{kk}'},
\end{aligned}
\end{equation*}
where $\mathcal{M}$ is an operator in the Schrodinger picture. We consider the time derivatives of such expectation values, to give the so-called injection rate associated with $\mathcal{M}$. The second and third terms in the above expectation value are expected to be oscillating rapidly and are therefore neglected, as we are interested in the underlying dc response that may be experimentally captured by measurement with electrodes \cite{Hache97}. The injection rate associated with $\mathcal{M}$ is then given by
\begin{align}
&\frac{d}{dt}\expval{\mathcal{M}(t)}{\psi(t)}=\label{injRate}\\
&\qquad\sum_{cc'vv'\boldsymbol{k}}\frac{d}{dt}\left(e^{i\omega_{c'v'}(\boldsymbol{k})t}e^{-i\omega_{cv}(\boldsymbol{k})t}\gamma_{c'v'}(\boldsymbol{k},t)^*\gamma_{cv}(\boldsymbol{k},t)\right)\nonumber\\
&\qquad\qquad\qquad\times\bra{c'v'\boldsymbol{k}}\mathcal{M}\ket{cv\boldsymbol{k}}\nonumber
\end{align}
under the approximations we take. To investigate the $\boldsymbol{k}$-space distributions of such injection rates, the above is rewritten, in the continuous $\boldsymbol{k}$ limit, as 
\begin{equation}
\begin{aligned}
&\frac{d}{dt}\expval{\mathcal{M}(t)}{\psi(t)}=\int_{BZ}\frac{d^Dk}{\left(2\pi\right)^{D}} \frac{d}{dt}\left\langle \mathcal{M}\left(\boldsymbol{k};t\right)\right\rangle,
\label{density1}
\end{aligned}
\end{equation}
where we have defined
\begin{equation*}
\begin{aligned}
&\frac{d}{dt}\left\langle \mathcal{M}\left(\boldsymbol{k};t\right)\right\rangle\equiv\\
&\qquad\sum_{cc'vv'}\frac{d}{dt}\left(e^{i\omega_{c'v'}(\boldsymbol{k})t}e^{-i\omega_{cv}(\boldsymbol{k})t}\gamma_{c'v'}(\boldsymbol{k},t)^*\gamma_{cv}(\boldsymbol{k},t)\right)\\
&\qquad\qquad\qquad\times L^D\bra{c'v'\boldsymbol{k}}\mathcal{M}\ket{cv\boldsymbol{k}},
\end{aligned}
\end{equation*}
making connection with (\ref{eq:density_rewrite}). 

Now, the adopted model for TMDs simplifies the above to considering a single $c$ and $v$ for electrons of a given spin about each valley. This, in combination with (\ref{epsilonLimit}), gives
\begin{align}
&\frac{d}{dt}\left\langle \mathcal{M}\left(\boldsymbol{k};t\right)\right\rangle=\nonumber\\
&\qquad 2\pi\big|\mathcal{R}_{cv}(\boldsymbol{k})\big|^{2}L^D\bra{cv\boldsymbol{k}}\mathcal{M}\ket{cv\boldsymbol{k}}\delta\big(\Omega-\omega_{cv}(\boldsymbol{k})\big),\nonumber
\end{align}
which is indeed independent of $t$. Furthermore, recalling (\ref{eq:Rscript_amplitudes}) and including only the dominant contributions to $\mathcal{R}^{(N)}_{cv}(\boldsymbol{k})$ as dictated by the QuIC condition, the above can be written in the form 
\begin{align}
&\frac{d}{dt}\left\langle \mathcal{M}\left(\boldsymbol{k};t\right)\right\rangle=\label{density2}\\
&\underset{\substack{a,...b;N\\a',...b';N'}}{\sum}\mu_{NN'}^{a'\cdots b'a\cdots b}(\Omega;\boldsymbol{k})E_{-\Omega/N'}^{a'}\cdots E_{-\Omega/N'}^{b'}E_{\Omega/N}^{a}\cdots E_{\Omega/N}^{b},\nonumber
\end{align}
where we have defined the response coefficient density
\begin{align}
\mu_{NN'}^{a'\cdots b'a\cdots b}(\Omega;\boldsymbol{k})&\equiv 2\pi \left(R^{(N')a'\cdots b'}_{cv}(\boldsymbol{k};\Omega/N',...,\Omega/N')\right)^*\nonumber\\
&\times R^{(N)a\cdots b}_{cv}(\boldsymbol{k};\Omega/N,...,\Omega/N)\nonumber\\
&\times L^D\bra{cv\boldsymbol{k}}\mathcal{M}\ket{cv\boldsymbol{k}}\delta\big(\Omega-\omega_{cv}(\boldsymbol{k})\big).\nonumber
\end{align}
The form of the above has been simplified by taking the two-band limit; this is all that is required for our analysis. However, it can be shown that for a system with many bands the response coefficient density takes the form \cite{Muniz18}
\begin{align}
\mu_{NN'}^{a'\cdots b'a\cdots b}(\Omega;\boldsymbol{k})&= 2\pi \sum_{cc'vv'}\left(R^{(N')a'\cdots b'}_{c'v'}(\boldsymbol{k};\Omega/N',...,\Omega/N')\right)^*\nonumber\\
&\times R^{(N)a\cdots b}_{cv}(\boldsymbol{k};\Omega/N,...,\Omega/N)\delta_{\omega_{cv}(\boldsymbol{k}),\omega_{c'v'}(\boldsymbol{k})}\nonumber\\
&\times L^D\bra{c'v'\boldsymbol{k}}\mathcal{M}\ket{cv\boldsymbol{k}}\delta\big(\Omega-\omega_{cv}(\boldsymbol{k})\big).\nonumber
\end{align}

In the main text $\mu_{NN'}^{a'\cdots b'a\cdots b}(\Omega;\boldsymbol{k})$ corresponds to a particular absorption process's contribution to the total $\xi(\Omega;\boldsymbol{k})$ or $\eta(\Omega;\boldsymbol{k})$, depending if $\mathcal{M}$ equals $n_c$ or $\boldsymbol{\mathcal{J}}$, respectively. For example, if $\mathcal{M}=n_c$, and if we consider 1+2 PA (such that $N,N'\in(1,2)$), then
\begin{equation*}
\begin{aligned}                                                           
\mu_{11}^{a'a}(\Omega;\boldsymbol{k})&\longleftrightarrow\xi_{1}^{a'a}(\Omega;\boldsymbol{k}),\\
\mu_{12}^{a'b'a}(\Omega;\boldsymbol{k})&\longleftrightarrow\xi_{1+2}^{a'b'a}(\Omega;\boldsymbol{k}),\\
\mu_{22}^{a'b'ab}(\Omega;\boldsymbol{k})&\longleftrightarrow\xi_{2}^{a'b'ab}(\Omega;\boldsymbol{k}),\\
\end{aligned}
\end{equation*}
and note that 
\begin{equation*}
\begin{aligned}
&\mu_{21}^{a'ab}(\Omega;\boldsymbol{k})E_{-\Omega}^{a'}E_{\Omega/2}^{a}E_{\Omega/2}^{b}=\\
&\qquad\qquad\qquad\qquad\left(\xi_{1+2}^{a'b'a}(\Omega;\boldsymbol{k})E_{-\Omega/2}^{a'}E_{-\Omega/2}^{b'}E_{\Omega}^{a}\right)^*;
\end{aligned}
\end{equation*}
this is the complex conjugate contribution in (\ref{eq:1+2carriers_full}). Finally, introducing response coefficients for the injection rate as
\begin{align}
&\frac{d}{dt}\expval{\mathcal{M}(t)}{\psi(t)}=\nonumber\\
&\underset{\substack{a,...b;N\\a',...b';N'}}{\sum}\mu_{NN'}^{a'\cdots b'a\cdots b}(\Omega)E_{-\Omega/N'}^{a'}\cdots E_{-\Omega/N'}^{b'}E_{\Omega/N}^{a}\cdots E_{\Omega/N}^{b},\nonumber
\end{align} 
we find the response coefficient for a particular absorption process can be written as
\begin{equation*}
\begin{aligned}
\mu_{NN'}^{a'\cdots b'a\cdots b}(\Omega)=\int_{BZ}\frac{d^Dk}{\left(2\pi\right)^{D}}\mu_{NN'}^{a'\cdots b'a\cdots b}(\Omega;\boldsymbol{k}).
\end{aligned}
\end{equation*}
For the first example above, one has
\begin{align}
\xi_{1}^{a'a}(\Omega)=\int_{BZ}\frac{d^Dk}{\left(2\pi\right)^{D}}\xi_{1}^{a'a}(\Omega;\boldsymbol{k}).\nonumber
\end{align}
\vspace{0.51in}
\begin{widetext}
\section{Independent Response Tensor Components}
\label{Appendix}
\subsection{Carrier injection rate}
The nonvanishing, independent, and valley- and spin-dependent response tensor components corresponding to the carrier injection rate are given for the indicated photon absorption processes. The relation of these components to the carrier injection rate is given in Sec. \ref{sec:3}, and as there we use the notation $\xi_{n}=\sum_{\tau s}\xi_{n;\tau s}$. \\
\textit{One-photon absorption}
\begin{equation} 
\begin{array}{rl}
\xi_{1; \tau s}^{xx}(2\omega) & =\frac{\Theta(2\omega-2\Delta_{\tau s})e^{2}}{16\hbar^{2}\omega}\big(1+\frac{\Delta_{\tau s}^{2}}{\omega^{2}}\big)\\
\xi_{1; \tau s}^{xy}(2\omega) & =-i\tau\frac{\Theta(2\omega-2\Delta_{\tau s})e^{2}}{8\hbar^{2}\omega}\frac{\Delta_{\tau s}}{\omega}
\end{array}
\end{equation}
\textit{Two-photon absorption}
\begin{equation}
\begin{array}{rl}
\xi_{2; \tau s}^{xxxx}(2\omega) & =\frac{\Theta(2\omega-2\Delta_{\tau s})e^{4}\Xi^{2}}{\hbar^{4}\omega^{5}}\big(1-\frac{\Delta_{\tau s}^{2}}{\omega^{2}}\big)\big(\frac{1}{4}+\frac{3}{4}\frac{\Delta_{\tau s}^{2}}{\omega^{2}}\big)\\
\xi_{2; \tau s}^{xxxy}(2\omega) & =-i\frac{\Theta(2\omega-2\Delta_{\tau s})e^{4}\Xi^{2}}{2\hbar^{4}\omega^{5}}\tau\big(1-\frac{\Delta_{\tau s}^{2}}{\omega^{2}}\big)\frac{\Delta_{\tau s}}{\omega}\\
\xi_{2; \tau s}^{xxyy}(2\omega) & =-\frac{\Theta(2\omega-2\Delta_{\tau s})e^{4}\Xi^{2}}{4\hbar^{4}\omega^{5}}\big(1-\frac{\Delta_{\tau s}^{2}}{\omega^{2}}\big)^{2}\\
\xi_{2; \tau s}^{xyxy}(2\omega) & =\frac{\Theta(2\omega-2\Delta_{\tau s})e^{4}\Xi^{2}}{4\hbar^{4}\omega^{5}}\big(1-\frac{\Delta_{\tau s}^{4}}{\omega^{4}}\big)
\end{array}
\end{equation} 

\textit{1+3 absorption} 
\begin{equation}
\begin{array}{rl}
\xi_{1+3; \tau s}^{xxxx}(3\omega) & =-\frac{e^{4}}{\hbar^{4}}\frac{\Xi^{2}\Theta(3\omega-2\Delta_{\tau s})}{18\omega^{7}}\bigl[\left(2+\frac{1}{2}\right)\Xi^{2}k_{\tau s}^{2}-\frac{1}{6}\left(2^{2}+\frac{1}{2}\right)\frac{\Xi^{4}}{\omega^{2}}k_{\tau s}^{4}-\frac{9}{8}\omega^{2}\bigr]\\
\xi_{1+3; \tau s}^{xxyy}(3\omega) & =-\frac{e^{4}}{\hbar^{4}}\frac{\Xi^{2}\Theta(3\omega-2\Delta_{\tau s})}{18\omega^{7}}\bigl[\frac{1}{3}\left(2+\frac{1}{2}\right)\Xi^{2}k_{\tau s}^{2}-\frac{1}{18}\left(2^{2}+\frac{1}{2}\right)\frac{\Xi^{4}}{\omega^{2}}k_{\tau s}^{4}-\frac{3}{8}\omega^{2}\bigr]\\
\xi_{1+3; \tau s}^{xxxy}(3\omega) & =i\frac{e^{4}}{\hbar^{4}}\frac{\Xi^{2}\Theta(3\omega-2\Delta_{\tau s})}{18\omega^{7}}\frac{2\tau\Delta_{\tau s}}{3\omega}\bigl[\left(2+\frac{1}{2^{2}}\right)\Xi^{2}k_{\tau s}^{2}-\frac{9}{8}\omega^{2}\bigr]\\
\xi_{1+3; \tau s}^{xxyx}(3\omega) & =-i\frac{e^{4}}{\hbar^{4}}\frac{\Xi^{2}\Theta(3\omega-2\Delta_{\tau s})}{18\omega^{7}}\frac{2\tau\Delta_{\tau s}}{3\omega}\bigl[\frac{1}{3}\left(2+\frac{1}{2^{2}}\right)\Xi^{2}k_{\tau s}^{2}-\frac{3}{8}\omega^{2}\bigr]
\end{array}
\end{equation} 
	
\noindent \textit{Three-photon absorption} 
\begin{equation}
\begin{array}{rl}
\xi_{3; \tau s}^{xxxxxx}(3\omega)=&\frac{e^{6}}{\hbar^{6}}\frac{3\Theta(3\omega-2\Delta_{\tau s})}{2^{4}\Xi^{2}\omega^{9}}\Big[\frac{1}{4}\Xi^{6}-\Xi^{4}\left(\frac{2\Xi^{2}}{3\omega}\right)^{2}\left(2+\frac{3}{8}\right)k_{\tau s}^{2}+\Xi^{2}\left(\frac{2\Xi^{2}}{3\omega}\right)^{4}\left(9+\frac{9}{32}\right)k_{\tau s}^{4}-\left(\frac{2\Xi^{2}}{3\omega}\right)^{6}\left(20+\frac{1}{4}\right)\frac{15}{48}k_{\tau s}^{6}\Big]\\
\xi_{3; \tau s}^{xxxxyy}(3\omega)=&\frac{e^{6}}{\hbar^{6}}\frac{\Theta(3\omega-2\Delta_{\tau s})}{2^{4}\Xi^{2}\omega^{9}}\Big[\frac{1}{4}\Xi^{6}\left(2-(\frac{2\Delta_{\tau s}}{3\omega})^{2}\right)-\Xi^{4}\left(\frac{2\Xi^{2}}{3\omega}\right)^{2}\left[3+\frac{3}{4}-(\frac{2\Delta_{\tau s}}{3\omega})^{2}(1+\frac{1}{8})\right]k_{\tau s}^{2}+\Xi^{2}\left(\frac{2\Xi^{2}}{3\omega}\right)^{4}(5+\frac{11}{32})k_{\tau s}^{4}\\
&\qquad\qquad\qquad\quad-\left(\frac{2\Xi^{2}}{3\omega}\right)^{6}(3+\frac{3}{4}+\frac{3}{64})k_{\tau s}^{6}\Big]\\
\xi_{3; \tau s}^{xxyxxy}(3\omega)=&\frac{e^{6}}{\hbar^{6}}\frac{\Theta(3\omega-2\Delta_{\tau s})}{2^{4}3\Xi^{2}\omega^{9}}\Big[\frac{1}{4}\Xi^{6}\left(5-4(\frac{2\Delta_{\tau s}}{3\omega})^{2}\right)-\Xi^{4}\left(\frac{2\Xi^{2}}{3\omega}\right)^{2}\left[4+\frac{15}{8}-(\frac{2\Delta_{\tau s}}{3\omega})^{2}(2+\frac{1}{2})\right]k_{\tau s}^{2}+\Xi^{2}\left(\frac{2\Xi^{2}}{3\omega}\right)^{4}(21+\frac{29}{32})k_{\tau s}^{4}\\
&\qquad\qquad\qquad\quad-\left(\frac{2\Xi^{2}}{3\omega}\right)^{6}(9+\frac{9}{4}+\frac{9}{64})k_{\tau s}^{6}\Big]\\
\xi_{3; \tau s}^{xxxxxy}(3\omega)=&\frac{e^{6}}{\hbar^{6}}\frac{\Theta(3\omega-2\Delta_{\tau s})}{2^{4}\Xi^{2}\omega^{9}}\left(-i\tau\frac{2\Delta_{\tau s}}{3\omega}\right)\Big[\frac{1}{4}\Xi^{6}-\Xi^{4}\left(\frac{2\Xi^{2}}{3\omega}\right)^{2}\left(2+\frac{1}{4}\right)k_{\tau s}^{2}+\Xi^{2}\left(\frac{2\Xi^{2}}{3\omega}\right)^{4}\left(6+\frac{3}{2}+\frac{3}{32}\right)k_{\tau s}^{4}\Big]\\
\xi_{3; \tau s}^{xxxyyy}(3\omega)=&\frac{e^{6}}{\hbar^{6}}\frac{3\Theta(3\omega-2\Delta_{\tau s})}{2^{4}\Xi^{2}\omega^{9}}\left(-i\tau\frac{2\Delta_{\tau s}}{3\omega}\right)\Big[\frac{1}{4}\Xi^{6}-\Xi^{4}\left(\frac{2\Xi^{2}}{3\omega}\right)^{2}\left(2+\frac{1}{4}\right)k_{\tau s}^{2}+\Xi^{2}\left(\frac{2\Xi^{2}}{3\omega}\right)^{4}\left(2+\frac{1}{2}+\frac{1}{32}\right)k_{\tau s}^{4}\Big]\\
\xi_{3; \tau s}^{xxyxyy}(3\omega)=&\frac{e^{6}}{\hbar^{6}}\frac{\Theta(3\omega-2\Delta_{\tau s})}{2^{4}3\Xi^{2}\omega^{9}}\left(i\tau\frac{2\Delta_{\tau s}}{3\omega}\right)\Big[\frac{1}{4}\Xi^{6}(\frac{2\Delta_{\tau s}}{3\omega})^{2}-\Xi^{4}\left(\frac{2\Xi^{2}}{3\omega}\right)^{2}\left(2\right)k_{\tau s}^{2}+\Xi^{2}\left(\frac{2\Xi^{2}}{3\omega}\right)^{4}\left(-8+\frac{1}{2}-\frac{3}{32}\right)k_{\tau s}^{4}\Big]
\end{array}
\end{equation}

\subsection{Current injection rate}
The nonvanishing, independent, and valley- and spin-dependent response tensor components corresponding to the current injection rate are given for the indicated photon absorption processes. The relation of these components to the current injection rate is given in Sec. \ref{sec:3}, and as there we use the notation $\eta_{n}=\sum_{\tau s}\eta_{n;\tau s}$. \\

\noindent\textit{1+2 absorption} 
\begin{equation}
\begin{array}{rl}
\eta_{1+2; \tau s}^{xxxx}(2\omega)  &=i\frac{\Theta(2\omega-2\Delta_{\tau s})e^{4}\Xi^{2}}{2\hbar^{3}\omega^{3}}\big(1-\frac{\Delta_{\tau s}^{2}}{\omega^{2}}\big)\big(\frac{1}{4}+\frac{3}{4}\frac{\Delta_{\tau s}^{2}}{\omega^{2}}\big)\\
\eta_{1+2; \tau s}^{xxyy}(2\omega) &=i\frac{\Theta(2\omega-2\Delta_{\tau s})e^{4}\Xi^{2}}{8\hbar^{3}\omega^{3}}\big(1-\frac{\Delta_{\tau s}^{4}}{\omega^{4}}\big)\\
\eta_{1+2; \tau s}^{xyyx}(2\omega)  &=-i\frac{\Theta(2\omega-2\Delta_{\tau s})e^{4}\Xi^{2}}{8\hbar^{3}\omega^{3}}\big(1-\frac{\Delta_{\tau s}^{2}}{\omega^{2}}\big)^{2}\\
\eta_{1+2; \tau s}^{xxxy}(2\omega)  &=\frac{\Theta(2\omega-2\Delta_{\tau s})e^{4}\Xi^{2}}{2\hbar^{3}\omega^{3}}\tau\frac{\Delta_{\tau s}}{\omega}\big(1-\frac{\Delta_{\tau s}^{2}}{\omega^{2}}\big)\\
\eta_{1+2; \tau s}^{xxyx}(2\omega)  &=-\frac{\Theta(2\omega-2\Delta_{\tau s})e^{4}\Xi^{2}}{4\hbar^{3}\omega^{3}}\tau\frac{\Delta_{\tau s}}{\omega}\big(1-\frac{\Delta_{\tau s}^{2}}{\omega^{2}}\big)\\
\eta_{1+2; \tau s}^{xyyy}(2\omega) &=0
\end{array}
\end{equation}
%
%

\textit{2+3 absorption} 
\begin{equation}
\begin{array}{rl}
\eta_{2+3; \tau s}^{xxxxxx}(3\omega) & =i\frac{\Theta(3\omega-2\Delta_{\tau s})e^{6}}{\hbar^{5}}\frac{2^{3}}{3^{6}}\frac{\Xi^{8}}{\omega^{11}}\bigl[-\frac{9}{2}\frac{\omega^{2}}{\Xi^{2}}k_{\tau s}^{2}+15k_{\tau s}^{4}-5\frac{\Xi^{2}}{\omega^{2}}k_{\tau s}^{6}\bigr]\\
\eta_{2+3; \tau s}^{xxxyxy}(3\omega) & =i\frac{\Theta(3\omega-2\Delta_{\tau s})e^{6}}{\hbar^{5}}\frac{2^{3}}{3^{6}}\frac{\Xi^{8}}{\omega^{11}}\bigl[-\frac{3}{4}\frac{\omega^{2}}{\Xi^{2}}k_{\tau s}^{2}+\frac{47}{12}k_{\tau s}^{4}-\frac{\Xi^{2}}{\omega^{2}}k_{\tau s}^{6}\bigr]\\
\eta_{2+3; \tau s}^{xyyyxy}(3\omega) & =i\frac{\Theta(3\omega-2\Delta_{\tau s})e^{6}}{\hbar^{5}}\frac{2^{3}}{3^{6}}\frac{\Xi^{8}}{\omega^{11}}\bigl[-\frac{9}{4}\frac{\omega^{2}}{\Xi^{2}}k_{\tau s}^{2}+\frac{11}{4}k_{\tau s}^{4}-\frac{\Xi^{2}}{\omega^{2}}k_{\tau s}^{6}\bigr] \\
\eta_{2+3; \tau s}^{xxxxyy}(3\omega) & =i\frac{\Theta(3\omega-2\Delta_{\tau s})e^{6}}{\hbar^{5}}\frac{2^{3}}{3^{6}}\frac{\Xi^{8}}{\omega^{11}}\bigl[\frac{1}{2}k_{\tau s}^{4}-\frac{\Xi^{2}}{\omega^{2}}k_{\tau s}^{6}\bigr]\\
\eta_{2+3; \tau s}^{xxyyyy}(3\omega) & =i\frac{\Theta(3\omega-2\Delta_{\tau s})e^{6}}{\hbar^{5}}\frac{2^{3}}{3^{6}}\frac{\Xi^{8}}{\omega^{11}}\bigl[(3+\frac{1}{6})k_{\tau s}^{4}-\frac{\Xi^{2}}{\omega^{2}}k_{\tau s}^{6}\bigr] \\
\eta_{2+3; \tau s}^{xxyyxx}(3\omega) & =i\frac{\Theta(3\omega-2\Delta_{\tau s})e^{6}}{\hbar^{5}}\frac{2^{4}}{3^{7}}\frac{\Xi^{6}}{\omega^{11}}\bigl[-\Delta_{\tau s}^{2}k_{\tau s}^{2}+2\Xi^{2}k_{\tau s}^{4}-\frac{3}{2}\frac{\Xi^{4}}{\omega^{2}}k_{\tau s}^{6}\bigr] \\
\eta_{2+3; \tau s}^{xxxxxy}(3\omega) & =\frac{\Theta(3\omega-2\Delta_{\tau s})e^{6}}{\hbar^{5}}\frac{2}{3^{4}}\frac{\Xi^{8}}{\omega^{11}}\left(\tau\frac{2\Delta_{\tau s}}{3\omega}\right)\bigl[-\frac{\omega^{2}}{\Xi^{2}}k_{\tau s}^{2}+3k_{\tau s}^{4}\bigr]\\
\eta_{2+3; \tau s}^{xyyyxx}(3\omega) & =-\frac{\Theta(3\omega-2\Delta_{\tau s})e^{6}}{\hbar^{5}}\frac{2^{2}}{3^{4}}\frac{\Xi^{8}}{\omega^{11}}\left(\tau\frac{2\Delta_{\tau s}}{3\omega}\right)\bigl[-\frac{\omega^{2}}{\Xi^{2}}k_{\tau s}^{2}+k_{\tau s}^{4}\bigr]\\
\eta_{2+3; \tau s}^{xxxyxx}(3\omega) & =-\frac{\Theta(3\omega-2\Delta_{\tau s})e^{6}}{\hbar^{5}}\frac{2^{2}}{3^{5}}\frac{\Xi^{8}}{\omega^{11}}\left(\tau\frac{2\Delta_{\tau s}}{3\omega}\right)\bigl[-\frac{\omega^{2}}{\Xi^{2}}k_{\tau s}^{2}+3k_{\tau s}^{4}\bigr]\\
\eta_{2+3; \tau s}^{xxyyxy}(3\omega) & =-\frac{\Theta(3\omega-2\Delta_{\tau s})e^{6}}{\hbar^{5}}\frac{2}{3^{5}}\frac{\Xi^{8}}{\omega^{11}}\left(\tau\frac{2\Delta_{\tau s}}{3\omega}\right)\bigl[\frac{\omega^{2}}{\Xi^{2}}k_{\tau s}^{2}+k_{\tau s}^{4}\bigr]\\
\eta_{2+3; \tau s}^{xxxyyy}(3\omega) & =\frac{\Theta(3\omega-2\Delta_{\tau s})e^{6}}{\hbar^{5}}\frac{2^{3}}{3^{5}}\frac{\Xi^{8}}{\omega^{11}}\left(\tau\frac{2\Delta_{\tau s}}{3\omega}\right)k_{\tau s}^{4}\\
\eta_{2+3; \tau s}^{xyyyyy}(3\omega) & =0
\end{array}
\end{equation}
\end{widetext}
\bibliographystyle{apsrev4-1}
\bibliography{2+3TMDs.bib}

\end{document}